\documentstyle[11pt,epsf]{article}

\def\beq{\begin{equation}}
\def\eeq{\end{equation}}
\def\eq{\end{equation}}
\def\ba{\begin{eqnarray}}
\def\ea{\end{eqnarray}}

\def\centeron#1#2{{\setbox0=\hbox{#1}\setbox1=\hbox{#2}\ifdim
\wd1>\wd0\kern.5\wd1\kern-.5\wd0\fi
\copy0\kern-.5\wd0\kern-.5\wd1\copy1\ifdim\wd0>\wd1
\kern.5\wd0\kern-.5\wd1\fi}}
\def\ltap{\;\centeron{\raise.35ex\hbox{$<$}}{\lower.65ex\hbox{$\sim$}}\;}
\def\gtap{\;\centeron{\raise.35ex\hbox{$>$}}{\lower.65ex\hbox{$\sim$}}\;}
\def\gsim{\mathrel{\gtap}}

\def\singleandthirdspaced{\baselineskip=\normalbaselineskip\multiply
    \baselineskip by 130\divide\baselineskip by 100}

\def\dslash{\not{\hbox{\kern-2pt $\partial$}}}
\def\Dslash{\not{\hbox{\kern-4pt $D$}}}
\def\Oslash{\not{\hbox{\kern-4pt $O$}}}
\def\Qslash{\not{\hbox{\kern-4pt $Q$}}}
\def\pslash{\not{\hbox{\kern-2.3pt $p$}}}
\def\kslash{\not{\hbox{\kern-2.3pt $k$}}}
\def\qslash{\not{\hbox{\kern-2.3pt $q$}}}
\def\epsilonslash{\not{\hbox{\kern-2.3pt $\epsilon$}}}

\newcommand{\newc}{\newcommand}
\newc{\qbar}{{\overline q}}
\newc{\Kahler}{K\"ahler }
\newc{\deltaGS}{\delta_{\rm GS}}
\begin{document}
\begin{titlepage}
\begin{flushright}
{\large hep-th/0201256 \\ SCIPP-01/40 \\ SU-ITP-01/54}

\end{flushright}

\vskip 1.2cm

\begin{center}

{\LARGE\bf
Brane World Susy Breaking from String/M Theory
}

\vskip 1.4cm

{\large Alexey Anisimov$^1$, Michael Dine$^1$, Michael Graesser$^1$, and
 Scott Thomas$^{2}$}
\\
\vskip 0.4cm
{\it $^1$Santa Cruz Institute for Particle Physics,
     Santa Cruz CA 95064  } \\ \vskip 1pt
{\it $^2$Physics Department, Stanford University, Stanford, CA 94305  } \\
 \vskip 1pt




\vskip 4pt

\vskip 1.5cm

\begin{abstract}

String and M-theory realizations of
brane world supersymmetry breaking scenarios are considered in which
visible sector
Standard Model fields are confined on a brane,
with hidden sector supersymmetry breaking isolated on a distant brane.
In calculable examples with an internal manifold of
any volume the Kahler potential generically contains brane--brane
non-derivative
contact interactions coupling the visible and hidden sectors
and is not of the no-scale sequestered form.
This leads to non-universal scalar masses and without additional
assumptions
about flavor symmetries may in general induce dangerous
sflavor violation even though the Standard Model and supersymmetry
branes are physically separated.
Deviations from the sequestered form are dictated by bulk supersymmetry
and can in most cases be understood
as arising from exchange of bulk supergravity fields
between branes or warping of the internal geometry.
Unacceptable visible sector tree-level tachyons arise in many
models but may be avoided in certain classes of compactifications.
Anomaly mediated and gaugino mediated contributions to scalar masses are
sub-dominant except in special circumstances such
as a flat or AdS pure five--dimensional bulk geometry
without bulk vector multiplets.

\end{abstract}

\end{center}

\vskip 1.0 cm

\end{titlepage}
\setcounter{footnote}{0} \setcounter{page}{2}
\setcounter{section}{0} \setcounter{subsection}{0}
\setcounter{subsubsection}{0}

\singleandthirdspaced


\section{Introduction}
\label{sec:intro}

The origin of supersymmetry breaking
must ultimately be addressed by any supersymmetric theory of nature.
A plethora of scenarios for breaking supersymmetry and for the messenger
interactions which must transmit the breaking to the Standard Model fields
have been proposed.
Brane world supersymmetry breaking (BWSB) scenarios utilize
the very old idea that the Standard Model is confined to a brane
in a higher dimensional space.
In this case supersymmetry breaking may be isolated on a distant hidden
sector brane which is not in direct contact with the visible sector
Standard Model brane.
The messenger interactions which couple the
visible and hidden sectors then arise from exchange of fields
which reside in the bulk of the higher dimensional space.

In this paper BWSB
is investigated in consistent string and M-theory backgrounds which have a
geometric interpretation.
The general form of visible sector scalar and gaugino masses
which arise in such a scenario are addressed.
Tree-level scalar masses comparable to the gravitino mass
in general arise from exchange of bulk supergravity fields between the
visible and hidden sector branes in addition to
universal radiative anomaly mediated soft masses.
In almost all cases fields in the minimal bulk supergravity multiplet
(required by higher dimensional bulk supersymmetry) are sufficient
to generate the brane--brane couplings which lead to tree-level masses.
The scalar masses turn out generally to be non-universal.
Without additional assumptions about flavor symmetries
the squark and slepton mass eigenstates then need not be aligned
with the quark and lepton mass eigenstates, and dangerous flavor
violating processes can result.
This implies that geometric separation of the visible
and hidden sector branes within extra dimensions alone
is not sufficient to
solve the supersymmetric flavor problem.
The phenomenology of BWSB
then is similar
to standard supergravity (SUGRA) scenarios with hidden sector breaking.

In any scenario for supersymmetry breaking,
the messenger interactions
which couple the visible and hidden sectors largely determine
the form of the soft supersymmetry breaking terms and in turn
the superpartner mass spectrum.
In almost any scenario 
there are irreducible contributions for both scalar and gaugino masses
which are related by supersymmetry to anomalous violations
of scale invariance.
Anomalous one-loop contributions to gaugino masses
were noted some time ago \cite{macintire}, but their origin
was not fully understood, and it was thought that these contributions
were too small to be phenomenologically interesting.
Subsequently, Randall and Sundrum \cite{randallsundrum} and
Giudice,
Luty, Murayama and Rattazzi \cite{lutyetal}
provided a theoretical understanding of these masses in terms
of the supersymmetric multiplet of anomalies and regulator
dependence of the infrared theory.
They showed, systematically,
that there
are not only anomalous contributions to gaugino masses but, at two
loops, to scalar
masses as well.  The one-loop gaugino and two-loop scalar
anomaly mediated masses are given by
\beq
m_g = -b_0 {g^2 \over 16 \pi^2} m_{3/2}~~~~~~~
\tilde m_q^2 = {1 \over 2} c_0 b_0 \left( {g^2 \over 16 \pi^2} \right)^2
  \vert m_{3/2} \vert^2
  \label{anomalymasses}
\eeq
where $b_0$ and $c_0$ are the leading beta function and
anomalous dimension coefficients respectively (for vanishing
Yukawa couplings), and $m_{3/2}$ is the gravitino mass.
These contributions
may be thought of as arising from gauge mediation through the
ultraviolet regulator fields for which supersymmetry is broken
by a non-vanishing auxillary component for the conformal compensator.
Further theoretical insight into the anomaly has been provided by
the work of \cite{baggeretal,gaillardnelson}.
These authors provided a thorough
understanding of the nature of the anomaly, and also gave certain
conditions under which the one- and two-loop formulas
(\ref{anomalymasses})
are applicable.

Anomaly mediated contributions to soft masses are loop suppressed
and therefore generally unimportant unless tree-level
masses vanish or are suppressed by some mechanism.
It has been argued that in BWSB, tree-level masses vanish and that anomaly
mediation gives the dominant contributions \cite{randallsundrum}.
The argument is based on the assumption that tree-level
brane--brane interactions are absent since the visible and
hidden sector branes are not in direct physical contact.
However, such contact interactions do in fact arise from exchange of bulk
fields.
And as shown below, in almost every case fields in the minimal bulk
supergravity multiplet are sufficient to generate
tree-level brane--brane couplings.
Such interactions are suppressed by the volume of the internal
manifold or separation scale between the branes, and therefore might
naively appear unimportant for large volume.
However, the four-dimensional gravitino mass, which sets the scale
for supersymmetry breaking effects in the low energy theory, is
also suppressed by the volume,
\beq
m_{3/2}^2 \sim {F^2 \over V},
\label{gravitinomass}
\eq
where $F$ is an auxiliary expectation value on the hidden sector
brane.
So the magnitude of the tree-level scalar masses
are not suppressed relative to the gravitino mass.
Said another way, from the perspective of the low energy
theory, there is no sense in which the visible and hidden sector
branes are far apart.


In order to study BWSB
we utilize string and M-theory backgrounds which contain
separated branes with world volume fields which model the visible
and hidden sectors.
While many phenomenological issues of supersymmetry breaking
are inaccessible with our present understanding of string
and M-theories, one feature which can be addressed
with present technology
is the form of the Kahler potential which couples the
visible and hidden sectors.
This determines the form of the visible sector soft masses
arising from hidden sector supersymmetry breaking.
So string and M-theory backgrounds are well suited to address
generic features of the squark and slepton spectra, in particular
for BWSB.

In the next section we present simple macroscopic arguments based
on extended supersymmetry and tree-level inheritance for the form
of the lowest order Kahler potential in a variety of brane world
backgrounds, including Horava-Witten models with end of the world
branes, D-brane models, and pure five--dimensional supergravity
with end of the world branes. In general the Kahler potentials are
not of the sequestered form, although the no-scale form can be
inherited at lowest order in special circumstances. This section
includes a review of the results presented in
\cite{previouspaper}. In section \ref{sec:locality}, Kahler
potentials for backgrounds
with extended supersymmetry are derived from the microscopic
point of view within the high energy theory.
Interactions between visible and hidden sector branes
contained within the Kahler potentials
can be understood in this langauge
as arising from exchange of bulk supergravity fields.
The volume
dependence of brane--brane interactions is illustrated in some D-brane
examples.
In section 4 we discuss the corrections to the Kahler potential
which arise in backgrounds with less than maximal supersymmetry
from warping of the internal geometry in both
D-brane models as well as Horava-Witten models with end of the
world branes.
Flavor violating corrections in the latter models are also shown to have
implications for standard heterotic string compactifications.
In the D-brane models the classical bulk
warping corrections are quantum effects
from the four--dimensional point of view and demonstrate that
the Kahler potential is in general not protected in brane
world backgrounds.
We also discuss the relation between co-dimension one
brane world models with an AdS bulk and
a dual boundary field theory description.
Section 5 presents the role of moduli masses in the low
energy theory.
In section 6 the problem of gaugino masses in brane world
realizations is addressed.
The implications of the general form of the Kaher potential on the
brane world mechanism of gaugino mediation is considered in section 7.
The implications for the supersymmetric flavor problem in BWSB scenarios is
summarized in section 8.
In Appendix A the form of the inherited Kahler potentials
for untwisted states in different classes of orbifold compactifications
are presented.
In Appendix B the soft masses which arise from hidden sector
supersymmetry breaking with these Kahler potentials
are derived.
In many cases the extended supersymmetry of the underlying Kahler
potential leads to sum rules for scalar masses which imply the
existence of unacceptable tree-level tachyons;
the conditions for avoiding these are given.

\section{Kahler Potentials from Simple Arguments}
\label{section1}

The form of the
couplings between the visible and hidden sector branes
determines the visible sector soft terms arising from
hidden sector supersymmetry breaking.
These couplings may be studied in a number of different frames.
In Einstein frame the coefficient of the gravitational
Einstein term is by definition independent of all fields
and has canonical normalization.
In this frame, couplings between the branes which determine
visible sector scalar masses reside in the Kahler potential.
We therefore focus on the leading form of the Kahler potential
which arises in consistent calculable examples.

Before addressing the form of the Kahler potential in
Einstein frame it is instructive to consider the
so-called supergravity or conformal frame which is
particularly convenient for displaying couplings between
the branes.
This frame is defined by a Weyl rescaling of the metric
from the Einstein frame by
$g^{SG}_{\mu \nu} = e^{K/3} g^{E} _{\mu \nu}$
where $K$ is the Einstein frame Kahler potential.
The auxiliary $F$-components are also rescaled as
$F^{SG} =
e^{-K/6} F^{E}$.
The supergravity frame has the advantage that the Lagrangian for
the auxiliary fields closely resembles the analogous expression in
global supersymmetry.
There, a visible sector scalar
field acquires a tree-level soft mass only if it has
a direct $D$-term coupling with
a field which breaks supersymmetry. The same is true in
local supersymmetry in the supergravity frame.
In the local case the relevant supergravity frame Lagrangian
is \cite{cremmer}
\ba
{\cal L}& = & {f \over 6} {\cal R}_{4}
-\sum_{i \bar{j}} f_{i \bar{j}} \partial_{\mu} \varphi_i \partial_{\mu}
\varphi^* _{\bar{j}} - {1 \over 4 f}
(\sum_i f_i \partial_{\mu} \varphi_i -\hbox{h.c.})^2 + \cdots
\nonumber \\
& & + \sum_{i \bar{j}} f_{i \bar{j}} F_i F^* _{\bar{j}} +
\vert F_{\Phi} \vert ^2 f
+\sum _i (W_i F_i + f _i F_{\Phi}^{*} F_i  + 3 F_{\Phi} W +\hbox{h.c.})
\label{cremmerlag}
\ea
where
$f$ is the field dependent supergravity function which multiplies
the four-dimensional Einstein term, and $W$ is the superpotential.
Subscripts on scalar fields, $\varphi_i$, and auxiliary
components, $F_i$, label the different fields, while
subscripts on $f$ and $W$ indicate derivatives with respect to the
corresponding
scalar field.  $F_{\Phi}$ is the auxiliary component of the
conformal compensator superfield.
Note that current--current couplings between matter fields
proportional to
$(1/f)(f_i \partial_{\mu} \varphi_i - \hbox{h.c.})^2$
depend on first derivatives of the $f$ function, while
non-derivative couplings between matter fields proportional
to $f_{i \bar j} F_i F_{\bar j}^*$ depend on
the field dependence of two derivatives
of the $f$ function.
Specifically, the latter non-derivative couplings give contributions to
the soft masses of visible sector fields $Q_i$
from hidden sector auxiliary expectation values $F_{\Sigma_j}$
of the form
\beq
{\cal L} \supset
f_{Q_i \bar{Q}_{\bar{j}} \Sigma_k \bar{\Sigma}_{\bar{l}} }
  ~Q^*_{\bar j} Q_i F^*_{\Sigma_{\bar{l}}}  F_{\Sigma_k}
\label{fquadratic}
\eq
and are the analogs of direct $D$-term couplings between the visible
and hidden sector fields in the globally supersymmetric case.
For general $f$ both current--current and non-derivative couplings
arise between visible and hidden sector fields.
Note that since the current-current coupling in
supergravity frame between
charged fields depends on $(f_i)^2$, its lowest order
form already has four matter fields.
Knowing the current-current coupling to
this order does not then
uniquely determine the higher order non-derivative
terms (\ref{fquadratic}) which contribute to the soft masses
in this frame.
The supergravity $f$ function and Einstein frame Kahler
potential are related by
\beq
K= - 3 \ln ( -f/3).
\eeq

A special class of supergravity $f$ functions is of
the separable form
\beq
f(T_i,Q_i,\Sigma_i) =
f_{\rm vis}(Q_i)+  f_{\rm hid}(\Sigma_i) - f_{\rm mod}(T_i+T_i^{\dagger})
\label{fsep}
\eeq
where $Q_i$ and $\Sigma_i$ are visible and hidden sector fields
respectively and $T_i$ are moduli.
The imaginary components of moduli often transform
non-linearly under Peccei-Quinn symmetries.
Tree-level invariance under these symmetries then fixes the moduli
functional dependence to be $T_i + T_i^{\dagger}$ in the classical
supergravity $f$ function.
With a non-vanishing auxiliary expectation value for either
hidden sector fields, $F_{\Sigma_i}$, or moduli, $F_{T_i}$,
visible sector scalar masses vanish for a supergravity $f$
function of the separable form (\ref{fsep}).
If $f$ were the Kahler potential of global supersymmetry this statement
would be
obvious since with the separable form there
are no non-derivative couplings between the sectors
from the $f_{i \bar j} F_i F_{\bar j}^*$ terms.
In supergravity this result is still true
even though the different
sectors are indirectly coupled through the conformal compensator
auxiliary field $F_{\Phi} \sim m_{3/2}$.
This follows from (\ref{cremmerlag}) with the separable
form (\ref{fsep}) after integrating out
$F_{Q_i}$ because there is
a cancellation in the expression for the visible sector
soft masses between four terms each
proportional $\vert F_{\Phi} \vert^2$.
This also follows from the fact that at tree-level
there is a basis where the
conformal compensator only couples to operators in $f$
which are not bilinear in fields,
and to operators in the superpotential which are not tri-linear
in fields \cite{lutyetal}.
Note that with the separable form (\ref{fsep})
current--current interactions between visible and hidden sector
fields do exist, even though non-derivative couplings are absent.
So even in this special case, the visible and hidden sectors
are not actually decoupled.


A supergravity $f$ function of the separable
form (\ref{fsep}) has been referred to as ``sequestered.''
The Kahler potential in Einstein frame associated with
this sequestered form is of the no-scale type.
With canonical tree-level kinetic terms for the visible
and hidden sector fields,
$f_{\rm vis}=3 \hbox{tr}Q^{\dagger}_i Q_i$ and
$f_{\rm hid}=3 \hbox{tr}\Sigma^{\dagger}_i \Sigma_i$,
and only a single modulus $T$, the no-scale Kahler potential is
\beq
K = -3 \ln (
  f_{\rm mod}(T+T^{\dagger}) - \hbox{tr}Q^{\dagger}_i Q_i
  -  \hbox{tr}\Sigma^{\dagger}_i \Sigma_i ).
\label{sequesteredkahler}
\eeq
A non-vanishing auxiliary component for either a hidden sector
field, $F_{\Sigma _i}$, or modulus, $F_{T}$, does
not give rise to visible sector scalar masses.
In Einstein frame, this vanishing of visible sector scalar masses
from (\ref{sequesteredkahler})
involves a seemingly miraculous cancellation depending crucially
on the functional form of the logarithm and prefactor of `3.'
However, as discussed above, this follows from the separable
form of the supergravity $f$ function in supergravity frame.

For the question of visible sector scalar masses in
BWSB, an understanding
of the form of the supergravity
$f$ function in supergravity frame,
or equivalently the Kahler potential in Einstein frame,
is clearly crucial.
In the brane world scenario, visible and hidden sector
fields reside on physically separated branes.
It has been argued \cite{randallsundrum,lutysundrum}
that the separable form
(\ref{fsep}) might then appear naturally since visible and hidden
sector fields are not in direct physical contact
in the microscopic theory.
In the following subsections we
investigate the form of the supergravity $f$ function
or equivalently the Kahler
potential which arises in string and M-theory
models of BWSB.
%
%
Even at the leading order
the no-scale sequestered form is not generally obtained, and
unsuppressed non-universal tree-level visible sector scalar masses result from
hidden sector supersymmetry breaking.
The lowest order supergravity $f$ function is generally not of the separable
form and gives unsuppressed non-derivative couplings between branes,
even though the branes are physically separated.
We also give a simple argument for the form of the leading Kahler
potential arising from pure five--dimensional supergravity
with end of the world branes.
Some corrections to the Kahler
potentials are discussed in section \ref{sec:corrections}.



\subsection{Horava-Witten Theory}
\label{sec:hwsimple}


The first brane world model was the Horava-Witten compactification
of M-theory on an $S^1/Z_2$ interval \cite{hv}.
In order to cancel gravitational anomalies, $E_8$ gauge degrees of freedom
must be introduced at the two fixed points which bound the interval.
The $E_8$ gauge supermultiplets then reside on end of the world
branes separated by a distance $R_{11}$,
and may be thought of as twisted states of this M-theory background.
These end of the world branes may be identified with the
visible and hidden sector branes.
In the limit $R_{11} \ll \ell_{11}$, where $\ell_{11}$ is the
eleven--dimensional Planck length, the Horava-Witten compactification
of M-theory reduces to weakly coupled $E_8 \times E_8$
heterotic string theory in ten dimensions.
The strongly coupled limit,
$R_{11} \gsim \ell_{11}$, provides a realization of the brane world picture.

Compactification of the Horava-Witten theory to four
dimensions on a Calabi-Yau
space has been studied extensively \cite{wittency,ovrutetal}.
But to explore the structure of four--dimensional
Kahler potentials we first consider the simpler case of toroidal
compactification on $S^1/Z_2 \times T^6$.
In this case, all of the states of the $S^1/Z_2$ Horava-Witten background,
including the gauge supermultiplets,
survive compactification to four dimensions.
While this is not a realistic phenomenological
compactification since it preserves $N=4$ supersymmetry in four dimensions,
it demonstrates that even with high degree of symmetry,
the sequestered form is not obtained.
It also illustrates the origin of certain features of inherited
Kahler potentials in more realistic compactifications presented below
which preserves only $N=1$ in four dimensions.

Compactification on $S^1/Z_2 \times T^6$ can be described by three complex
coordinates $z_i$, $i=1,2,3$.
States may be organized in four--dimensional $N=1$ multiplets.
At the level of two derivatives the four dimensional theory
has a $SU(4)$ global $R$-symmetry.
In an $N=1$ description only a $SU(3) \times U(1)_R$ subgroup
is manifest.
In this description the $SU(3)$ is a manifest global flavor symmetry
of the low-energy theory.
The four--dimensional chiral multiplets include
geometric moduli $T_{i \bar j}$ transforming as
${\bf 8}_0 \oplus {\bf 1}_0 \in SU(3) \times U(1)_R$ and
$T_{ij}$ transforming as
${\bf 6} \oplus \overline{{\bf 3}} \in SU(3) \times U(1)_R$,
and the dilaton $S$ which is invariant.
These moduli are related to the geometric parameters of the
compactification in eleven--dimensional Planck units by
$S + S^{\dagger} \sim V_6$ and
$T \equiv [{\rm det}(T_{i \bar j} + T_{i \bar j})]^{1/3}
\sim R_{11} V_6^{1/3}$.
The eleven--dimensional gravitational multiplet
reduces to the four--dimensional $N=4$ gravitational
multiplet plus 6 $U(1)$ $N=4$ vector multiplets.
The $E_8 \times E_8^{\prime}$ gauge multiplets reduce to
visible and hidden sector brane fields $Q_i$ and $\Sigma_i$
transforming as
${\bf 3}_+ \in SU(3) \times U(1)_R$ and
${\bf 248} \in E_8$, and ${\bf 248} \in E_8^{\prime}$
respectively.

The four--dimensional Kahler potential is exact and not renormalized
with $N=4$ supersymmetry \cite{kahlera,kahlerb,polchinski}.
The Kahler potential is
therefore exact for the case of toroidal compactification.
The four--dimensional Kahler potential in the strongly coupled
brane world limit, $R_{11} \gsim \ell_{11}$, is then
identical in this case to that of the weakly coupled heterotic string
compactified on $T^6$.
Non-renormalization of the Kahler potential
can also be obtained by considering a strongly coupled
ten--dimensional limit
$R_i \gg R_{11} \gsim \ell_{11}$, where $R_i$ are the $T^6$ radii.
The Kahler potential is determined by two derivative
terms in the low energy theory.
It can therefore be obtained from the lowest order ten--dimensional
supergravity Lagrangian with gauge group $E_8 \times E_8$.
This is identical to that obtained in the weakly coupled ten--dimensional
limit, $R_i \gg  \ell_{11} \gg R_{11}$,
since the ten--dimensional supergravity Lagrangian is unique
at the level of two derivatives.
By any of these methods the four--dimensional
Kahler potential is \cite{kahlera,kahlerb,polchinski},
\beq
K= -\ln \hbox{det}(T_{i \bar j} + T^{\dagger}_{i \bar j} -
\hbox{tr} Q_i Q_{\bar j}^{\dagger} -
\hbox{tr} \Sigma_i \Sigma_{\bar j}^{\dagger}) -\ln(S + S^{\dagger}
)
\label{tdeteqn}
\eeq
where the two traces are over $E_8$ and $E_8^{\prime}$
gauge groups respectively and
the dependence on the $T_{ij}$ moduli is suppressed. Note that
this result is explicitly $SU(3)\times U(1)_R$ invariant.
The microscopic origin of this expression is explained
in section \ref{hw10d}.
The supergravity $f$ function associated with the Kahler
potential (\ref{tdeteqn}) is
\beq
f = -3 \left[ (S + S^{\dagger})
 \hbox{det}  (T_{i \bar j} + T^{\dagger}_{i \bar j} -
 \hbox{tr} Q_i Q_{\bar j}^{\dagger} -
\hbox{tr} \Sigma_i \Sigma_{\bar j}^{\dagger})
 \right]^{1/3}  ~.
 \label{nfourf}
\eq
The Kahler potential (\ref{tdeteqn}) is not of the no-scale
sequestered form and the $f$ function (\ref{nfourf}) is clearly not
separable.
This is true even ignoring the dilaton.
So we see that in this highly symmetric brane world model
the sequestered intuition breaks down even without the inclusion
of corrections which would generically be
present in more realistic models with less supersymmetry.

The Kahler potential (\ref{tdeteqn}) or supergravity
$f$ function (\ref{nfourf}) imply the existence
of non-derivative couplings, proportional to
$f_{i \bar j} F_i F^*_{\bar j}$ in (\ref{cremmerlag}), between visible
and hidden sector fields which reside on the two end of the
world branes.
These interactions are tree-level and unsuppressed in any way
in the four--dimensional theory.
The visible sector soft masses arising from hidden sector
auxiliary expectation values with the Kahler potential
(\ref{tdeteqn}) are derived in Appendix B.
Assuming that the moduli are stabilized with vanishing
auxiliary components by superpotential
interactions and that the cosmological constant vanishes,
the eigenvalues of the visible sector scalar mass squared
matrix in the $3 \times 3$ $SU(3)$ flavor space in this
case are
\beq
m^2_{Q_i} = m^2_{3/2} (-2,1,1) ~.
\label{nfoursoft}
\eeq
These soft masses are proportional to the gravitino mass,
$m_{3/2}$, and are not suppressed by the compactification
volume or brane separation (relative to the gravitino mass).
In addition the tree-level masses are not degenerate.
So degeneracy and universality can not be guaranteed
simply by separating visible and hidden sectors on different branes.
The soft scalar masses squared, in this case, satisfy $\hbox{Tr}~m^2=0$.
As described in Appendix B, this special sum rule is the result
of the $N=4$ supersymmetry of the Kahler potential and is not
necessarily obtained in more realistic models with less
supersymmetry.

Phenomenologically viable models should possess at lowest order only $N=1$
supersymmetry in four dimensions.
The extended supersymmetries of critical string or M-theory
must therefore be broken in compactification to four dimensions.
This can be achieved by compactifying on a manifold which
preserves only 4 supersymmetries.
For simplicity we consider compactifications of the
Horava-Witten M-theory background of the form $S^1/Z_2 \times {\cal M}$,
where ${\cal M}$ is a Calabi-Yau manifold or orbifold,
although the conclusions may be applicable to more general
compactifications on $G_2$ manifolds for example.


With only $N=1$ supersymmetry in four dimensions
the lowest order form of the Kahler potential can
in general receive corrections.
In some cases, however, the corrections to the superpotential
and Kahler potential might be argued to be small in the brane world
limit, $R_{11} \gsim \ell_{11}$.
To illustrate this consider first
the case in which ${\cal M}$ is $K3 \times T^2$
which preserves 8 supersymmetries which gives $N=2$ supersymmetry in
four dimensions.
In this case the Kahler potential is derivable from a
holomorphic prepotential.
Holomorphy and the classical Peccei-Quinn shift symmetries
imply that there are no $S$ or $T$ dependent corrections
to the prepotential at any order in perturbation theory,
where here $T$ refers generically to any Kahler moduli.
However, the shift symmetries allow non-perturbative
corrections which are exponentials of functions of $-S$ and $-T$.
In the perturbative heterotic string limit these arise from
gauge instantons proportional to ${\rm exp}(-S)$ and string
world sheet instantons wrapping two-cycles proportional
to ${\rm exp}(h(-S,-T))$, where the function
$h(-S,-T)$ depends on the shift symmetries.
The perturbative Kahler potential
then receives exponentially small corrections
in the weakly coupled heterotic string theory limit corresponding to
$R_{11} \sim T/ S^{1/3} \ll \ell_{11}$,
with $S \gg T^3 \gg 1$,
where the perturbative string coupling is $g_s^{2/3} \sim R_{11}$.
It is then possible to move to the strongly coupled brane world
limit, $R_{11} \gg \ell_{11}$ with
$T^3 \gg S \gg 1$, while keeping the corrections exponentially small.
So the perturbative prepotential receives small corrections
also in this strongly coupled limit \cite{banksdine}.
In this limit the non-perturbative corrections arise
from M5 branes wrapping $K3 \times T^2$ proportional to
${\rm exp}(-S)$ and M2 branes wrapping two cycles of
$K3 \times T^2$ and extended in $S^1/Z_2$
proportional to ${\rm exp}(h(-S,-T))$.
For $S,T \gg 1$
the lowest order form of the Kahler potential obtained from
perturbative heterotic string theory on $K3 \times T^2$
in this case is then identical,
up to exponentially small corrections, to that for
M-theory on $S^1/Z_2 \times K3 \times T^2$ in the brane
world limit.
This illustrates that the leading form of the
Kahler potential for Horava-Witten
brane world backgrounds can be obtained from weakly coupled heterotic
string theory in certain regions of moduli space, at least
for compactifications which preserve 8 supersymmetries.

For an ${\cal M}$ which preserves only 4 supersymmetries a
similar arguement utilyzing holomorphy and symmetries
can be made for the lowest order form of the superpotential
in a Horava-Witten brane world limit.
The Kahler potential can, however, in principle receive
non-holomorphic corrections which are not restricted by
holomorphy and symmetries.
In fact, continuously connecting the weakly coupled heterotic
limit, $R_{11} \ll \ell_{11}$ to the strongly coupled
brane world limit, $R_{11} \gg \ell_{11}$, necessarily
involves passing through a region in the which the pertrubative
string coupling,
$g_s^{2/3} \sim R_{11}$, is not small.
In this region string non-holomorphic corrections are not
suppressed in any way with only 4 supersymmetries.
So in this case the weakly and strongly coupled
Kahler potentials can not be connected directly by any path
through moduli space.
However, for the case in which
${\cal M}$ is a Calabi-Yau manifold, the Kahler potential
at string tree-level to zeroth-order in
$\alpha^{\prime}$ is determined in the large volume limit,
$S \gg T^3 \gg 1$ by the classical geometry of the Calabi-Yau
manifold.
Likewise, in the large volume brane-world limit,
$T^3 \gg S \gg 1$ with $S \gg T$, the Kahler potential is also determined
by classical geometry up to volume suppressed quantum M-theory
corrections.
In this region of moduli space $V_6^{1/6} \gg R_{11} \gg \ell_{11}$.
Below the scale $R_{11}^{-1}$ the theory is effectively
ten--dimensional supergravity on a large Calabi-Yau, and so the
Kahler potential is determined by classical geometry.
So in this region of moduli space the lowest order
form of the Kahler potential obtained from perturbative heterotic
string theory on $CY$ is the same, up to power suppressed
corrections,
as that for M-theory on $S^1/Z_2 \times CY$ in the brane world
limit.

Now the tree-level Kahler potential
for heterotic string theory on a Calabi-Yau manifold
to zero-th order in $\alpha^{\prime}$ and
expanding in the fluctuating matter fields takes the general form
\beq
K = -\ln(S+ S^{\dagger}) + K(T,T^{\dagger}) +
  Z_{i \bar j} ~\varphi_i \varphi^{\dagger}_{\bar j}
 + {1 \over 2} Z_{i \bar{j} k \bar{l}}
   ~ \varphi_i \varphi_{\bar j}^{\dagger} \varphi_k
   \varphi_{\bar l}^{\dagger}
 + \cdots
\label{generalK}
\eq
where here $T$ refers generically to any of the $(1,1)$
Kahler moduli.
The matter fields $\varphi_i = Q_i$ or $\Sigma_i$ have
wave functions $Z_{ i \bar j} = Z_{ i \bar j}(T)$ and
quartic couplings
$Z_{i \bar{j} k \bar{l}} = Z_{i \bar{j} k \bar{l}} (T)$
which at tree-level are general functions of the (1,1) moduli.
For $(2,2)$ Calabi-Yau compactifications with the spin connection
embedded in the gauge connection
the tree-level form of the Kahler potential
is exact to all orders in perturbation theory, while
for $(0,2)$ compactifications with torsion general corrections
are allowed so that
at the perturbative level the matter field wave functions and
quartic couplings acquire $S$ dependence,
$Z_{ i \bar j} = Z_{ i \bar j}(S,T)$ and
$Z_{i \bar{j} k \bar{l}} = Z_{i \bar{j} k \bar{l}} (S,T)$.
The supergravity $f$ function for the Kahler potential (\ref{generalK})
contains a coupling between the visible and hidden sector
fields
\beq
f \supset (S + S^{\dagger})^{1/3} e^{-K(T, T^{\dagger})/3}
 \left(
  Z_{ Q_i \bar{Q}_{\bar{j}}  \Sigma_k \bar{\Sigma}_{\bar{l}}}
  - {1 \over 3}
  Z_{Q_i  \bar{Q}_{\bar{j}}}   Z_{\Sigma_k \bar{\Sigma}_{\bar{l}}}
  \right)
  ~ Q_i Q^{\dagger}_{\bar{j}} \Sigma_k \Sigma^{\dagger}_{\bar{l}}
  \label{generalf}
\eq
As long as
$Z_{ Q_i \bar{Q}_{\bar{j}}  \Sigma_k \bar{\Sigma}_{\bar{l}}}
\neq {1 \over 3}
 Z_{Q_i  \bar{Q}_{\bar{j}}}   Z_{\Sigma_k \bar{\Sigma}_{\bar{l}}}$
non-derivative couplings exist between the visible and hidden
sectors and  
non-vanishing visible sector soft masses arise from hidden sector
supersymmetry breaking.
The matter wave function $Z_{i \bar{j} } (T)$
can be calculated to all orders in perturbation theory from the tree-level
result for classes of (2,2) compactifications
and are generally moduli dependent.
The quartic couplings $Z_{i \bar{j} k \bar{l}} (T)$
have not been calculated but should also reasonably be
expected to be moduli dependent.
The combination $Z_{ Q_i \bar{Q}_{\bar{j}}  \Sigma_k \bar{\Sigma}_{\bar{l}}}
- {1 \over 3}
 Z_{Q_i  \bar{Q}_{\bar{j}}}   Z_{\Sigma_k \bar{\Sigma}_{\bar{l}}}$
is then likely to be non-vanish except perhaps
at isolated points on moduli space.
This should also generally be the case for $(0,2)$ compactifications.
So in the region of moduli space in which the corrections
to the zeroth-order values of the
wave and quartic couplings functions in (\ref{generalK})
are small in the Horava-Witten brane world limit, the visible
sector soft masses arising from hidden sector supersymmetry breaking
are very likely to be non-vansihing on a generic Calabi-Yau manifold.
Outside these regions of moduli space there will in general be
further corrections to these couplings and therefore further
corrections to the soft masses.
The leading corrections to
the zeroth-order Kahler potential
in the strongly coupled brane world limit are
proportional to $(T+T^{\dagger})/(S + S^{\dagger})$,
and are discussed in section \ref{sec:corrCY}.
These corrections generally violate flavor, so even
if the lowest order Kahler potential is flavor conserving,
at generic points on moduli space flavor violation can arise.


The Kahler potential
simplifies considerably for (2,2) Calabi-Yau compactifications
with $h^{1,1}=1$ and therefore only one Kahler modulus $T$.
In this case the Kahler potential including the $T$ modulus, dilaton,
and matter fields is fixed by the extra world sheet
supersymmetry to be the no-scale form \cite{polchinski}
\beq
K =  \ln(T + T^{\dagger}
- \hbox{tr} Q_{}^{\dagger} Q_{}
  )
 - \ln (S + S^{\dagger})
\label{nosimple}
\eeq
However, since these compactifications have the spin connection
embedded entirely in the visible sector gauge connection,
there are no hidden sector matter fields.
So this class of models is not useful for hidden sector
BWSB, but might be applicable to dilaton or moduli dominated
scenarios.


Another possibility for $S^1/Z_1 \times {\cal M}$ brane world
compactifications of M-theory is for ${\cal M}$ an orbifold.
The general conditions for consistent singular M-theory backgrounds
are not known, although in specific examples evidence for consistent
backgrounds with orbifold singularities and possibly M2 and M5 branes can
be found \cite{morb}.
However, for $S^1/Z_2 \times {\cal M}$ compactifications
in the weakly coupled heterotic string limit, $R_{11} \ll \ell_{11}$,
the consistency conditions based on moduli invariance of
the perturbative string description are well established,
and reviewed in Appendix A.
Although we can not demonstrate that a generic consistent
orbifold compactification
of heterotic string theory lifts to M-theory compactifications,
it seems reasonable that this is in fact the case,
and that chiral symmetry protects all chiral states
which are massless in the heterotic string theory
from gaining a mass in the strongly coupled brane world limit.
So with these caveats
we consider orbifold compactifications of Horava-Witten theory
of this type.

In an orbifold construction the states which survive in the low
energy theory are invariant under the orbifold action.
In general this action is non-trivial in both compact geometric
directions as well as in the gauge group of the underlying theory.
In addition, twisted states which reside at orbifold fixed
points also appear in the low energy theory.
For M-theory backgrounds $S^1/Z_2 \times {\cal M}$, a subset of
the $E_8 \times E_8^{\prime}$
gauge supermultiplets which reside on the end of the world
visible and hidden sector branes
survive in the low energy theory,
$Q_i \subset {\bf 248}_i \in E_8$ and
$\Sigma_i \subset {\bf 248}_i^{\prime} \in E_8^{\prime}$ respectively,
where $i=1,2,3$ labels the internal complex
coordinates of ${\cal M}$. From the weakly coupled heterotic string
point of view these
fields are untwisted states, and will be referred to as such below.
The lowest order tree-level Kahler potential for these
states is inherited directly from the $N=4$ Kahler
potential (\ref{tdeteqn}) by simply removing non-invariant states.
This can be obtained from the eleven-dimensional
supergravity solution for bulk fields
as discussed in Section \ref{sec:inhmorb}.
There are in general additional twisted states in the low energy theory
which reside at fixed points of ${\cal M}$.
The dependence of the Kahler potential on these twisted states
is not restricted by extended symmetries since ${\cal M}$ preserves
only $N=1$ supersymmetry.
Here we focus only on the $Q_i$ and $\Sigma_i$ visible and
hidden sector fields which do inherit a lowest order Kahler potential
from the ten--dimensional theory.

For an orbifold which preserves 4 supersymmetries,
the inherited Kahler potential for the untwisted states
receives corrections even in the weakly coupled heterotic limit.
It will also receive additional corrections in the brane world limit.
Unlike the case of a Calabi-Yau manifold in which the
latter corrections can be studied perturbatively in certain regions
of moduli space as described above, the general form of the corrections
are unknown in this case.
So the lowest order form of the Kahler potentials given below
may not literally be applicable to the brane world limit in these backgrounds.
However, since corrections are very likely to give yet additional contributions
to soft masses, the inherited Kahler potentials are instructive
in indirectly illustrating couplings between the visible and hidden sector
branes and the associated
lowest order irreducible contribution to the scalar masses.

The form of the inherited Kahler potential
for an $S^1/ Z_2 \times {\cal M}$ compactification which
preserves $N=1$ supersymmetry in four dimensions
is restricted by the moduli and
gauge quantum numbers of the matter fields.
The simplest case arises with
$T_{i \bar j}$ moduli and three generations
of visible and hidden sector
matter,
$Q_i$ and $\Sigma_i$, which then
have identical gauge quantum
numbers for each $i$, where $i=1,2,3$ are the complex coordinates
of ${\cal M}$ and also the flavor index.
A simple example of this type with ${\cal M}$ a symmetric $Z_3$
orbifold is given in the Appendix A.
In this class of compactifications the
off-diagonal combinations
of fields $Q_i Q^{\dagger}_{\bar j}$ and
$\Sigma_i \Sigma^{\dagger}_{\bar j}$ are gauge invariant and
can appear in the Kahler potential.
With all the $T_{i \bar j}$ moduli the inherited Kahler potential
and supergravity $f$ function are then identical to
(\ref{tdeteqn}) and (\ref{nfourf}) respectively.
Note that the inherited Kahler potential has a $SU(3)$
flavor symmetry in this case.
As discussed in Appendix B,
with any supersymmetry breaking
hidden sector auxillary expectation values, $F_{\Sigma_i} \neq 0$,
and assuming the moduli are stabilized with vanishing
auxillary expectation values,
the non-universal tree-level visible sector masses
given in (\ref{nfoursoft}) are obtained.

Another class of $S^1/Z_2 \times {\cal M}$ compactifications
have only diagonal $T_{i \bar i}$ moduli and
visible and hidden sector matter, $Q_i$ and $\Sigma_i$,
which have different gauge quantum numbers for each $i$.
In this case off-diagonal combinations are not
guaranteed to be gauge invariant, and in fact are not
if a given representation under the unbroken subgroup
arises only once.
A simple example of this type with ${\cal M}$ a $Z_6$ orbifold
which does not respect any permutation symmetries
is given in Appendix A.
In this class of compactifications, since the off-diagonal
combinations $Q_i Q_{\bar j}^{\dagger}$ are not gauge invariant
they can not appear in the Kahler potential.
The tree-level Kahler potential inherited from (\ref{tdeteqn})
in this case is a sum of logarithms
\beq
K = - \sum_i \ln(T_i + T_i^{\dagger}
- \hbox{tr} Q_{i}^{\dagger} Q_{i} -
  \hbox{tr} \Sigma_{i}^{\dagger} \Sigma_{i})
 - \ln (S + S^{\dagger}),
\label{kahlersumln}
\eeq
where the traces are over gauge qauntum numbers and are
in general different for each $i$.
This Kahler potential is invariant under an $S_3 \times U(1)_R$
global symmetry.
The supergravity $f$ function associated with the
Kahler potential (\ref{kahlersumln}) is
\beq
f = -3 \left[ (S + S^{\dagger})
 \prod_i (T_i + T_i^{\dagger} - {\rm tr}Q_i Q_i^{\dagger}
   -   {\rm tr}\Sigma_i \Sigma_i^{\dagger} )
  \right]^{1/3}
\label{fsumln}
\eq
The Kahler potential (\ref{kahlersumln}) is not of the no-scale
sequestered form and the $f$ function (\ref{fsumln}) is not
seperable even ignoring the dilaton.

The visible sector scalar masses arising from hidden
sector supersymmetry breaking with the Kahler potential
(\ref{kahlersumln}) are discussed in Appendix B.
Assuming the moduli are stabilized with vanishing auxillary
expectation values the scalar masses squared for each $i$ are
\beq
m_i^2 = m_{3/2}^2 (1-x_i)
\label{sumlnmass}
\eq
where $x_i = 2 {\rm Re} T_i | F_i|^2 / |W|^2$.
Vanishing of the cosmological constant implies $\sum_i x_i =3$
as shown in Appendix B.
Again in this case the soft masses are proportional
to the gravitino mass and are non-universal with the
detailed spectrum depending on the $x_i$.
The condition on the $x_i$ in this case implies that the sum of the
three mass squared eigenvalues vanish
\beq
m_1^2 + m_2^2 + m_3^2 =0
\label{sumlnsum}
\eq
However, since the multiplicities for each $i$ are not
necessarily the same, ${\rm Tr}~m^2 \neq 0$ in general.
The condition (\ref{sumlnsum}) implies the existence of
unacceptable tree-level visible sector tachyons.
However, in more realistic compactifications these
can be avoided by, for example, projecting out the dangerous
states.
A special case of the condition (\ref{sumlnsum}) is obtained
for both hidden sector fields with diagonal auxillary
expectation values, $F_i = F$, and
diagonal moduli expectation values, $T_i = T$.
Vanishing cosmological constant and unbroken $S_3$ then implies
$x_i=1$ and the scalar masses vanish, $m_i^2=0$.
It is important to note in this case,
however, the vanishing masses result from
an unbroken $S_3$ flavor symmetry rather than the form of the
Kahler potential.


As a final example consider a class of $S_1/Z_2 \times {\cal M}$
orbifold compactifications
which have moduli $T_{i \bar{j}}$ for $i=1,2$,
and $T_{3 \bar{3}}$, and visible and hidden sector chiral matter
$Q_{i}$, $\Sigma_{i}$.
Further, suppose there is a
$S_2$ permutation symmetry
for $i=1,2$, so that the low-energy theory has two generations
in $Q_{i=1,2}$ and $\Sigma_{i=1,2}$ ($Q_i$ and $\Sigma_i$ are charged
under different groups), and states $Q_3$ and $\Sigma_3$
which have different quantum numbers from the first two
generations.
A class of orbifold $Z_6$ orbifold
examples leading to this spectrum
is provided in the Appendix A.
In this class of compactifications
the off-diagonal elements $Q^{\dagger} _i Q_3$ for
$i=1,2$ are not gauge invariant and do not appear
in the Kahler potential.
The lowest order inherited tree-level Kahler
potential is then
\beq
K = - \ln \det_{i=1,2} \left(T_{i \bar{j}}+
T^{\dagger}_{i \bar{j}}  -
\hbox{tr} Q^{\dagger} _i Q_{\bar{j}}
- \hbox{tr} \Sigma^{\dagger} _i
\Sigma _{\bar{j}} \right)
- \ln \left(T_{3 \bar{3}} + T^{\dagger}_{3 \bar{3}} -
\hbox{tr} Q^{\dagger} _3 Q_{\bar{3}} - \hbox{tr}
\Sigma^{\dagger} _3 \Sigma_3
\right)  ~.
\eeq
Assuming that only $T_{i \bar{i}}$ acquire vevs
and that all the $T_{i \bar{j}}$ moduli
are stabilized with vanishing auxillary components,
then with
hidden sector supersymmetry breaking
the soft masses are
\beq
m^2= m^2_{3/2}(1,-2+x_3,1-x_3)
\eeq
where $x_3 \equiv 2 \hbox{Re}T_3 |F_{\Sigma_3}|^2/|W|^2 \leq 3$.
It is important
to note that the first two states have the same
gauge quantum numbers. Inspecting the mass eigenvalues
indicates that with
hidden sector supersymmetry breaking the
two generations are not generically
degenerate, and
one may be tachyonic.
Degeneracy occurs with unbroken or approximate $SU(2)$ flavor
symmetry in the hidden sector, and is lifted for
$x_3 = 3 -\epsilon$.
For small enough $\epsilon$ the first two generations are
approximately degenerate, but results from an approximate
$SU(2)$ flavor symmetry.

So we see that in general without unbroken flavor symmetries
tree-level non-universal scalar masses arise from
Horava-Witten BWSB.


\subsection{D-Branes}
\label{sec:Dbranesimple}

Large classes of brane world models can in principle be
constructed using perturbative string theory
D-branes on compact manifolds.
Examples of this type are instructive since,
unlike compactifications of the Horava-Witten theory,
the co-dimension of the internal manifold can be larger than one.

In order to illustrate a simple D-brane world model
consider first a toroidal compactification of type I string theory with
gauge group $SO(32)$ on $T^6$.
This preserves $N=4$ supersymmetry in four dimensions,
and is therefore of course not phenomenologically realistic
but again illustrates that even in this highly symmetric case
the sequestered intuition breaks down.
In the absence of Wilson lines, the four--dimensional Kahler
potential, in $N=1$ notation, is determined by $N=4$ supersymmetry
to be
\beq
K= -\ln
\det(T_{i \bar j} + T^{\dagger}_{i \bar j} -
{\rm tr}\varphi_i^{\dagger}
\varphi_{\bar j}) - \ln (S^{\dagger} + S),
\label{typeonek}
\eeq
where
$T_{i \bar j}$ are the geometric moduli,
$\varphi_i$ are the $N=1$ chiral matter fields arising
from the compactification of the ten dimensional gauge supermultiplets,
and the trace is over $SO(32)$ indices.
This result may also be obtained directly by compactification of the
ten--dimensional type I theory with
$T_{i \bar j}  = g^{\rm I}_{i \bar j} / \lambda_{\rm I}$ and
$S = V^{\rm I}_6 /  \lambda_{\rm I}$, where
$g^{\rm I}_{i \bar j}$ and $V^{\rm I}_6$ are
the $T^6$ complex metric and volume in type I string
frame, and $\lambda_{\rm I}$ is the type I string coupling.
Alternately it can also be obtained from the heterotic $SO(32)$
string theory.
This theory compactified on $T^6$ has the Kahler potential
(\ref{typeonek}) with $T_{i \bar j} = g^h_{i \bar j}$
and $S = V^h_6 / \lambda_h^2$, where $\lambda_h$ is the heterotic
string coupling.
Under type I--heterotic duality the
string couplings and metrics of the ten--dimensional Lagrangians
are related by $\lambda_h = 1/ \lambda_{\rm I}$ and
$g^h_{i \bar j} = g^{\rm I}_{i \bar j} / \lambda_{\rm I}$.


A brane world model in
type II theory with D-branes may be obtained
by a T--duality transformation of the type I theory.
Under T-duality on all the $T^6$ directions
\beq
R_i \rightarrow 1/R_i ~, ~~~ \lambda_I \rightarrow
\lambda _I /V
~.
\eeq
The $N=4$ supersymmetry and the invariance of
the low energy theory under this transformation
uniquely determines
the Kahler potential in the type I$^{\prime}$
theory compactified on $T^6$ to be identical to (\ref{typeonek}).

The type IIB theory has (including images)
32 D3 branes and 16 O3 orientifold planes
in a ${1 \over 2}$ BPS configuration which preserves
16 supersymmetries in the four dimensional theory.
Motions of the D3 branes away from one of the orientifold
planes correspond to Wilson lines in the type I description
which break the $SO(32)$ gauge symmetry to a product group.
Separating the D3 branes into two groups provides a model of
visible and hidden sector branes in the type I$^{\prime}$
description.
For example, 16 D3 branes (including images)
at each orientifold plane
gives $SO(16) \times SO(16)$ gauge group.
In this case, the four--dimensional matter fields break up into
visible and hidden sector fields, $\varphi_i = Q_i$ or $\Sigma_i$,
which reside on each group of D-branes.
The Kahler potential (\ref{typeonek}),
in this case, then has the same form (\ref{tdeteqn})
as that of the Horava-Witten theory as dictated by $N=4$ supersymmetry.
As discussed in the previous subsection, even in this
highly symmetric case the Kahler potential is not of the
no-scale sequestered form.
Supersymmetry breaking isolated in matter fields on the hidden
sector brane would in general then lead to unsuppressed tree-level
visible sector scalar masses of order the gravitino mass.
Note that the branes in this example are co-dimension six, unlike
the Horava-Witten example of the previous subsection
in which the end of the world branes
are co-dimension one.

The Kahler potential with $N=4$ supersymmetry is uncorrected.
So the inherited Kahler potential (\ref{typeonek}) is exact
for compactification of type I or I$^{\prime}$ theories on $T^6$.
However, in a more realistic model
compactified on a manifold ${\cal M}$ which preserves only
$N=1$ supersymmetry, the inherited Kahler potential is not
protected and in general receives corrections in the low energy theory.
In the type I picture there potentially are
one-loop quantum corrections proportional to inverse powers of the
Wilson line which breaks to the product gauge group.
Alternately, in the type I$^{\prime}$ picture these one-loop quantum
corrections may be understood as arising from integrating out
massive string states which stretch between the visible
and hidden sector branes.
In the closed string channel this one-loop quantum amplitude
amounts to tree-level interaction of the visible and hidden
sector branes through exchange of bulk closed string states.
In the limit of large separation this is dominated by
exchange of bulk supergravity fields.
The explicit form of Kahler potential corrections in a
${1 \over 4}$ BPS D-brane configuration which preserves
8 supersymmetries is illustrated in section \ref{kahlerD}.
For general compactifications with only $N=1$ supersymmetry,
the corrections should be expected to
give further contributions
to the tree-level masses.

As in the heterotic Horava-Witten example, visible sector
scalar masses arising from supersymmetry breaking on the hidden
sector brane in D-brane realizations of
BWSB
are generally already present at tree-level, are non-universal,
and are parameterically of the same order as the gravitino mass,
$m_Q^2 \sim m_{3/2}^2$.
This would hold generally for D-brane models of any co-dimension.


\subsection{Pure Five--Dimensional Supergravity}
\label{sec:puresimple}

A BWSB situation which can not at present by analyzed directly within the
framework of string or M-theory backgrounds is pure five--dimensional
supergravity on an interval with end of the world branes on which the
visible and hidden sectors reside.
The pure five--dimensional example turns out to be very special
for a number of
reasons.
First, the minimal five--dimensional supergravity multiplet contains
only a single gauge field.
The only other higher dimensional case in which this arises
is that of eleven dimensions.
Geometric compactifications of M-theory generally result
in a number of vector and antisymmetric
tensor fields as well as scalar moduli and therefore cannot lead to the pure
five-dimensional supergravity theory.
A more promising possibility for obtaining a five dimensional BWSB
model might be to consider asymmetric orbifold compactifications
in type II string theory since these generally possess far fewer moduli.
Still, these compactifications always contain a dilaton.
At best one could hope that the
pure five dimensional theory is the strongly coupled description
of an asymmetric orbifold compactification to four dimensions, with the
dilaton equivalent to the radion of the five dimensional theory.
At present, however, no realization of this possibility
has been constructed \cite{harvey}.

The most important feature of pure five--dimensional supergravity
is that in compactification of
five to four dimensions, the single real radius modulus, $R$,
for the compact volume turns out not to possess a kinetic term.
This is not the case for compactification of higher dimensions
down to four dimensions.
The imaginary component partner of the four--dimensional radius
modulus is the periodic Wilson line for the single $U(1)$ gauge
boson of the five--dimensional supergravity multiplet.
The absence of a kinetic term for the real component,
along with the classical Peccei-Quinn symmetry for the imaginary
component, implies that the Einstein frame tree-level four--dimensional
Kahler potential for the volume modulus without any brane matter is
\beq
K = - 3 \ln (T + T^{\dagger}),
\label{kahlerfive}
\eeq
or equivalently that $f_{\rm mod} = -3 ( T + T^{\dagger})$ in
supergravity frame,
where $T+T^{\dagger}= 2 R$.
As an aside, note also that
the supergravity frame, in this case, happens to be
the same as the geometric frame obtained by simply
dimensionally reducing the five dimensional theory
(or in string theory language
the string frame) since $f=-6R$, which again is special to the case
of compactification of five to four dimensions.

Compactification of pure minimal five--dimensional supergravity
on $S^1$ gives pure $N=2$ supergravity in four dimensions.
This theory possess an SL$(2,R)$ non-compact symmetry
at the classical level which acts on the radius modulus and implies
that the moduli space metric is quaternionic.
The Kahler potential (\ref{kahlerfive}) is consistent with this
requirement.
For the orbifold compactification $S^1/Z_2$, with projections
to give $N=1$ in four dimensions, the radius modulus (which
necessarily survives the orbifold projection) inherits
at lowest order the tree-level Kahler potential (\ref{kahlerfive}).

For a brane world realization, compactification from five to
four dimensions on an $S^1/Z_2$ interval allows the introduction
of four--dimensional $N=1$ chiral multiplet matter fields on
end of the world branes at the orbifold fixed points which bound the
interval.
The end of the world branes may be interpreted as the visible
and hidden sector branes.
The four--dimensional tree-level Kahler potential including
hidden and visible sector brane matter cannot be derived
from consistency of the low-energy four--dimensional
theory alone.
In principle it could be derived microscopically
from an underlying string
or M-theory background as in the Horava-Witten and D-brane examples
of the previous subsections if a compactification
to pure five--dimensional supergravity (plus perhaps a multiplet
which contains the dilaton)
on the interval $S_1 / Z_2$ were known.
However, if an example of this type exists
in which the visible and hidden sector brane matter are
remnants of $N=4$ states in the underlying theory, then
as described in Appendix B, the lowest order
tree-level Kahler potential inherited from the
$N=4$ form (\ref{tdeteqn}) would presumably
contain
\beq
K \supset -3 \ln(T + T^{\dagger} - \hbox{tr} Q_i^{\dagger} Q_i -
 \hbox{tr} \Sigma_i^{\dagger} \Sigma_i)  ~.
\label{fivenoscale}
\eq
This is the no-scale sequestered form of the Kahler potential.
The associated supergravity $f$ function, in this case,
is of the separable form
\beq
f = - 3 ( T + T^{\dagger}  - \hbox{tr} Q_i^{\dagger} Q_i -
 \hbox{tr} \Sigma_i^{\dagger} \Sigma_i )   ~.
\label{noscalef}
\eq
Even with the separable form (\ref{noscalef})
the branes are not decoupled since there exist current--current
couplings in (\ref{cremmerlag}) proportional to
$(1/f)(f_{Q_i} \partial_{\mu} Q_i - \hbox{h.c.})
(f_{\Sigma_j} \partial_{\mu} \Sigma_j - \hbox{h.c.})$
which couple fields on the two branes \cite{lutysundrum}.
As mentioned in section \ref{section1},
the lowest order form of the current--current couplings
in supergravity frame are not sufficient to fix the non-derivative
couplings in this frame.
The no-scale form can therefore not be derived from pure five--dimensional
supergravity with end of the world brane matter by simply
matching the lowest order supergravity frame
current-current couplings to the effective
four--dimensional theory without additional assumptions
about the bulk--brane couplings.
Here, the Kahler potential (\ref{fivenoscale}) and
$f$ function (\ref{noscalef}) follow at leading
order from inheritence of the underlying theory
with flat Kahler metric.

With supersymmetry breaking isolated on the hidden sector
brane, the no-scale Kahler potential (\ref{fivenoscale})
gives rise to vanishing tree-level visible sector masses,
$m_i^2=0$.
The separable form
for the radion modulus supergravity $f$ function,
or equivalently the Kahler potential (\ref{kahlerfive})
and the presumed inherited form of the Kahler potential
with end of the world brane matter (\ref{fivenoscale})
is a direct result of the fact that the volume modulus
for compactification from five to four dimensions does not
possess a kinetic term.
This occurs only for compactification from five
to four dimensions.
Therefore, vanishing tree-level soft masses arising with generic
hidden sector
supersymmetry breaking from the no-scale Kahler
potential (\ref{fivenoscale}) should clearly be thought
of as a property
of a particular hypothetical model rather than a general
feature of brane world supersymmetry breaking.

With $N=1$ supersymmetry in four dimensions the Kahler potential
is not protected from corrections. It has been argued that with
pure five--dimensional supergravity BWSB corrections to the
no-scale form of the Kahler potential (\ref{fivenoscale}) occur
only at one--loop and are suppressed by additional powers of the
brane separation \cite{randallsundrum,lutysundrum}. This is only
true if the brane tensions vanish, which seems a rather strong
additional assumption. In the general case there are tree-level
corrections to the Kahler potential from bulk warping and
therefore to the tree-level soft masses, as discussed in section
\ref{fivedwarp}. This also occurs in the Horava--Witten theory
compactified on a Calabi-Yau three--fold discussed in section
\ref{sec:corrCY}. In contrast, a brane world model with bulk
warping but a sequestered Kahler potential occurs in the
supersymmetric Randall-Sundrum model with two branes (and pure
anti-de-Sitter bulk) \cite{lutysundrumads}. This sequestering  can
be understood from the AdS/CFT correspondance \cite{ls2001}. We
return to this example in Section 4, where we speculate that the
sequestered form can be spoiled by the presence of additional bulk
vectors.

So we conclude from the above simple arguments for the form of
Kahler potentials
that the sequestered intuition and associated vanishing
tree-level scalar masses are not generally realized in BWSB.
Anomaly mediated supersymmetry breaking therefore does not
seem to be a generic or robust feature of BWSB
scenarios, but might be obtained in very special models,
or as the result of unbroken flavor symmetries.


\section{Locality and the Low Energy Effective Action}
\label{sec:locality}

Couplings between branes which are physically separated
within a compact manifold manifest themselves as
contact interactions in the low energy theory below the
compactification scale.
As mentioned in section \ref{sec:intro}, such interactions
do not in fact violate notions of bulk locality since
brane--brane interactions can arise already at tree-level
from exchange of bulk fields between the branes.
We have seen that
these brane--brane tree-level interactions in general violate the
naive sequestered expectation that the supergravity
$f$ function is separable or equivalently that the
Kahler potential is of the no-scale form.
In this section the microscopic origin of brane--brane
interactions from exchange of bulk fields is presented
in a number of calculable, controlled examples.
We begin with examples with $N=4$ supersymmetry in four dimensions.
The Kahler potentials for both the heterotic and type I or I$^{\prime}$
examples derived in the previous section from general arguments are
shown to arise directly from exchange of bulk
supergravity fields.
The existence of these bulk fields is guaranteed by
higher dimensional supersymmetry.
Even though these $N=4$ examples are not phenomenologically
realistic they have the virtue
that explicit computations are straightforward,
and already the sequestered intuition breaks down at tree-level.
As discussed in section \ref{section1} these features are
inherited by the lowest order tree-level Kahler potential of
general $N=1$ compactifications.
Additional tree-level corrections to the brane Kahler potential
from exchange of bulk fields in examples with less supersymmetry
are discussed in section \ref{sec:corrections}.


\subsection{The Strongly Coupled Heterotic Theory in the Ten Dimensional
Limit}
\label{hw10d}


The microscopic origin of brane--brane interactions from the
exchange of bulk fields is well illustrated by the
$S^1/Z_2$ Horava-Witten M-theory orbifold background.
The $S^1/Z_2$ M-theory interval has length $R_{11}$
with $E_8$ gauge supermultiplets localized on end of the world
brane boundaries.
Consider first the theory
at energy scales below the inverse interval length $R_{11}^{-1}$.
The action of the low energy ten--dimensional
supergravity theory coupled to $E_8 \times E_8^{\prime}$
super Yang-Mills at these scales includes a
kinetic term for the NS field strength proportional to
\beq
\int d^{10} x ~e^{-2 \phi_{10}}
~H^2 = \int d^{10} x ~e^{-2 \phi_{10}}
~(dB + \omega_{E_8} + \omega_{E_8^\prime} -\omega_L)^2,
\label{Heq}
\eeq
where
$\omega_{E_8}$ and $\omega_{E_{8^\prime}}$ are the Chern-Simons
terms
associated with each $E_8$, and where $\omega_L$ is
the Lorentz Chern-Simons form.  From an eleven-dimensional
point of view, $E_8$ and $E_8^{\prime}$
lie on different end of the world
branes, separated
by the distance $R_{11}$.
At low energies this theory must reduce to the
ten-dimensional theory above.
This low energy theory clearly has tree-level
contact interactions which couple the Chern-Simons forms
on each brane.

The way in which the brane--brane contact interaction
comes about was explained in
\cite{ovrutten}.  In eleven dimensions, a non-vanishing
Chern-Simons term on one of
the brane serves as a source for the three-index
antisymmetric tensor potential, $C_{ABC}$, in the bulk. This appears
as a modification of the Bianchi identity for the four--form
field strength $G_{ABCD}$ \cite{hv},
\beq
(d G)_{ABCD11} = \lambda \left(J^{hid}_{ABCD} \delta(y_{11}-y_{hid})
+ J^{vis}_{ABCD} \delta(y_{11}-y_{vis}) \right),
\label{modifieddG}
\eeq
where
\beq
J^{i}= \hbox{tr} (F^{i} \wedge F^{i})
 -{1 \over 2} \hbox{tr}( R \wedge R),
\label{jsource}
\eq
and $\lambda \equiv (\kappa_{11} / 4 \pi) ^{2/3}/( 2 \sqrt{2} \pi^2)$.
This implies that
$G_{ABCD}$ is in general non-vanishing in the bulk
\cite{ovrutten}
\beq
G_{ABC11}=3 dB_{[ABC]}+
{\lambda \over 2 } (\omega_{E_8}+ \omega_{E^{\prime} _8}
- \omega _L )_{ABC} ~.
\label{gsol}
\eeq
where here $x^{11} \in [0,1]$.
Note that the four-form field strength is constant in the bulk
between the branes.
Now the eleven--dimensional Lagrangian
contains a kinetic term for the field strength proportional to
\beq
\int d^{11}x \sqrt{g} ~G_{ABC11} G^{ABC11}
\label{fourkinetic}
\eq
Inserting
the expression (\ref{gsol}) for the four-form field strength
into this action and integrating over the
eleventh dimension gives precisely the ten--dimensional
action (\ref{Heq}).

Two features are important to the form of the
brane--brane tree-level interaction in this case.
First, and most crucially, fields on the
branes are sources for bulk fields.
This is a generic feature of the all the examples addressed
here.
Second, the bulk field strength sourced by the brane fields in this
geometry is constant and does not fall with distance.
This feature is special to a bulk with co-dimension one,
and as described below, is not crucial to the existence
of unsuppressed brane--brane interactions in examples
with larger co-dimension.

Next consider the further reduction to four dimensions.
The Chern-Simons
terms discussed above play a crucial role in yielding the Kahler potentials
described in the previous section.
Consider first the
special compactification $T^2 \times T^2 \times T^2$ where each
$T^2$ is a symmetric torus with both radii given by $R_i$ for
$i=1,2,3$.
This compactification preserves a complex structure with
complex metric $g_{i \bar j}$.
After compactification to four dimensions, but
before Weyl rescaling to Einstein frame, the
four-dimensional kinetic terms for visible and hidden sector scalars,
$\varphi_i = Q_i$ or $\Sigma_i$
which arise from the ten--dimensional
gauge boson kinetic terms in string frame are
\beq
{\cal L}_{kin} =\sum_i e^{-2 \phi_{10}}  {V_6  \over R_i^2}
\hbox{tr}
\vert {\partial_{\mu} \varphi_i}\vert^2,
\eeq
where $V_6 = {\rm det}(g_{i \bar j})^{1/2}$
is the volume of the compact space,
$\phi_{10}$ is the ten--dimensional dilaton, and the trace
is over the gauge and flavor indices.
After Weyl rescaling to the Einstein frame, where
the string and Einstein frame metrics are related by
\beq
g_{\mu \nu} ^{ST} = {e^{2 \phi_{10}} \over  V_6} g_{\mu \nu} ^{E}
\eeq
the kinetic terms become
\beq
{\cal L}_{kin}=\sum_i {1 \over R_i^2} \hbox{tr}\vert{\partial_{\mu}
\varphi_i}\vert^2 ~.
\eeq
In  Einstein frame the moduli
terms for the radii and four-dimensional dilaton are
\beq
{\cal L}_{rad}=
\sum_i {1 \over R_i^2} (\partial_{\mu} R_i)^2 +(\partial \phi)^2
= \sum_i {1 \over 4 R_i^4}  (\partial_{\mu} R_i^2)^2
+(\partial_{\mu} \phi)^2,
\eeq
where the four-dimensional dilaton is a combination of the
ten-dimensional dilaton and the volume modulus
\beq
\phi=\phi_{10} -
{1 \over 4} \ln(\det g_{i \bar j}) =
\phi_{10} - {1 \over 2} \ln V_6.
\eeq
In addition, the four-dimensional
Einstein frame lagrangian contains a term coming from the
reduction of the Chern-Simons squared terms (\ref{Heq})
which arise from
the eleven-dimensional theory by
integrating out $G_{\mu a b 11}$,
\beq
{\cal L}_{cs}=\sum _i {1 \over R^4_i }
\left(\partial_\mu a_i  -{i \over \sqrt{2}} \hbox{tr}( \varphi_i^*
\partial_\mu \varphi_i -  \varphi_i \partial_\mu
\varphi_i^*)\right)^2,
\label{cs2}
\eeq
where $a_i$ is the pseudoscalar partner of radii,
$a_i =B_{i \bar i}=C_{11 i \bar i} \equiv {\rm Im}(T_i)$.
The trace is
over the gauge degrees of freedom which includes a sum over
{\it both} visible and hidden sector
fields, $\varphi_i =Q_i $ or $\Sigma_i$, which from
an eleven dimensional point of view reside on different boundaries.
Note that the radii, $R_i$, do not have derivative couplings to
the brane matter fields.

Now in order to write the entire action in a
manifestly supersymmetric form in terms of $N=1$ supermultiplets,
it is necessary to define the scalar component of the
chiral multiplet fields $T_i$ which contain the radii moduli
as \cite{wittensimple}
\beq
T_i = R_i^2 + i a_i + {1 \over 2}
\hbox{tr} Q^* _i Q_i +{1 \over 2} \hbox{tr}
\Sigma^{*} _i \Sigma _i  ~,
\label{fieldredef}
\eeq
and the scalar component of the dilaton chiral multiplet as
\beq
S= e^{-2 \phi} + i \sigma,
\label{fieldredef2}
\eeq
where $\sigma$ is the model independent axion.
With these definitions, the Kahler potential for these four-dimensional
$N=1$ chiral multiplets is
\beq
K = -\ln (S+S^{\dagger}) -\sum _i \ln(T_i + T_i^{\dagger} -
\hbox{tr} \Sigma^{\dagger}_i \Sigma_i -
\hbox{tr} Q^{\dagger}_i Q_i )
~.
\label{kahlerst}
\eeq
The appearance of the logarithm in
(\ref{kahlerst}) might at first sight seem puzzling
since this non-linear function might be expected to generate,
first, an infinite series of high dimension couplings between
the brane fields, and second, derivative couplings between $R_i$
and the brane fields.
In contrast, the classical ten-dimensional Lagrangian discussed above
has operators which involve at most six fields (arising
from the Chern-Simons squared interactions),
and does not involve derivative couplings of $R_i$
to matter fields.
However, the field redefinitions (\ref{fieldredef}) and
(\ref{fieldredef2}) with the Kahler potential (\ref{kahlerst})
gives precisely the four--dimensional Lagrangian terms identified above
including the Chern-Simons squared terms, and no others.
The supergravity $f$ function is obtained by a Weyl rescaling
from the four--dimensional Einstein frame in which supersymmetry is manifest,
and not by a matching of the original four--dimensional
Einstein frame with current-current couplings (\ref{cs2})
directly to supergravity frame.
It should be noted that the Einstein frame Chern-Simons Lagrangian
(\ref{cs2}) is of the form of a current-current interaction,
and does not contain non-derivative interactions between the
branes.
The non-derivative couplings appear only after the field
redefinitions (\ref{fieldredef}) to obtain a manifestly
supersymmetric action in terms of $N=1$ supermultiplets.
Since these field redefintions are so important in obtaining
in the correct form of the Kahler potential and supergravity
$f$ function it is instructive to consider them in more detail.
First, the field redefinitions (\ref{fieldredef})
indicate that the arguments of the logarithms in the Kahler potential
(\ref{kahlerst}) depend only on the geometric radii, and not
the brane fields,
$T_i + T_i^{\dagger} -
\hbox{tr} \Sigma^{\dagger}_i \Sigma_i -
\hbox{tr} Q^{\dagger}_i Q_i = 2 R_i^2$.
The corresponding supergravity $f$ function
then also only depends on the compact volume
and not the brane fields, $f = -3 (R_1 R_2 R_3)^{2/3} (S+S^{\dagger})^{1/3}
= -3 V_6^{1/3} (S+S^{\dagger})^{1/3}$.
This in fact must be the case for direct compactification from ten dimensions
since the coefficient of the Einstein term in the geometric
frame just depends on the compact volume $V_6$
and the dilaton, independent of any brane
matter.
A Weyl rescaling from geometric to Einstein to supergravity frame
then results in the $f$ function given above which depends
only on $V_6$ and $S$.

The field redefinitions (\ref{fieldredef}) can be
derived in this case from the underlying theory.
The fermionic partners of the scalar components of the $T_i$ moduli
are components of the ten-dimensional gravitino.
In the four-dimensional theory the supersymmetric variation
of a fermion is proportional to the derivative of its
scalar partner.
The ten-dimensional local supersymmetric variation of the
gravitino is \cite{gsw}
\beq
\delta_{\eta} \psi_M = D_M {\eta } + {\sqrt{2}  \over 32} \eta
 e^{- \phi_{10}/2}
(\Gamma_{M} ^{NPQ} - 9 \delta ^{N} _{M} \Gamma ^{PQ} )
H_{NPQ}
\label{gravvar}
\eeq
where a $\Gamma$ matrix with $n$ indices is the antisymmetric product of
$n$ Dirac matrices that satisfy the (field-dependent)
ten-dimensional Clifford algebra.
The variation (\ref{gravvar}) gives rise to two types of contributions
in the four-dimensional theory.
The first is from the covariant derivative appearing in the first
term.
The covariant derivative includes derivatives of $R_i^2$.
The second type of term arises from the the three-form field
strength, which contains the Chern-Simons form.
Upon compactification to four dimensions, the Chern-Simons
form includes derivatives of two matter fields as described above.
In this way the supersymmetric variation of $\psi_{T_i}$  is the derivative
of a sum of two terms, the geometric radius squared and a
composite of matter fields.
The scalar partner of $\psi_{T_i}$ is then precisely this combination
of scalar fields, as given in (\ref{fieldredef}).
The four-dimensional superpotential and Kahler
potential are then functions of this combination of fields, rather
than the original geometric variables.

For a general $T^6$ compactification which preserves
$N=4$ and includes
off-diagonal fields, $T_{i \bar j}$ with
$i \neq \bar j$, we may use the $SU(3)$ symmetry
to infer from (\ref{kahlerst}) the full result
\beq
K=  -\ln (S+S^{\dagger})
-\ln \det(T_{i \bar j} + T_{i \bar j}^{\dagger} -
\hbox{tr} \Sigma_i \Sigma_{\bar j}^{\dagger} -
\hbox{tr} Q_i Q_{\bar j}^{\dagger} ) ~,
\label{logdet3}
\eeq
which agrees with the known result for $N=4$.
As discussed in section \ref{section1},
this Kahler potential is not of the sequestered
form.
In supergravity frame the supergravity function
$f$ of the low energy four-dimensional theory
is not separable and there are unsuppressed tree-level
interactions between the visible and hidden sector branes.
In
an $S_3$ preserving $N=1$ orbifold
the lowest tree-level Kahler potential for the untwisted states
is the same as in (\ref{logdet3}) but where the non-invariant
states are projected out. With hidden sector supersymmetry
breaking this
in general gives rise to unsuppressed non-degenerate
visible sector soft scalar masses
which are of order the gravitino mass.

An interesting feature of the $N=4$ Kahler potential as well
as those obtained by inheritance is that the overall
scale for the associated soft masses is independent
of the moduli and can not be be made parametrically small in some
region of moduli space, as discussed in Appendix B.
This feature can be easily seen in the strongly coupled Horava-Witten
limit derivation given here with the rescalings between frames given above.
The coefficient of the
eleven--dimensional Einstein term
${1 \over 2}\int d^{11}x \sqrt{g} {\cal R}$ reduced
to geometric frame in four dimensions is proportional the total
internal volume $R_{11} V_6 \sim TS^{2/3}$,
where $R_{11} \sim T/S^{1/3}$ and $V_6 \sim S$.
The coefficient of the ten-dimensional Horava-Witten brane
gauge fields ${1 \over 4} \int d^{10}x \sqrt{g} F_{\mu \nu} F^{\mu \nu}$
reduced to geometric frame in four dimensions gives brane
matter field kinetic terms proportional to
$V_6 g^{i \bar i} \sim S^{2/3}$
where $S \sim V_6$ is the coefficient of the four--dimensional gauge
coupling and the inverse metric
$g^{i \bar i} \sim 1/ V_6^{1/3} \sim 1/S^{1/3}$
arises from contraction of the brane gauge fields indices on the internal
space.
Weyl rescaling from geometric to Einstein frame then gives brane
field kinetic terms proportional to $1/T$.
Now the coefficient of the
brane--brane quartic couplings arising form integrating
out the bulk four-form field strength kinetic term
$\int d^{11}x \sqrt{g} G_{ABC11}G^{ABC11}$
with the solution (\ref{gsol})
reduced to geometric frame in four dimensions is proportional
to $R_{11} V_6 g^{11} g^{i \bar i}g^{j \bar j}
\sim V_6^{1/3}/R_{11} \sim S^{2/3}/T$,
where $R_{11} V_6$ is the internal volume,
$g^{11} = 1/R_{11}^2$ is the inverse metric
with the $x^{11} \in [0,1]$ coordinates of (\ref{gsol}),
and the inverse
metric $g^{i \bar i} g^{j \bar j} \sim 1/ V_6^{2/3}$ arises from
contraction of the brane gauge field indices on the internal space.
A Weyl rescaling to Einstein frame then gives four--dimensional
matter quartic couplings proportional to
$1/T^2$.
Finally, in order to obtain the physical quartic couplings
and physical soft masses, the matter fields
must rescaled to canonical normalization by multiplying
each matter field by $T^{1/2}$.
This gives quartic couplings and soft masses which are
independent of both $T$ and $S$,
and therefore also of $R_{11}$ and $V_6$ holding a four--dimensional
scale such as the gravitino mass fixed.
All these rescalings are of course implicitly contained in the
above Kahler potentials and eleven--dimensional derivation given
above.


\subsubsection{Inheritance for M-theory Orbifolds}
\label{sec:inhmorb}

The microscopic derivation of the Kahler potential
for $N=4$ compactifications from Horava-Witten theory
can be modified to
provide a justification for the inheritance principle
determining the lowest order tree-level
Kahler potential
for the untwisted states in an
$N=1$ orbifold compactification
$T^6 /\Gamma \times S_1 / Z_2$ in the brane world limit.
General arguements were presented in Section 2.1
that the lowest order Kahler potential
for these states is determined
from the $N=4$ expression
by deleting the non-invariant states.
This inheritance can justified in the large volume
brane world limit, $T^3 \gg S \gg 1$
 by inspecting the
explicit supergravity solution for the
four-form field strength (\ref{gsol}), which is valid at generic
points in the compact space in this limit up to small
power suppressed quantum M-theory corrections.

If
the fluctuating fields appearing on
the right hand side of (\ref{gsol}) are restricted o those
massless states that survive the orbifold
projection, then the Chern-Simons form
\beq
\omega_{\mu i \bar{j}} = \hbox{tr}Q^{\dagger}_j \partial
_{\mu} Q_{i} - \hbox{h.c.} +{\cdots}
\label{wcsexpan}
\eeq
contains some states $Q^{A}_i$ and
$Q^{A} _j$ transforming
under a representation $\{A\}$ of the unbroken gauge group.
The ellipses denote terms involving other
representations and also terms involving three
fields. The $i=1,2,3$ denotes the $SU(3)$ label,
where under an abelian orbifold action
the complex coordinate
of the torus $z_i
\rightarrow \gamma_i z_i$ with $\gamma^{|\Gamma|}_i =1$.
The orbifold is also embedded into the gauge group,
so that each field transforms under the orbifold
group as
\beq
Q^{A} _i \rightarrow
e^{2 \pi i r_i} e^{-2 \pi i \beta_{A}} Q^{A}_i = Q^{A} _i
\label{orbtranssec3}
\eeq
where the last equality follows from restricting to
invariant
states.
More details on the notation can
be found in Appendix A.

The argument for inheritance appearing
in the subsequent paragraph utilizes
the following fact: if
after the orbifold projection
the Chern-Simons terms in (\ref{wcsexpan}) appear,
then necessarily  $\gamma_i \equiv \gamma_j$.
This is almost obvious, since the
states are invariant and it naively appears
that the gauge transformation
appearing in (\ref{orbtranssec3}) cancels
between $Q_i$ and $Q_j$.
But there is a loophole that
could allow two states
$Q$ and $Q^{\prime}$ with the same quantum
numbers to have
different transformations under the embedding of
the orbifold into the gauge group, namely
$e^{2 \pi i \beta}  \neq e^{2 \pi i \beta^{\prime}}$.
This could in principle occur
if in the decomposition
of a represention ${\cal R}$ of ${\cal G}$ into
representations $\{{\cal R}^{\prime}\}$ of
a regular subalgebra ${\cal G}^{\prime}$
the same representation ${\cal R}^{\prime}$ appeared
more than once. Inspecting the branching tables
appearing in Slanksy \cite{slansky} indicates
that this does occur but rarely and only for
very high dimensional representations ${\cal R}$.
But
for an adjoint representation
this does not occur.
This is because the $U(1)$ charges of a charged root
of ${\cal G}$
are enough to
completely specify the root. Thus two states
$Q$ and $Q^{\prime}$ with the same quantum numbers must
have the same orbifold
gauge transformation, $\beta = \beta^{\prime}$, since
in the higher dimensional
theory
the quantum numbers of these states originate
from
the adjoint representation of $E_8$.
This combined together with the requirement
that $r_i \equiv \beta_A
\hbox{mod} |\Gamma|$ for invariant states
is enough to imply that if
gauge invariant terms appear
in (\ref{wcsexpan}), then necessarily
$\gamma_i = \gamma_j$. These arguments can
also be directly applied to $\omega_{\mu ij}$ with
the conclusion that if these
terms appear then $\gamma_i = \gamma^{*} _j$ .

With the above information it is now possible to
to argue that (\ref{gsol}) restricted to
invariant states remains a
solution in the orbifold background at tree-level.
The point is that
the solution for the four-form
field strength on the covering space
must respect
the transformatation properties specified by the
orbifold group.
Thus (\ref{gsol}) is still the correct solution
at lowest order
provided it satisfies the boundary conditions
\beq
G_{\mu i \bar{j} 11}(x_{torus},x_{11})
= q_{Z_2} \gamma_i \gamma^* _j G_{\mu i \bar{j} 11}
({\cal U} x_{torus},-x_{11}) ~,
\eeq
where ${\cal U}$ is the orbifold rotation acting
on the internal coordinates of the torus and $q_{Z_2}$
is the $M-$theory orbifold charge.
There is a similar condition for $\bar{j} \rightarrow j$.
To describe the massless untwisted states
we restrict the fluctuating fields
appearing on the right hand side of (\ref{gsol})
to those
massless states that survive
the orbifold projection.
Then the Chern-Simons term appearing
in (\ref{gsol}) is necessarily constant across the $T^6$ orbifold.
Thus in order that (\ref{gsol}) restricted to
the invariant states remains
a solution, the boundary conditions for
$G_{i \bar{j} \mu 11}$ and $G_{i j \mu 11}$  given
above
must allow for constant solutions
across
the
$T^6$ orbifold. This will be the case provided
that $\gamma_i \gamma^{*} _j \equiv 1 $ and
 $\gamma_i \gamma _j \equiv 1 $ respectively.
The argument of the previous paragraph
demonstrated that this is always the case.
The remaining arguments
leading from (\ref{gsol}) to the Kahler potential
(\ref{kahlerst})
then follow except that
now in (\ref{kahlerst}) or in (\ref{logdet3})
the non-invariant states are projected out.
Thus the solution (\ref{gsol}) for the toroidal
compactification can be trivially extended to obtain
the lowest order tree-level Kahler potential
for the untwisted states of the
the orbifold by simply deleting the non-invariant
states. This provides
a supergravity
justification for the use of the inheritance
principle in the large volume brane world limit.
The expression (\ref{kahlerst}) is then appropriate
for an abelian orbifold that does not preserve
any permutation symmetry, whereas (\ref{logdet3})
applies for an abelian orbifold that preserves
an $S_3$ permutation symmetry.

\subsection{The Strongly Coupled Heterotic Theory in the Five
Dimensional Limit}
\label{sec:hwfivelimit}

The Horava--Witten theory compactified
on a generic six--manifold preserving at least $N=1$ supersymmetry
in four dimensions leads, as discussed in section
\ref{sec:hwsimple},
to unsuppressed non-universal
tree-level visible sector scalar masses comparable to
the gravitino mass, even with supersymmetry breaking
isolated on a hidden sector brane.
The microscopic origin of these masses in the ten--dimensional
limit in which the M-theory direction is smaller than
any of the compact six--manifold
directions, $R_i \gg R_{11}$, was explained in the
previous subsection.  The required
contact terms arise from exchange in the M-theory direction
of the bulk four-form field strength between the branes.
In this subsection we consider the five--dimensional limit
in which the length scales
of the six--dimensional compact manifold are much smaller
than the M-theory direction, $R_{11} \gg R_i$.
This yields a five--dimensional theory below the compactification
scales of the six--dimensional compact manifold.
An $S_1/Z_2$ orbifold projection of the M-theory direction in
this five--dimensional theory then yields
the five--dimensional analog of the Horava--Witten theory with
end of the world branes just as in eleven dimensions.
If the compact manifold is $T^6$ then $N=4$ supersymmetry
is preserved in the low energy four--dimensional theory
obtained by the $S^1/Z_2$ orbifold of the five--dimensional
theory.
The Kahler potential of a
four--dimensional $N=4$ theory is not renormalized.
Because of this, the Kahler potential in the
ten--dimensional limit is identical to the Kahler potential
in the five--dimensional limit.
Non-universal tree-level scalar masses then also arise
in this limit.
The non-universal nature of the tree-level soft masses
survives generic projections of the $T^6$ compactification
which preserve only $N=1$ supersymmetry in four dimensions.

The origin of the non-universal visible sector scalar masses
is simple to understand in the five--dimensional
limit.
In particular, the five--dimensional theory
obtained by compactifying from eleven dimensions contains
a number of $U(1)$ gauge bosons which
couple non-universally to the matter living
on the boundaries.
Visible and hidden sector matter on the branes are neutral
under the bulk $U(1)$'s, but appear as sources in a
modification of the Bianchi identities for the bulk
gauge bosons.
Integrating out these gauge bosons to obtain
the four--dimensional effective theory precisely generates
the Chern-Simons squared contact interactions (\ref{cs2})
and Kahler potentials described
in the previous sections.

To see the origin of the
non-universal brane--brane interactions in this limit, note that
the five--dimensional decomposition of the eleven--dimensional
three--form potential with two compact indices and one non-compact
index gives five--dimensional vector bosons.
The number of these gauge bosons depends on properties of the
compact six--manifold.
For toroidal compactification on $T^6$
the low energy five--dimensional
theory has an $SO(6) \cong SU(4)$ symmetry at leading order in the
low energy derivative expansion.
The five--dimensional vector bosons arising from the three--form potential
in this case transform as ${\bf 15} \in SU(4)_R$.
Under the $N=1$ decomposition of the $N=4$ $R$-symmetry
the vector bosons transform as ${\bf 8}_0 \oplus {\bf 1}_0 \oplus {\bf 3}_+
 \oplus \overline{{\bf 3}}_- \in
SU(3) \times U(1)_R \subset SU(4)_R$.
For definiteness we focus on the ${\bf 8} \oplus {\bf 1} \in SU(3)$ gauge
bosons with four-- and five--dimensional components
\begin{eqnarray}
A^{(i \bar{j})} _{\mu} & \equiv  C_{i \bar{j} \mu } \nonumber \\
A^{(i \bar{j})} _{11} & \equiv  C_{i \bar{j}11}
\end{eqnarray}
where $i,\bar{j}$ are complex coordinate indices on the compact
six-manifold, and where the M-theory direction
(indicated by a subscript 11)
is now the fifth dimension.
These $U(1)$ gauge bosons are the ones
which would survive the $Z_3$
orbifold projection which preserves $N=1$ described
in section \ref{sec:hwsimple}.
The $SU(3)$ global symmetry is a flavor symmetry
under which matter on both the visible and hidden sector
end of the world branes transform.
The bulk five--dimensional $U(1)$ gauge bosons
then also transform under this global flavor symmetry.
Most importantly, the field strengths for these gauge bosons have a
modified Bianchi identity that is inherited from the
eleven--dimensional Bianchi identity (\ref{modifieddG}) for the four--form
field strength \cite{hv}
\beq
(d F^{(i \bar{j})} )_{\mu \nu 11} = \delta (x_{11}-x_{11,hid})
J^{(hid)}_{\mu \nu i \bar{j}}(x^{\mu})
+ \delta(x_{11}-x_{11,vis}) J^{(vis)} _{\mu \nu i \bar{j}}(x^{\mu}) ~.
\label{fivebianchi}
\eeq
The sources appearing on the right hand
side of the Bianchi identity depend on the
$i \bar j$  flavor for both the visible and hidden sector
fields and also transform under the global flavor
symmetry as
${\bf 8} \oplus {\bf 1} \in SU(3)$.
The solution to (\ref{fivebianchi}) is the same as the
equation (\ref{gsol}) for the four--form field
strength in the previous section,  but where here
the components $G_{\mu i \bar{j} 11}$ are relevant.
The dimensional reduction of the action from eleven to
five dimensions contains a term
\beq
\int dx_{11} d^{4} x ~V_6 e^{-2 \phi_{10}} {(F_{\mu 11} ^{(i \bar{j})})^2
\over  R^2_i R^2_j}.
\eeq
Integrating out $F_{\mu 11} ^{i \bar{j}} $
in the five--dimensional theory then yields
precisely the brane--brane contact interactions discussed above.
In this limit the $SU(3)$
flavor symmetries on each brane are coupled by exchange of the
bulk gauge bosons.
This arises in the five--dimensional theory by integrating
out the Kaluza-Klein tower of five-dimensional
bulk gauge boson states even though the four-dimensional zero
modes of these fields are projected out by the $S^1/Z_2$ orbifold
in the M-theory direction.
So in this limit the brane--brane contact interactions are generated by
exchange of massive bulk fields without any
(exponential) suppression of the coupling.

Note that since the bulk gauge bosons exist
in the five--dimensional theory, their mass is
protected
by a gauge invariance.
It is then not possible to
suppress the couplings they generate
by giving them a large mass in the $S^1/Z_2$ projection
to the four--dimensional theory.
Eliminating the brane--brane contact interactions
in this limit would require projecting out of the five--dimensional
theory all the bulk $U(1)$ gauge bosons which have
brane sources through Bianchi identities of the form (\ref{fivebianchi}).
In the compactification from eleven to five dimensions this
would require lifting all of the three--form potentials with
indices in the compact directions.
This is in fact not possible with a symmetric Abelian orbifold
projection which does not reduce the rank and
necessarily leaves invariant at
least three five--dimensional vector bosons with
diagonal internal indices $i = \bar{j}$.
These three bulk gauge bosons have Bianchi identity brane sources
for the $U(1)^3 \subset SU(3) \times U(1)_R$ preserved by any
Abelian orbifold.
Removing all vector bosons from the five--dimensional theory
(except of course the one in the gravitational supermultiplet) might
in principle be possible with asymmetric and/or non-Abelian
projections which reduce the rank of the global symmetry.

Since the flavor symmetries on each brane
are coupled through exchange of bulk gauge bosons,
non-universal visible sector tree-level scalar masses arise for
generic supersymmetry breaking on the hidden sector brane.
Universal vanishing tree-level visible sector masses require
flavor symmetric hidden sector supersymmetry breaking.
However, in this case, universality is the result
of unbroken flavor symmetries
and not simply from physically separating the visible and
hidden sector branes.


\subsection{Open Strings}
\label{sec:open}

The Kahler potential for D-brane world models
of BWSB are generally
not of the no-scale sequestered form, as described in
section \ref{sec:Dbranesimple}.
This implies that brane--brane contact interactions exist in the
low energy four--dimensional theory.
Microscopically, bulk locality implies that
these interactions must arise from tree-level
exchange of bulk fields between the branes.
In this subsection we illustrate the origin of these
interactions in simple D-brane world models which preserve
$N=4$ supersymmetry.
While not phenomenologically realistic, these models illustrate
that the
sequestered intuition breaks down even
in situations with a high degree of supersymmetry.  There is no
reason to think that things will be different in models with
less supersymmetry to protect the form of the Kahler potential,
and we will illustrate such effects in examples with $N=2$ and
$N=1$ supersymmetry later.
D-brane models are also instructive in illustrating the
dependence on the internal volume in examples with brane
co-dimension larger than one.
The volume dependence
of brane--brane interactions leads to unsuppressed
visible sector soft scalar masses from hidden sector supersymmetry
breaking as compared with the gravitino mass, just as in the co-dimensions one
cases.

The simplest example of a D-brane world model is provided by
type I string theory with gauge group $SO(32)$
in ten dimensions compactified
on a circle of radius $R$.
Interaction terms between visible and hidden sector fields
are easily exhibited, as
above, in Chern-Simons squared couplings.
The low energy nine--dimensional theory
at energy scales below $R^{-1}$ contains, among other terms,
the dimensional reduction of the Chern-Simons term
proportional to
$R \omega_{MNO}^2$, where upper case Latin indices
denote the non-compact dimensions.
In particular, the nine--dimensional action
includes a term proportional to
\beq
\int d^9 x   R \left( d C_{(2)} - \omega _{CS} \right)^2,
\label{typei}
\eeq
where $C_{(2)}$ is the type I RR two--form potential.
With Wilson lines turned on the Chern-Simons form
factorizes into a sum
over the Chern-Simons forms of the unbroken subgroups.
Thus (\ref{typei}) includes quadratic terms involving the
product of different
Chern-Simons forms.

Now consider the type I$^{\prime}$ description obtained by
T-duality on the compact direction.
In this theory there are 2 O8 planes separated by a distance
$R^{\prime}= 1/R$ and (including images)
32 D8 branes transverse to the interval between the
O8 planes in a ${1 \over 2}$ BPS configuration which preserves
16 supersymmetries.
A brane world model may be obtained by arranging the
D8 branes in two physically separated groups along the interval.
For example, if (including images) 16 D8 branes are
placed on each O8 plane an $SO(16) \times SO(16)$ gauge theory
is obtained.
The bulk between the branes is co-dimension one in this example.
Generalization to other co-dimensions will be analyzed below.
The T-dual of the low energy nine-dimensional
action (\ref{typei})
may be written
\beq
\int d^9 x \left({1 \over R^{\prime} } (d C_{(2)} )^2 -
{2 \over R^{\prime} } dC_{(2)} \cdot  \omega _{CS} + {\omega ^2_{CS}
\over R^{\prime} }\right),
\label{typeip}
\eeq
where dependence on the type I RR two-form field has been retained.
An important observation is
that for separated branes
the Chern-Simons form (as opposed to its square)
appearing in (\ref{typeip})
is a sum of the individual
Chern-Simons forms
from each of the groups of D-branes.
This follows by T-duality from the type I description
with Wilson lines turned on.

We have focused on terms where all the
indices are in the non-compact
direction and they are raised and lowered using the non-compact metric.
Since the low-energy nine--dimensional theory is invariant under T-duality,
these interactions must also be present in the type I$^{\prime}$
description.
Cross terms between fields localized on the
visible and hidden sector D-branes in the type I$^{\prime}$
description must then arise microscopically from
exchange of bulk fields between the visible and hidden sector
D-branes.
Of importance
are the volume suppressed Chern--Simons squared interactions.

The first two terms in (\ref{typeip})
may be easily understood in the type I$^{\prime}$ D-brane picture.
This theory has a RR three-form potential $C_{(3)}$ in the bulk which
couples to gauge fields propagating on the branes in a manner
described below.
The first term in (\ref{typeip})
is just the dimensional reduction of the bulk kinetic
term for
this three--form
\beq
\int d^{10} x \vert d C_{(3)} \vert ^2 \rightarrow
\int d^9 x R^{\prime} g^{99} \vert d C_{(3) \mu \nu 9} \vert ^2 .
\label{Cthreekinetic}
\eeq
Under T-duality
the type I RR two-form potential becomes
$C_{(2) \mu \nu} \rightarrow C_{(3) \mu \nu 9}$ in the type I$^{\prime}$
theory.
Then with $g^{99}=1/R^{\prime 2}$
the kinetic term (\ref{Cthreekinetic}) agrees with the
first term of (\ref{typeip}) in terms of type I fields.
The second term in (\ref{typeip}) is obtained in the type I$^{\prime}$
description  from Wess--Zumino
couplings between the brane
fields and the bulk RR potentials \cite{polchinski},
\beq
S_{WZ} = \int_9 \hbox{Tr}\left( e^{2 \pi \alpha^{\prime} F} \right)
\wedge \sum _q C_{(q)}
\label{WZaction}
\eeq
where the sum is over all the RR potentials.
The existence of these
interactions may be understood from the observation that
instantons on the brane world volume act as sources for
RR forms of lower rank \cite{douglas}.
Note that the interactions (\ref{WZaction})
do not depend on either the dilaton or volume
of the compact space.
The expansion of the exponential
in the type I$^{\prime}$ theory
gives a coupling of brane instanton number to the bulk
five--form potential
\beq
S_{WZ} \supset \int _{9}   F \wedge F \wedge C_{(5)}    ~.
\eeq
With $d \omega _{CS} = F \wedge F$, this interaction may be written
as
\beq
\int_9 \omega _{CS} \wedge d C_{(5)}    ~.
\eeq
Using Poincare duality $d C_{(5)}$ can be written as a four--form
field strength, $F_{(4)} = ^*d C_{(5)} = d C_{(3)}$.
However, Poincare duality utilizes the Levi-Civita
$\epsilon$-tensor, which, since one of its indices is in the
compact
direction, introduces a factor of
$1/ R^{\prime}$. To see this, note that the above coupling is
proportional to
\beq
\omega_{(CS) [\mu_1 \mu_2 \mu_3 }  F_{(6) \mu_4 \cdots
\mu_9]}
\eeq
and the Poincare duality gives $F_{(6) \mu_4 \cdots \mu_9}
= \epsilon _{\mu_4 \cdots \mu_9} ^{~~~~~~~
a _1 \cdots a _{4}} F_{(4) a_1
\cdots a_4}$.
Now all of $\mu_1, \cdots \mu_9$ are along the D8 brane
directions, so one of the components of $F_{(4)}$ and
an upper component of the $\epsilon$-tensor is in the compact
direction. If the metric is used to express the $\epsilon$-tensor
with mixed
components in terms of an $\epsilon$-tensor with all lower components,
then three of the indices of $F_{(4)}$ are contracted with the
indices of the Chern-Simons form, with one index left out.
Since the non-zero value of the
$\epsilon$-tensor with all lower indices is $\pm
\sqrt{-g}$, where $g$ is the determinant of the bulk metric,
the dependence of this coupling
on the metric is $g^{99} \sqrt{-g} = 1/R^{\prime}$.
After dimensional reduction the remaining component of
$F_{(4)}$ in the compact direction cannot involve a derivative,
so it must also be a component of
$C_{(3)}$.

Putting this together, the Wess-Zumino interactions, together
with Poincare duality, imply a coupling of the brane gauge fields
to the bulk three-form of the form
\beq
\int d^9 x   {1 \over R^{\prime}} \omega _{CS} \cdot d C_{(3)},
\eeq
with $C_{(3)} = C_{(3) \mu \nu 9}$, and the three indices of
$\omega_{CS}$ are contracted with the three non-compact indices
of $d C_{3}$ using the lower dimensional metric.
The generalization to any number of tori or circles transverse
to the brane implies that this interaction is in general
suppressed by the
volume of the compact space transverse to the brane.
Notice that the above dependence on the volume and the dilaton
agrees with the second term (\ref{typeip}) since
$C_{(2) \mu \nu} \rightarrow C_{(3) \mu \nu 9}$
under T-duality.


Now the full action of the compactified nine--dimensional theory
should be invariant under T-duality.
But since the low-energy Lagrangian in the type I theory contains
the $R \omega_{CS}^2$ interaction in (\ref{typei}),
we infer from the
invariance under T-duality that the low energy type I$^{\prime}$
D-brane world theory contains not only the
first two terms of (\ref{typeip})
discussed above from the point of view of brane field interactions with
bulk RR fields, but also
the third  term of (\ref{typeip}) which is the
volume suppressed contact interaction
\beq
\int d^9x~ {1 \over R^{\prime}} \omega ^2 _{CS},
\label{ninecs}
\eeq
which couples branes that are physically separated in the
microscopic type I$^{\prime}$ theory.
More generally, in any number of dimensions, this coupling is
suppressed by a factor of the compact volume transverse
to the brane as discussed
below.


In the closed string channel general brane--brane interactions are
obtained from integrating out the Kaluza--Klein tower of
RR fields.
In the D8 brane example above, brane gauge instanton number,
$F \wedge F$,  appears as a source
in the bulk five--form potential equation of motion.
By using Poincare duality
the sources appear in a modified Bianchi identity for the four--form
field strength. The contact brane--brane
interaction (\ref{ninecs})
is then generated from integrating out the Kaluza-Klein tower
of the three--form potential, much as for the one--form in the
five--dimensional limit of the Horava-Witten model discussed
in the previous subsection.
This can be generalized to branes of lower dimension,
but the modified Bianchi identity that follows is
now more intricate and a solution is not presented here.

In the open string
channel the brane--brane interactions are generated
by the one--loop amplitude of open string states stretching between
the branes.
By modifying the RR amplitude for
the force
between two parallel groups of D branes \cite{polchinski} to include
the gauge boson vertex operators, it is possible to
obtain the brane--brane interaction (\ref{ninecs}) directly with
correct dependence on the volume. Further
work on this open string perspective is in
progress \cite{michaelotherstudent}.

The general volume dependence of brane--brane interactions for
brane co-dimensions larger than one and resulting Kahler
potentials in four dimensions may be illustrated by additional
T-duality transformations of the type I theory.
Consider first type I theory on a symmetric $T^2$ of radius $R_1$.
The T-dual description in the $T^2$ directions of this background
is type IIB string theory with (including images) 32 D7
branes and 4 O7 orientifold planes on $T^2/Z_2$ with
radius $R^{\prime}=1/R$.
This configuration is ${1 \over 2}$ BPS
and preserves 16 supersymmetries.
Following arguments similar to those preceding (\ref{typeip})
and (\ref{ninecs})
the low energy eight--dimensional action contains
the Chern-Simons squared volume suppressed contact interactions
between D7 brane fields
\beq
\int d^8x~ { \omega_{CS}^2 \over R^{\prime 2} }.
\eq
By T-duality, similar transverse volume suppressed brane--brane
contact interactions arise for type IIB configurations with
32 D5 branes and 8 O5 planes or with 32 D3 branes and 16 O3 planes.
The Kahler potential derived below applies to these
cases also.

%

Now
consider the further toroidal compactification of the eight dimensional
type IIB configuration with 32 D7 branes and four O7 planes
on a product of two symmetric tori $T^2 \times T^2$ with radii
$R_2$ and $R_3$, and we relabel $R^{\prime} \rightarrow
R_1$.  
The low energy four--dimensional theory
has $N=4$ supersymmetry.
Before Weyl rescaling the Lagrangian in geometric frame obtained
by direct compactification contains the couplings
\ba
& & \int d^4 x \left\{ V_6 e^{-2 \phi_{10}} {\cal R}_{(4)}
+ e^{- \phi_{10}}
(R^2_2 R^2 _3) {1 \over 4} \hbox{tr} F^2_{\mu \nu} \right.  \nonumber \\
& & + R^2_2 R^2_3 e^{- \phi_{10}}
\left( \hbox{tr}{ (\partial_{\mu} \varphi_2)^2  \over 2 R^2_2} +
{e^{\phi_{10}}
\over 8 R^2_1 R^4_2} (\hbox{tr} \varphi_2^* \partial_{\mu} \varphi_2 -
\hbox{h.c})^2
\right. \nonumber \\
& & \left. \left. + {\hbox{tr} (\partial_{\mu} \varphi_3)^2  \over 2 R^2_3} +
{e^{\phi_{10}}
\over 8 R^2_1 R^4_3} (\hbox{tr} \varphi^* _3 \partial_{\mu} \varphi_3 -
\hbox{h.c})^2
\right) \right\}
\label{typeiiblag}
\ea
where $V_6=R_1^2 R_2^2 R_3^2$ is the compact volume,
$\phi_{10}$ is the type IIB ten-dimensional dilaton, and
$\varphi_{2}$
and $\varphi_{3}$ are complex matter fields that are the zero modes
of the
brane gauge field with components along $T^2 \times T^2$ directions.
The Lagrangian (\ref{typeiiblag}) includes the Einstein--Hilbert action, the
gauge kinetic function, and the matter action obtained
from the dimensional reduction of the higher dimensional
gauge kinetic terms localized on the branes, and also the
volume suppressed dimensionally reduced Chern-Simons squared interactions.
The dependence on the radii arises from the internal metric, except for the
quartic terms which contain an additional volume suppression $R^2_1$.
After a Weyl rescaling to Einstein frame
the above action for the matter fields becomes
\beq
\int d^4 x ~{e ^{\phi_{10}} \over R^2 _1} \left
( { (\partial_{\mu} \varphi_2)^2 \over 2 R^2_2} + {e^{\phi_{10}}
\over 8 R^2 _1 R^4 _2} (\hbox{tr} \varphi_2^* \partial_{\mu} \varphi_2 -
\hbox{h.c})^2
+(2 \rightarrow 3)  \right)   ~.
\eeq
It is important to note that now the dilaton and radii
dependence of the coefficient of
the quartic term is the square of the coefficient of the quadratic term.
Finally, there is also an adjoint scalar
from the collective coordinates of the D7 branes.
The kinetic terms for these collective coordinates
come from the lowest terms in the Nambu-Goto action
for the D7 branes
\beq
T_7 \int d^8 x ~e^{-\phi_{10} }
\sqrt{h_{\parallel}} h_{\perp ij} {1 \over 2}
\hbox{tr} \partial_{\mu} x^i \partial_{\mu} x^j
\rightarrow  \int d^4 x ~ R^2 _1 R^2 _2 R^3 _2 e^{-\phi_{10}}
\hbox{tr}  \vert \partial_{\mu} \varphi_1 \vert ^2,
\label{nambu}
\eeq
where $T_7$ is the string frame D7 brane tension,
${h_{\parallel}}$ is the determinant of the induced longitudinal metric,
and $h_{\perp ij}$ is the induced transverse metric with $i,j=8,9$.
In the second expression in (\ref{nambu}) the D7 branes are compactified on
$T^2 \times T^2$ in geometric frame and
the collective coordinate scalars properly normalized in
geometric string frame are
\beq
\varphi_1 = \sqrt{T_7} ~{
  \sqrt{h_{\perp}} (x^8 + i x^9) \over  \sqrt{2} ~R_1 }.
\eq
After the same Weyl rescaling to Einstein frame as above
these kinetic terms become
\beq
\int d^4 x~
 e^{\phi_{10}} \hbox{tr} \vert \partial_{\mu} \varphi_{1} \vert ^2
~.
\eeq

In order to write the action in a manifestly supersymmetric form
it is necessary to redefine the scalar components of the four--dimensional
chiral supermultiplets in terms of the geometric radii,
ten--dimensional dilaton, and brane matter fields.
With the field redefinitions for the scalar components
of the four--dimensional
dilaton
\beq
S = e^{- \phi_{10}} R^2_2 R^2 _3   ~,
\label{typeifds}
\eeq
and the other moduli
\ba
T_1 &=& e^{-\phi_{10}} + {1 \over 2}
  \hbox{tr} \varphi_1^* \varphi_1~, \nonumber \\
T_2 &=& e^{-\phi_{10}}
R^2_1 R^2_2  + {1 \over 2}  \hbox{tr} \varphi_2^* \varphi_2~, \nonumber \\
T_3 &=& e^{-\phi_{10}}
R^2_1 R^2_3  + {1 \over 2}  \hbox{tr} \varphi_3^* \varphi_3~,
\label{typeifd}
\ea
the Lagrangian terms described above are
obtained from the Kahler potential
\beq
K= -\ln ( S+ S^{\dagger} )
-\sum_i \ln( T_i + T_i^{\dagger} -
  \hbox{tr} \varphi^{\dagger}_i \varphi_i ).
  \label{opensum}
\eq

For a general $T^6$ compactification, inclusion of the off-diagonal
moduli and interactions yields the $N=4$ Kahler potential
\beq
K= -\ln ( S+ S^{\dagger} )
-\ln \hbox{det}( T_{i \bar j} + T_{i \bar j}^{\dagger} -
  \hbox{tr}  \varphi_i \varphi^{\dagger}_{\bar j} )    ~.
  \label{openfour}
\eq
In order to obtain a brane world model, the D-branes may be
separated into two groups which model the visible and
hidden sectors, $\varphi_i = Q_i$ or $\Sigma_i$.
Due to $N=4$ supersymmetry there are no corrections to the Kahler
potential as one turns on the branes (i.e. turns on expectation
values for certain scalar fields on the brane.  One can check
that integrating out massive fields does not generate such
couplings at tree level).
The four--dimensional
scalar fields describing the D-brane separation may then be
replaced by their expectation values, and the brane fields in (\ref{openfour})
break up into a sum of visible and hidden sector fields,
$\hbox{tr} \varphi_i \varphi_{\bar j}^{\dagger} =
\hbox{tr} Q_i Q_{\bar j}^{\dagger} +
\hbox{tr} \Sigma_i \Sigma_{\bar j}^{\dagger}$.
This is then of course the same Kahler potential given above for an
$N=4$ theory with visible and hidden sectors.

The
 Kahler potential (\ref{openfour}) is identical
to the one found in
the toroidal compactification of
heterotic theory in ten dimensions since both of
these low-energy theories preserve $N=4$.
Since the ten--dimensional Lagrangian for the dilaton and metric are identical
in both the type I and the heterotic theories,
it may then appear strange that the type I$^{\prime}$ definitions
(\ref{typeifds}) and (\ref{typeifd}) of the moduli
in terms of the string coupling and torus radii are different
than the heterotic definitions
and (\ref{fieldredef}) and (\ref{fieldredef2}).
This is, however, not a puzzle since these two sets
of definitions are in fact
mapped into each other by performing
T-duality in the $T^2$ direction with radius
$R_1$ and a subsequent type I--heterotic
duality, both of which are
symmetries of the low energy theory.

For a compactification which preserves only $N=1$ in four
dimensions
and for which all the off-diagonal moduli and interactions are
lifted, such as one of the $Z_6$ orbifold compactifications discussed
in Appendix A
the lowest order tree-level inherited Kahler potential would be
of the sum of logarithms form (\ref{opensum}).
For a compactification preserving
a $S_2$ or $S_3$ permutation symmetry,
such as one of the $Z_6$  and $Z_3$ orbifolds discussed
in Appendix A
having
2 or 3 generations in the untwisted sector,
their lowest-order inherited Kahler potential would
be related to (\ref{openfour})
Again, a parallel
brane world model can be constructed by simply separating
the D-branes into two groups.
In this case, $N=1$ supersymmetry does not forbid corrections to the
Kahler potential (\ref{opensum}).
The origin of some leading corrections
are described in the next section.
However, the lowest order tree-level Kahler potential is indeed inherited
in this case from (\ref{opensum}) or (\ref{openfour}).

In either case, with generic
hidden sector supersymmetry breaking with
stabilized moduli,
the Kahler potentials
(\ref{opensum}) and (\ref{openfour})
lead to unsuppressed non-degenerate tree-level visible sector scalars
and potentially undesirable tachyons.
As suggested above, the brane--brane contact interactions contained
in these Kahler potentials arise microscopically
from exchange of bulk RR fields
between visible and hidden sector brane fields.
The interactions are suppressed by the compact volume transverse
to the branes.
In terms of the underlying theory with the microscopic
Planck scale held fixed this leads to
four--dimensional visible sector scalar masses suppressed by
the total internal volume
$m_Q^2 \sim F^2/V$, where $F$ is a hidden sector auxiliary expectation
value.
However, the four--dimensional gravitino mass is determined by the
four--dimensional Newton constant, which is also volume suppressed,
$m_{3/2}^2 \sim F^2 / V$.
So the visible sector scalar masses are unsuppressed with respect
to the gravitino mass, $m_Q^2 \sim m_{3/2}^2$.
The volume dependence for both the brane--brane interactions and
four dimensional Newton constant is of course
implicitly contained within the Kahler potentials derived
above.

Finally, it is worth noting that all the brane world models discussed
here with $N=4$ supersymmetry in four dimensions are actually
related by dualities.
The Horava-Witten M-theory background is the strongly
coupled limit of type IIA on $S_1/Z_2$ which is T-dual to
to the type I description,
and this is in turn related back to the Horava-Witten
background by type I-heterotic duality in the strongly coupled
limit.
Since these symmetries survive compactification to four
dimensions these dualities are also sufficient to show that the
Kahler potentials in these examples are identical.
These are also sufficient to show that the brane--brane interactions
of these examples for any co-dimension are
suppressed by one power of the internal volume, just
as the four--dimensional gravitino mass.  The analysis here
exhibits the different microscopic mechanisms which explain the
seeming non-locality in each of these pictures.



\section{Corrections to the Inherited Kahler Potential}
\label{sec:corrections}

In many theories
the visible and hidden sector matter fields are remnants of extended
supermultiplets of the high energy theory.
In these cases the lowest order four--dimensional
Kahler potential which couples these sectors is inherited from the
form dictated by the extended supersymmetry of the microscopic theory.
As described from a number of points of view
in the previous sections, this occurs for many
brane world backgrounds of string and M-theories.
This generally leads to model dependent relations among the visible sector
tree-level soft masses, as discussed in section \ref{sec:hwsimple}
and Appendix B.
However, with only $N=1$ supersymmetry in four dimensions, the Kahler
potential is not protected from generic corrections.
The inherited Kahler potential, and therefore the tree-level relations
among the visible sector scalar masses, is then generally modified in the
full low-energy theory.
It is interesting to address what form and magnitude
these corrections may take in BWSB scenarios.

With standard hidden sector supersymmetry breaking the Kahler potential
is believed to obtain generic corrections.
However,
BWSB differs in that the visible and hidden
sector fields are physically separated in the microscopic theory.
It has been argued that this feature implies that the
separable form of the supergravity $f$ function is stable, that the
leading corrections are suppressed by additional powers of the compact
volume, and are therefore controllably small \cite{randallsundrum}.
The argument starts with the observation that corrections
which are quantum one-loop from the bulk point of view
are suppressed by two powers of the internal volume.
Since the four--dimensional gravitino mass is suppressed by only
one power of the internal volume, these corrections are small in the
large volume limit.
In addition, quantum corrections arising from point--like bulk
fields are ultraviolet finite since the four--dimensional brane--brane
correlators are point split (and therefore regulated)
in the higher dimensional theory.
These features were used to argue that the sequestered hypothesis was
natural and stable in brane world realizations \cite{randallsundrum}.
As we have seen, the sequestered hypothesis breaks down
even at tree level in an expansion in powers of the brane
separation,
and that there are generically
brane--brane contact interactions which give rise to tree-level
soft masses.
However, one might wonder if the above arguments could be modified
to demonstrate that the inherited form of the Kahler potential in brane
world models is stable with controllably small corrections.
In this section we show that in fact this is not the case.
There is generally no sense in which the lowest order form of the
Kahler potential is protected from generic corrections in
BWSB models, although the corrections may be small in some
corners of moduli space.

The origin of corrections to the inherited Kahler potential in
brane world theories is easy to understand. From the
bulk point of view, brane--brane contact
interactions responsible for visible sector soft scalar
masses coming from hidden sector supersymmetry breaking are
generated by tree-level exchange of bulk supergravity fields.
With $N=4$ supersymmetry in four dimensions these interactions
are fixed to have a particular form, as discussed in the previous sections.
With compactifications which preserve only $N=1$ supersymmetry these
interactions take a more general form.
This in turn gives rise to more general tree-level Kahler potentials
with additional contributions to scalar masses beyond those
of the lowest order inherited Kahler potential.
In addition, warping of the internal space by non-vanishing
brane tensions which generally arise even with $N=2$
supersymmetry.
This also modifies the form of the Kahler potential.
If moduli acquire auxiliary components, which commonly occurs
in mechanisms which stabilize the moduli,
additional contributions to tree-level masses arise.

The leading
corrections to the inherited Kahler potential may also be understood
in a slightly different manner in weakly coupled D-brane models.
In this case
couplings between the visible and hidden sectors
are generated quantum mechanically by integrating out
heavy states which are charged under both the hidden and visible gauge
groups.
In a brane world model there are no localized point-like states which
couple these sectors directly since they are physically separated.
However, there are massive string states which stretch between
the hidden and visible sector branes.
The amplitude for integrating out these states at one-loop
is just the open-string channel of the annulus diagram
with boundaries on the visible and hidden sector branes.
In the closed string channel this amplitude is simply tree-level
exchange of bulk closed-string modes between the branes.
For large volume this is dominated by exchange of massless bulk
supergravity fields, and
is therefore only suppressed by one power (rather than two as for
point-like quantum amplitudes) of the internal volume.
This gives rise to unsuppressed corrections to visible
sector soft masses from hidden sector supersymmetry breaking.
We see that the
low-energy local effective field theory reasoning described above
breaks down in this picture
because of the existence of physically extended states in the
full underlying theory.

In the next subsection corrections to the inherited Kahler
metric are illustrated in simple ${1 \over 4}$ BPS
D-brane models which preserve only 8 supersymmetries.
In the following subsections warping of the internal
space by non-vanishing brane tensions which result in
generic $N=1$ compactifications of the Horava-Witten
and pure five--dimensional supergravity examples are shown
to generally give corrections
to the lowest order inherited Kahler potential.
These in turn generally give additional unsuppressed tree-level
contributions to scalar masses which are not small except
in corners of moduli space.


\subsection{Kahler Potential Corrections in the D-Brane Picture}
\label{kahlerD}

It is instructive to consider how corrections to the
lowest order inherited Kahler potential arise in D-brane world models.
This is easily illustrated in simple D-brane configurations.
In BPS configurations which preserve 16 supersymmetries,
specifically parallel Dp-branes, the
Kahler metric is flat and the Kahler potential is exact.
But in configurations with less supersymmetry the Kahler metric
is modified by brane--brane interactions.

To illustrate the modification of the metric
in more general D-brane configurations
consider type II uncompactified string theories in ten dimensions with a
source Dp$^{\prime}$-brane and a probe Dp-brane.
We follow and slightly extend an analysis due to Brodie \cite{brodie}.
The source Dp$^{\prime}$-brane may be thought of as the hidden
sector and the Dp-brane probe as the visible sector.
The metric line element and dilaton backgrounds of the source
Dp$^{\prime}$-brane at distances large compared to the string
scale are
\beq
ds^2 = f(r)^{-{1 \over 2}} dx_{\|}^2 + f(r)^{1 \over 2} dx_{\perp}^2
~~~~~~e^{-2\phi} = f(r)^{p^\prime -3 \over 2},
\eeq
with
\beq
f(r) = 1 + g_s \left(
  {\sqrt{\alpha^\prime} \over r} \right)^{7-p^{\prime}}.
\eeq
On the visible sector probe Dp-brane world volume, these background bulk fields
yield possible corrections to the
potential and visible sector kinetic terms.
Evaluating the Dp-brane Dirac-Born-Infeld action in these background
fields
\beq
S_p =- T_p\int d^{p+1}x ~e^{-\phi} \sqrt{{\rm det}~ (h_{\mu \nu}
+ {1 \over 2} F_{\mu \rho} F^{\rho} _{\nu})},
\eeq
where $h_{\mu \nu}$ is the induced metric,
yields
\beq
S_p \sim  \int d^{p+1} x f(r)^{p^\prime -3 \over 4} f(r)^{-({p+1 \over
4})}
 \left[ 1+f(r)
\left({1 \over 2} \partial_{\mu} X^i \partial^{\mu} X_i - {1 \over 4}
F_{\mu \nu} F^{\mu \nu}.
\right)+ \cdots \right]
\label{dbranecorr}
\eeq
with indices now raised and lowered using the Minkowksi metric.
For $p=p^\prime$, the D-branes are in a ${1 \over 2}$ BPS configuration and
preserve 16 supersymmetries.
In this case, from (\ref{dbranecorr}), it is apparent that the
Dp-brane world volume kinetic terms receive no corrections.
The Kahler metric is flat and the Kahler potential is exact as
required with 16 supersymmetries.
(The correction to the potential term in (\ref{dbranecorr})
is canceled by the exchange of
the RR p-form antisymmetric tensor field for $p=p^\prime$).
For $p=p^\prime-4$, the configuration is ${1 \over 4}$ BPS and
preserve 8 supersymmetries.
In this case, the dilaton contribution cancels the gravitational
contribution to the potential, but there is a correction to the
Dp-brane world volume kinetic terms.
The flat inherited Kahler metric
for an isolated Dp-brane, which in the absence of other D-branes
would preserve 16 supersymmetries,
is modified by the background fields generated
by the Dp$^{\prime}$-brane.
The inherited Kahler potential is therefore modified by brane--brane
interactions in the configuration with only 8 supersymmetries.

The background fields generated by the Dp$^{\prime}$-brane which
modify the Dp-brane Kahler metric are a condensate
of closed string states.
In terms of brane--brane interactions this amounts at the perturbative
level to tree-level exchange of closed string states between the branes.
This classical exchange amplitude
may also be interpreted in the crossed channel
as a one-loop quantum amplitude over open string states which stretch
between the branes.
In the open string language this corresponds to a one-loop
correction to the Dp-brane kinetic terms.
This is not surprising since one-loop corrections to the Kahler
potential are allowed with 8 supersymmetries.


Modification of the Kahler metric and associated Kahler potential
in the above example illustrates
that corrections to the effective Kahler potential
generally exist in D-brane configurations with less than 16
supersymmetries.
{}From the expression for the modification of the kinetic
terms in (\ref{dbranecorr}) there is clearly
a direct coupling between the brane--brane separation $r$ and the
brane world volume fields.
If this modulus acquires an $F$-component,
then barring a cancellation, this contact interaction
leads to tree-level visible sector scalar masses.

The breakdown of the sequestered intuition for the leading
decoupling of separated branes may be understood in general
D-brane world models as illustrated above in two equivalent ways.
First, in the closed string channel tree-level couplings between
the branes can in fact give rise to non-derivative brane--brane
interactions.
Second, in the open string channel, the
quantum one-loop amplitude which involves extended
string states stretching between the branes is not part
of the low-energy local effective field theory description.
The low energy arguments for volume suppression
of point--particle quantum amplitudes therefore
do not apply.
In either channel,
the leading inherited Kahler potential is also seen not to be protected from
unsuppressed corrections in generic configurations.

In the closed string language warping of the background metric and
dilaton also affects the propagation of the RR
or vector potentials that generate the brane--brane
contact interactions discussed in section \ref{sec:locality}.
In theories with 8 or fewer supersymmetries this provides yet
another source of modification to the Kahler potential.
An example of this effect is presented in the next subsection.




\subsection{The Strongly Coupled Heterotic Theory On a Calabi-Yau
Space}
\label{sec:corrCY}

The Horava-Witten background of M-theory provides a brane world
model with end of the world branes with transverse co-dimension
one.
As discussed in section \ref{sec:hwsimple}, for a Horava-Witten
compactification of M-theory on $S^1/Z_2 \times T^6$
the 16 unbroken supersymmetries fix the form of the Kahler
potential.
This Kahler potential includes brane--brane couplings
which at low energy lead to non-derivative contact interactions between
matter fields residing on the two end of the world branes.
The origin of these brane--brane couplings may be understood
in, for example, the eleven--dimensional limit as arising
from exchange of the M-theory bulk four-form field strength,
as described in section \ref{hw10d}, or in the five-dimensional
limit as arising from the exchange of vector bosons as
described in section \ref{sec:hwfivelimit}.
In this description, it is reasonable to expect
that in backgrounds with less supersymmetry
brane--brane couplings arising from exchange of bulk fields
persists in general.
And since the Kahler potential is not as constrained with
less supersymmetry the specific form of the brane--brane
couplings may be more general.
In the brane world picture this can arise because
the form of the background and fluctuating bulk
fields and metric are also less constrained with less supersymmetry.

In this subsection the leading form of the Kahler potential
for a class of Calabi-Yau fibrations over $S^1/Z_2$
M-theory backgrounds which preserve only 4 supersymmetries is illustrated.
Calculable (in principle)
flavor violating
modifications to the lowest order Kahler potential
and tree-level soft masses
arise and may be traced to warping and distortion
of the bulk space between the branes.
 From
the bulk point of view these corrections are tree-level effects
and not suppressed in all regions of moduli space where
the brane world limit is obtained.
So even if the leading tree-level contributions to the soft scalar
masses happen to vanish in specific models, tree-level flavor
violation generally arises at generic points on moduli space.
And as discussed below these effects
are not likely to be small in a Horava-Witten model of nature.

For simplicity we consider Calabi-Yau compactifications
in which the spin connection is embedded in the $SU(3) \subset E_8$
visible sector gauge connection \cite{wittency}.
This leaves an unbroken $E_6 \times E_8^{\prime}$ gauge symmetry
with
$h^{1,1}(h^{2,1})$ generations(antigenerations)
of ${\bf 27}(\overline{\bf 27}) \in E_6$
chiral matter on the visible sector brane and pure $E_8^{\prime}$
super Yang-Mills on the hidden sector brane.
Horava-Witten compactifications of this type have been
extensively studied \cite{ovrutetal,ovrutten,losw}.
Since the hidden sector does not contain any chiral matter,
this class of compactifications is not a viable model of
BWSB arising from hidden sector matter.
However, these examples are instructive in illustrating
corrections to the lowest order form of the Kahler potential
in brane world backgrounds with only 4 supersymmetries.
Warping of the bulk metric in these examples
induces corrections to the tree-level Kahler potential
at order $T/S$.
These corrections are also interesting in that they do
not arise at any order in
perturbation theory in the weakly coupled heterotic string limit.

More realistic models with hidden sector chiral matter can arise
in Calabi-Yau fibrations over $S^1/Z_2$ in which the
spin connection is not embedded solely in the visible sector
gauge connection.
This requires in general
turning on a hidden sector gauge connection
and bulk four-form field strength.
Corrections to the Kahler potential appearing at
order $T/S$ are also expected in these classes of models,
as explained below.

For
toroidal compactification of the Horava-Witten background,
the $S^1 / Z_2$ M-theory bulk interval separating the branes
is flat.
This is guaranteed by the 16 unbroken supersymmetries and remains
exact in the full interacting theory.
Physically, this follows from the fact that the end of world
branes in a background with this
much supersymmetry do not have any tension
and therefore do not distort the bulk.
For compactifications with less supersymmetry the
brane tensions need not vanish and the
bulk is not guaranteed to be flat.
If the brane background sources are small, a systematic
expansion for the bulk fields may be developed
by expanding about the solutions for a flat interior
\cite{wittency}.


For the class of Calabi-Yau compactifications considered here
the brane tensions and background sources for the bulk four-form
field strength vanish to
zeroth order in an expansion in powers of $\kappa^{2/3}
R_{11}/V^{2/3}_{6} \sim T/S$, where $\kappa^2$ is the eleven-dimensional
Newton's constant and $V_6$ is the Calabi-Yau volume.
At this order the $S^1/Z_2$ interval is flat and the internal
space is a direct product $S^1/Z_2 \times CY$.
At higher orders in the $\kappa^{2/3}
R_{11}/V^{2/3}_{6}$ expansion, non-vanishing
brane fields act as sources for the metric and
bulk four-form field strength
through the modified Bianchi identity (\ref{modifieddG}).
In order to determine the four--dimensional Kahler potential
it is important to keep both background $(B)$ brane fields
resulting from the gauge connection embedding
as well as fluctuating $(F)$ brane fields which represent visible sector
fields. For the standard embedding,
$\hbox{tr}(F \wedge F)^{({\rm B})}_{E_8} = \hbox{tr}(R \wedge R)$ and
$\hbox{tr}(F \wedge F)^{({\rm B})}_{E_8^{\prime}} =0$,
the solution for the bulk
three-form potential associated with the four-form field strength
\cite{ovrutetal} with both background and fluctuating
brane sources using the unperturbed flat metric is
\beq
C_{ABC} = {\lambda \over 12} [
{1 \over 2} \omega_{E_8}^{({\rm B})}
+(1-x^{11})
(\omega_{E_8}^{({\rm F})} - {1 \over 2} \omega_L^{({\rm F})})
-x^{11}
(\omega_{E_8^{\prime}}^{({\rm F})} - {1 \over 2} \omega_L^{({\rm F})})
  ]_{ABC}
\label{cycsol}
\eq
where $x^{11} \in [0,1]$ and
$\hbox{tr}(F \wedge F)_{E_8} = d \omega_{E_8}$ and
$\hbox{tr}(R \wedge R)= d \omega_L$.
Note that the background brane gauge fields only induce a constant
three-form potential in the bulk at this order.
In fact with this embedding the background brane sources do not induce a
background four-form field strength in the bulk at any order in the
expansion, although the
fluctuating fields do give rise to a fluctuating field strength.

The background brane fields do, however, lead to finite brane
tensions.
Four--dimensional $N=1$ supersymmetry along with a vanishing
bulk four-form field strength resulting from the standard embedding
imply that one brane has positive tension while the other has
negative tension of equal magnitude.
This gravitational source warps the internal space
which becomes a fibration of Calabi-Yau over the $S^1/Z_2$ interval.
At the first non-trivial order
the induced perturbation of the eleven-dimensional metric grows
linearly between the boundaries.
The solution for the metric can
be expressed in terms of
$R^0_{11}$ and the $h^{1,1}$ Kahler moduli
appearing in the
low-energy theory.
The metric for the Calabi-Yau fiber
can
be expanded in a fixed basis $\{\omega^{n}_{i \bar{j}}\}$ for the
$(1,1)$ Kahler forms
as
\beq
g_{i \bar{j}} = V^{1/3} _6
\sum ^{h^{1,1}} _{n=1} b^n \omega^{n}_{i \bar{j}}
\label{cymetric}
\eeq
where $V_6=V_6(x^{11})$ and the $b^n=b^n(x^{11})$
are the $x^{11}$ dependent volume
and Kahler moduli, related to the standard normalization by
$\hbox{Re} S = V_{6}$ and
$\hbox{Re} T_n = R_{11} b^0_n $.
The $h^{1,1}$ moduli $b^n$ have a normalization which is
independent of $V_6$, $R_{11}$, and $x^{11}$, and are constrained
to satisfy
$d_{ijk} b^i b^j b^k =6$ (and therefore only describe
$h^{1,1}-1$ independent degrees of freedom)
where $d_{ijk}$
are the Calabi-Yau intersection numbers.
In terms of these moduli
the metric for the Calabi-Yau fiber is given by
(\ref{cymetric}) with moduli and volume varying
along $x^{11}$ as
\cite{ovrutetal,losw}
\ba
b^i  &=& b^i_0 -
  {R^0_{11} \over \sqrt{2} \hbox{Re} V^0_6 } \left( \gamma_i
-{2 \over 3} b^i _0 \sum^{h^{1,1}} _{k=1} \gamma_k b^k _0 \right)
(x^{11}-{1 \over 2})
\nonumber \\
V_6 &=& V^0 _6 \left( 1
+ {3 \over 2}\sum^{h^{1,1}} _{n=1} \alpha_n {\hbox{Re}T_n \over \hbox{Re} S}
(x^{11}-{1 \over 2})\right)
\label{moduliwarp}
\ea
where $V_6^0$ and $b_0^i$ are the unperturbed values and
$\alpha_n$ and $\gamma_n= 3 \alpha_n/ 2 \sqrt{2} $ are
numerical constants that depend topological data of the
Calabi-Yau and may be found in \cite{ovrutetal,losw}
and $x^{11} \in [0,1]$.
The four-dimensional metric in geometric frame, and
$S^1/Z_2$ interval metric along the $x^{11}$ direction,
are given by
\ba
g_{\mu \nu} &=& \left( 1+
  \sum^{h^{1,1}} _{n=1} \alpha_n {\hbox{Re}T_n \over \hbox{Re} S}
(x^{11}-{1 \over 2})  \right)
\eta_{\mu \nu}
\nonumber \\
g_{11,11} & = & \left(  1+ 2 \sum^{h^{1,1}} _{n=1}
\alpha_n
{\hbox{Re} T_n \over \hbox{Re} S}
(x_{11}-{1 \over
2}) \right) ({R^0}_{11})^2  \nonumber \\
\label{gwarpsol}
\ea
Because the brane tensions are equal in magnitude and opposite
in sign the metric perturbations averaged over the $S^1/Z_2$
M-theory direction vanish at this order for this class of compactifications
with standard embedding.
The zero-mode components of the four-dimensional
moduli are therefore equal to the unperturbed values,
$\langle V_6 \rangle = V_6^0$ and
$\langle b^i \rangle = b_0^i$.
It is important to note that
the bulk warping not only modifies the Calabi-Yau
volume along the M-theory direction, but also distorts
its geometry along this direction in a manner which
depends on the
four-dimensional moduli.

The magnitude of the bulk warpings (\ref{moduliwarp}) and
(\ref{gwarpsol})
induced by the brane tensions depends on the ratios
of moduli $ \epsilon_i  = T_i /S$.
This moduli dependence
is easily understood from the typical magnitude of the
brane sources $R \wedge R \sim 1 / V_6^{2/3}$.
Since for co-dimension one
the magnitude of the perturbation grows with the brane separation,
the total effect is proportional to $R_{11} / V_6^{2/3} = T/S$.
Since this dependence is a consequence of co-dimension one,
warpings for more general compactifications with
unequal brane tensions resulting from the spin
connection not embedded entirely in the visible sector gauge
connection should have the same parametric dependence.

In the Horava-Witten theory unification of four--dimensional
gauge and gravity couplings is obtained for an average Calabi-Yau volume
of order $S \simeq (3 \times 10^{16} ~\hbox{GeV})^{-6}$.
This, along with the value of the unified gauge coupling, imply
that numerically $S^{1/6} \simeq 2 \kappa^{2/9}$
and that the brane separation is
of order $R_{11} \simeq 8 \kappa^{2/9}$ \cite{wittency,banksdine}.
As advocated by Witten, one might hope that, given these
parameters, a supergravity approximation to the bulk physics is roughly
valid.
The expansion parameter $\epsilon = T/S$
for the overall volume modulus $T=R_{11} V_6^{1/3}$
is then of order one \cite{banksdine}.
The other moduli may generally be expected to have similar magntidue;
and therefore none of the expansion parameters may be
particularly small in a Horava-Witten model with gauge coupling
unification.
Aside from the $T_i/S$ dependence
the numerical factors in  (\ref{moduliwarp}) and
(\ref{gwarpsol}) are all order one numbers
which depend on the Calabi-Yau topology.
So distortion of the Calabi-Yau and the associated
modification of the inherited Kahler potential discussed below,
is likely to be
a significant effect in the Horava-Witten brane world theory.

The four--dimensional Kahler potential for this class of
compactifications with standard embedding can in principle
be obtained from the brane field action
arising in the perturbed background
given above, just as for the unperturbed case discussed in
section \ref{hw10d}.
There are two classes of corrections to the Kahler potential
which are important in determining
the visible sector soft masses arising from hidden sector
supersymmetry breaking.
First there are
corrections to the chiral matter kinetic term
wave functions, parameterized by $Z_{i \bar j}$
in (\ref{generalK}),
arising from the perturbations of the metric.
These have been computed from the warping deformations
(\ref{moduliwarp}) of the Calabi-Yau metric
(\ref{cymetric}) for the class of standard embeddings
considered here \cite{ovrutetal,losw}.
There are also
corrections to terms involving four
chiral matter fields, parameterized by $Z_{ i \bar j k \bar l}$
in (\ref{generalK}).
These arise from integrating out
the bulk four-form in the warped background metric,
and to our knowledge have not been computed.

The warping of the Calabi-Yau metric (\ref{cymetric})
at the position of the visible sector brane
modifies the normalization of the brane field kinetic terms
and therefore corrects the Kahler potential.
Since the magnitude of the metric perturbations averaged over
the $S_1/Z_2$ M-theory direction vanishes to ${\cal O}(\epsilon_i)$
with standard embedding, to this order Weyl rescaling from
geometric frame (obtained from compactification to four dimensions)
to Einstein frame is the same as in the unperturbed flat bulk case.
As discussed in section \ref{hw10d} the coefficients of the brane
matter kinetic terms in the flat case
are proportional to $1/T$ in four--dimensional Einstein frame.
So warping of the internal metric at the position of the visible
sector brane proportional to $T/S$ modifies the
visible sector brane matter kinetic terms in Einstein frame
by an amount proportional to $1/S$.
Since this is independent of $R_{11}$, this implies that
corrections to the quadratic terms in the
Einstein frame Kahler potential
must be proportional to $1/S$ and involve ratios of
$T_j/T_i$. In fact,
Lukas, Ovrut, Stelle and Waldram find
corrections \cite{losw} of just this form for the brane
matter wave functions
$$
Z_{i \bar{j}}(S,T_k) = Z^0 _{i \bar{j}}(T_k)
 + \delta Z_{i \bar{j}}(S,T_k / T_l) ~,
$$
\beq
\delta Z_{i \bar{j}}(S,T_k / T_l) = {1 \over S + S^{\dagger} }
h_{i \bar{j}}(T_k/T_l)
\label{nextorder}
\eeq
where $Z^0_{i \bar j}(T_k)$ is the wave function in the unperturbed
case without warping and is a function of the
Kahler moduli but not the dilaton as discussed in section \ref{sec:hwsimple}.
The function $h_{i \bar j}(T_k / T_l)$ is an order one function
determined by the lowest order Kahler potential of the (1,1) moduli
to be a function of ratios of these moduli.
The explicit forms for these functions for the class of compactifications
considered here have been computed
in terms of topological data $\alpha_n$ and intersection
numbers $d_{ijk}$ of the Calabi-Yau fiber \cite{losw}.
Note that since the wave function corrections do not
depend on $R_{11}$ in four-dimensional Einstein frame
they do not depend on the
overall volume modulus $T=R_{11} / V_6^{1/3}$.



A determination of the visible sector soft masses arising from
hidden sector supersymmetry breaking also
requires the corrections to the quartic terms
in the warped background,
$\delta Z_{i \bar{j} a \bar{b} }$,
appearing in the Kahler potential (\ref{generalK})
where $i,j$ refer to visible sector and $a,b$ refer to hidden sector.
In principle these couplings are determined by integrating
out the four-form field strength with brane Yang-Mills
Chern-Simons sources localized on the branes.
However, since the background bulk four-form field strength
vanishes exactly for compactifications in which the spin
connection is embedded in the gauge connection
the quartic couplings can arise only from integrating out
the fluctuating part of the bulk four-form field strength
sourced by brane field fluctuations.
This involves integrating over
the internal metric.
But since with standard embedding
the metric perturbations averaged over the
$S^1/Z_2$ interval vanish to ${\cal O}(\epsilon_i)$
the quartic couplngs are identical to those of the unperturbed
case at this order.
Corrections could however arise at ${\cal O}(\epsilon_i^2)$.
For a compacitification with more general embedding with
unequal brane tensions and non-vanishing background
bulk four-form field strength, corrections to the lowest
order quartic couplings would however be expected to arise at order $T_i/S$
since the metric perturbations do not average to zero in this case.
As discussed in section \ref{hw10d} the coefficients of the
brane--brane quartic Kahler potential couplings in the flat case
are proportional to $1/T^2$ in four--dimensional Einstein frame.
So for a compactification with general embedding
the brane--brane interactions in
Einstein frame should be modified by an amount proportional to $1/TS$.
For a standard embedding these corrections would vanish at this
order, and the next order corrections would modify the
quartic couplings by an amount proprotional to $1/S^2$.
For a general embedding the Einstein frame quartic Kahler potential couplings
should then be modified by warping and distortion of the internal
metric by an amount
$$
 Z_{i \bar{j} a \bar b}(S,T_k) = Z^0_{i \bar{j} a \bar b}(T_k)
 + \delta Z_{i \bar{j} a \bar b}(S,T_k / T_l) ~,
$$
\beq
\delta Z_{i \bar{j} a \bar b}(S,T_k / T_l) =
{1 \over (S + S^{\dagger})(T + T^{\dagger}) }
j_{i \bar{j} a \bar b}(T_k/T_l)
\label{nextquad}
\eeq
where $Z^0_{i \bar j a \bar b}(T_k)$ is the Kahler potential
quartic coupling in the unperturbed
case without warping and is a function of the
Kahler moduli but not the dilaton as discussed in section \ref{sec:hwsimple}.
The function $j_{i \bar j a \bar b}(T_k / T_l)$ is determined
by the lowest order Kahler potential of the (1,1) moduli
to be a function of ratios of these moduli.
For general embeddings it should be an order one function which
could in principle be determined from topological data of the
Calabi-Yau manifold.
For standard embedding $j_{i \bar j a \bar b}(T_k / T_l)$
includes at least an explicit factor
of $ \epsilon $.
To our knowledge the lowest order quartic terms
even without warping have not been calculated for a Calabi-Yau
fibration with any embedding.



With hidden sector
supersymmetry breaking, the visible--hidden quartic couplings
in Einstein frame for the general Kahler potential
(\ref{generalK})
to lowest order in the expansions described above
for a general Calabi-Yau fibration over $S^1/Z_2$ are,
from the associated supergravity $f$ function (\ref{generalf}),
proportional to
\beq
{1 \over 3} \left( Z^0_{i \bar{j}} Z^0 _{a \bar{b}} +
{ 1 \over S + S^{\dagger} }
 (  h_{i \bar{j}} Z^0_{a \bar b} + Z^0_{i \bar j} h_{a \bar{b}} )
\right)
 - \left( Z^0_{i \bar j a \bar b}  +
    {1 \over (S + S^{\dagger})(T + T^{\dagger}) }
          j_{i \bar{j} a \bar b}
 \right)
\label{zepsmass}
\eeq
In order to obtain the physical mass squared matrix
the scalar kinetic term wave functions factors
\beq
Z^0_{i \bar j} + {h_{i \bar j} \over S + S^{\dagger} } ~~~~~~~,~~~~~~~
Z^0_{a \bar b} + {h_{a \bar b} \over S + S^{\dagger} }
\eeq
must be rescaled to give canonically normalized fields.
As discussed in section \ref{hw10d},
in four--dimensional Einstein frame the coefficients of the brane
matter kinetic terms in the flat case
are proportional
to $1/T$ while the quartic couplings are
proportional
to  $1/T^2$.
To lowest order, extracting this dependence and rescaling
the fields by a compensating amount,
the dependence on the
expansion parameter $(T+T^{\dagger})/(S+S^{\dagger})$ is
apparent.
In order to obtain the physical masses, the mass squared matrix
must be diagonalized and canonically normalized,
which in general requires a unitary rotation and rescaling.
In a basis which both the visible
and hidden sector are diagonalized
at zero-th order in the
expansion, the mass squared matrix is proportional to
\beq
\left( {1 \over 3} \delta_{i \bar j} -
    \tilde{Z}_{i \bar j a \bar b} X_{a \bar b} \right)
  + { T + T^{\dagger} \over S+S^{\dagger} } \left[
   {1 \over 3} \tilde{h}_{i \bar j} +
     \left( {1 \over 3} \tilde{h}_{a \bar b} -
       \tilde{j}_{i \bar j a \bar b} \right)
       X_{a \bar b}  \right]
\eq
where the tilde functions are related to the previous ones by
rotation and rescaling, and
$X_{a \bar b} = F_a F^*_{\bar b} / {\rm tr}_{c \bar d}(F_c F^*_{\bar d})$.
Even if the lowest order masses arising
from the first terms in paranthesis are universal, as would occur
with a no-scale Kahler potential,
they are not universal in general.
The leading ${\cal O}(\epsilon_i)$ effects arise by
diagonalization at this order through rescalings  which
lead to non-degenerate masses at the same order.
Given that the functions inside the square paranthesis are unrelated
in any simple way, there
is no reason for the non-degenerate contributions
from these terms to cancel.
The upshot is that generically there are irreducible
non-degenerate
contributions to the visible sector masses appearing
at ${\cal O}({\epsilon}_i)$.
However, without a theory of flavor this rescaling need not
be aligned with the quark and lepton mass eigenstates,
and would in general introduce dangerous supersymmetric
flavor violation, since the expansion parameters $T_i/S$ are not
particularly small in Horava-Witten models.
Geometrically the flavor violation arises
because the bulk
warping (\ref{moduliwarp}) distorts the Calabi-Yau geometry
(\ref{cymetric}) at the position of the brane
in a manner depending on the four-dimensional
moduli
and is therefore in general not aligned with the lowest
order metric or matter field wave functions.

For Calabi-Yau manifolds with the standard embedding of the
spin connection in the visible sector gauge connection
with $h^{1,1}=1$ the Kahler potential including the
lowest order effects of warping can be obtained in a simple form
\cite{ovrutetal}
\beq
K= -\ln(S+S^{\dagger} - \epsilon Q^{\dagger} Q) - 3 \ln (T + T^{\dagger}
- Q^{\dagger} Q) ~.
\label{owkahler}
\eeq
where $\epsilon$ is the fixed background value
$(T + T^{\dagger})/(S + S^{\dagger})$ evaluated on the unperturbed
background, while $S$ and $T$ appearing (\ref{owkahler}) including
fluctuating pieces.
With the standard embedding the corrections to the quartic
terms vanishes to this order as discussed above, and the
only warping effect is the overall normalization
of the visible sector matter wave function.
The lowest order no-scale form is modified
at first non-trivial order.
Although with standard embedding this class of backgrounds
is not useful for hidden sector BWSB since the hidden sector
does not contain any chiral matter, this modification of the Kahler
does demonstrate that the lowest order Kahler is modified
in a non-trivial way by warping as expected.
The tree-level warping and distortion
corrections to the Kahler potential also
have implications for a dilaton dominated scenario
where
only the dilaton auxillary component is non-vanishing
(or dominant).
To lowest order the
visible sector scalar fields
acquire from the Kahler potential
(\ref{generalK}) with (\ref{nextorder}) a
universal
soft mass. But to next leading order
the second term in
(\ref{nextorder}) with non-vanishing
dilaton auxillary component generates a
general matrix in flavor
space proportional to $ h _{i \bar{j}} |F_S|^2 $ and this is
only suppressed only by $T_i/S$. Although the
kinetic terms given by $Z_{i \bar{j}}$
are also not diagonal, the point is that generically
these
two matrices are not simultaneously
diagonalizable.
Again this flavor violation is not small.
So we see in the Horava-Witten brane world theory
without additional assumptions about flavor symmetries, supersymmetric flavor
violation is not particularly suppressed.
Because the bulk is co-dimension one the warping of the Calabi-Yau
grows with transverse distance and these effects become larger
if the brane separation $R_{11}$ is increased, but become smaller
if $V_6$ is increased.



This again illustrates that brane world realizations of
hidden sector supersymmetry breaking alone do not provide a
solution of the supersymmetric flavor problem.
In the absence of assumptions about flavor symmetries, supersymmetric
flavor violation is not necessarily suppressed over much of moduli space,
just as for
standard hidden sector supersymmetry breaking scenarios.
In addition, the flavor violating effects are tree-level
effects from the bulk point of view.
So the lowest order
tree-level Kahler potential is not protected in any way from recieving
tree-level flavor violating corrections over much of moduli space.
In the Horava-Witten theory with moduli expectation values
which allow standard gauge coupling unification
these effects are significant.

Finally, it should be emphasized that the lowest order form of
the brane--brane
couplings discussed here as well as the corrections
arising from warping and distortion of the internal
geometry are obtained in the eleven-dimensional supergravity limit.
In this limit the order $T_i/S$ corrections result from the finite
M-theory interval separating the branes.
Now effects associated with the M-theory
direction do not appear at any order
in perturbation theory in the heterotic string theory limit.
So this class of non-holomorphic corrections to the Kahler potential are not
visible in perturbative heterotic string theory.
As long as the brane separation $R_{11}$ and Calabi-Yau volume
$V_6$ and inverse curvature
are large in eleven-dimensional Planck units, the supergravity
approximation should be good and quantum M-theory
corrections should be at least power law suppressed in these
parameters.
There are in additional semi-classical non-perturbative effects
due to M2 and M5 brane instantons which are exponentially suppressed.
These are visible in the perturbative heterotic limit as
world sheet and gauge theory instantons as discussed in section
\ref{sec:hwsimple}.


\subsection{Kahler potential Corrections with
Pure Five--dimensional Supergravity}
\label{fivedwarp}

Pure five--dimensional supergravity with
a flat unwarped interval separating end of the world branes
can give a low energy four--dimensional no-scale Kahler potential
of the separable sequestered
form \cite{randallsundrum,lutysundrum},
at least in the case
of inheritance from the $N=4$ form,
as argued in section \ref{sec:puresimple}.
A flat bulk is only obtained, however, for vanishing brane and bulk
tensions.
Non-vanishing brane tension will generically
warp the bulk metric.
The presence of a stabilizing mechanism
will also contribute to the warping due to non-zero
bulk stresses.
So in general the four--dimensional metric is expected to be
a function of the coordinate, $y$, transverse to the branes
\beq
g_{\mu \nu}(y) = (1+ \xi(y)) \eta _{\mu \nu},
\label{pmetric}
\eeq
where $\xi(y)$ is a model dependent function.
The warping of the metric (\ref{pmetric}) at the positions of the
branes gives a universal modification of the brane field kinetic
terms.
For perturbations about a flat interior the warping
will typically grow linearly with the
brane separation due to the one--dimensional geometry.
This warping induced modification will be reflected in a
modification of the four--dimensional Kahler potential.
The lowest order no-scale Kahler potential (\ref{noscalef}) inherited from
the five--dimensional theory is then generally modified.
This is consistent with the fact that a general warped compactification
only preserves $N=1$ supersymmetry in four dimensions, so the
inherited form of the Kahler potential is not protected in any way.
Corrections to the inherited no-scale Kahler potential are
small if the brane tensions are small.


The pure five--dimensional supergravity with end of the world
branes scenario is very similar to the five--dimensional
limit of the Horava-Witten theory discussed in section \ref{sec:hwfivelimit}.
Here, however, there is by assumption no dilaton.
In addition there is only a single bulk vector
boson which is part of the five--dimensional gravitational
supermultiplet, whereas in the Horava-Witten theory on a generic
background there are a number of additional vector bosons.
In pure five--dimensional supergravity, exchange of the single
vector boson between the branes can be understood as giving
rise to brane--brane current--current couplings contained
in the no-scale Kahler potential \cite{lutysundrum},
while in the Horava-Witten theory exchange of
the additional vector bosons can be understood as giving rise
to the more general non-sequestered form of the inherited
Kahler potential which contains non-derivative
brane--brane contact interactions, as discussed
in section \ref{sec:hwfivelimit}.
Just as in the Horava-Witten theory, warping of the bulk
geometry should lead to corrections to the lowest order inherited
form of the Kahler potential.
And as described below,
in the pure five--dimensional case in fact warping
generally does leads to visible
sector tree-level scalar masses, even though the masses vanish
in the flat case.

In the Horava-Witten theory with bulk warping discussed in
the previous subsection, it was possible to argue for the form
of the four--dimensional Kahler potential to first order in
perturbation about a flat background by inspecting the form of
the brane field interactions in the underlying theory.
In this pure five--dimensional supergravity scenario this is
not possible without a consistent underlying theory.
We therefore assume an
{\em ansatz} for the four--dimensional Kahler
potential with bulk warping of the form
\beq
K= -3  \ln
\left(
 f_{\rm mod}(T+T^{\dagger}) -
(1+h_{\rm vis}(T+T^{\dagger}))
{\hbox{tr}Q_i^{\dagger}Q_i  }
 -(1+h_{\rm hid}(T+T^{\dagger})) \hbox{tr}{\Sigma^{\dagger}_i
\Sigma_i } ,
\right)
\label{guess}
\eeq
where $h_{\rm vis}$ and $h_{\rm hid}$ are assumed to be universal
factors for each field localized on a given brane,
and $f_{\rm mod}$ is the total compact volume,
and $T$ is the radion modulus.
These  are model dependent functions of the brane tensions
and bulk stress-energy momentum tensor.
For a flat bulk $h_{\rm vis}(x)= h_{\rm hid}(x) =0$
and $f_{\rm mod}(x)=x$, resulting in the no-scale form.
Kahler potentials of this type have been obtained
\cite{baggerpok,lutysundrumads} in the supersymmetric version of
the Randall-Sundrum model with two branes \cite{randallsundrumi},
where the warping is obviously an important effect.


The Kahler potential (\ref{guess}) contains a direct coupling
between the radion or volume modulus, $T$, and visible sector
brane fields $Q_i$.
In the geometric frame obtained by direct compactification to
four dimensions this follows from the warping induced dependence
of the visible sector kinetic terms on the compact volume.
This survives unchanged in the supergravity frame since these frames
coincide for compactification from five to four dimensions.
If the radion obtains an auxiliary expectation value,
$F_T \neq 0$, this direct coupling leads to a tree-level
visible sector scalar mass.
As discussed below in more detail this is in fact the case in
most scenarios for stabilizing the bulk
geometry. From (\ref{guess}) the ratio of visible sector scalar masses to
the gravitino mass for both radion and hidden sector
auxiliary fields, $F_T \neq0$ and $F_{\Sigma} \neq 0$,
is
\beq
 {m^2 _Q \over m^2_{3/2}}
={ (T+ T^*)^2 \over 9 (1+h_{\rm vis})
  (f_{\rm mod}^{\prime 2} - f_{\rm mod} f_{\rm mod}^{\prime
\prime})^2}
\left( {(h_{\rm vis} ^{\prime} )^2 \over 1+h_{\rm vis} } -
   h_{\rm vis} ^{\prime \prime}\right)
\left({\vert F_T \vert ^2  \over \vert W \vert ^2}
  (T+ T^*)^2  \right)
\label{masswarp}
\eeq
where here $^{\prime} \equiv \partial/ \partial \hbox{Re} T$.

The expression for the soft masses (\ref{masswarp})
has a number of interesting features.
First the masses are universal since by assumption of
the ansatz the warping modification of kinetic terms is the same
for all fields on the visible sector brane.
Universality would, however, not be preserved for visible
sector fields propagating on different branes.
And the
$h$ functions in general need not even be universal for fields
propagating on the same brane.
Second, the overall magnitude of the tree-level mass (\ref{masswarp})
is clearly model dependent, but can be sizable.
The vanishing cosmological constant condition gives
an upper limit on the last term in parenthesis of
$ 2\hbox{Re}T |F_T|  / |W| \leq 3$.
The masses do, however, vanish even with $F_T \neq 0$ if
the condition condition $(h_{\rm vis}^{\prime})^2 =
(1 + h_{\rm vis}) h_{\rm vis}^{\prime \prime}$
is satisfied.
This arises for a flat interior $h_{\rm vis}(x)=1$ as well
as for an AdS intertior $h_{\rm vis}(x)=e^{-x}-1$
(a boundary interpretation of the latter case is presented
in subsection \ref{sec:conseq}).
However, the requisite dynamics which stabilizes the $T$ modulus
will generically create stresses which lead to warping of the bulk.
In general this will not preserve the flat interior metric or
result in a pure AdS warping.
Third, even though the auxillary component for the $T$ modulus,
$F_T$, contributes directly to the scalar masses, auxillary
hidden sector auxiliary components, $F_{\Sigma}$,
do not give rise directly to tree-level visible sector masses, as a
consequence of the assumed ansatz for the Kahler potential (\ref{guess}).
This is apparent in supergravity frame where
the supergravity $f$ function associated with (\ref{guess})
is separable and therefore does not give rise to non-derivative
brane--brane couplings between visible and hidden sector fields.
So in this sense the hidden sector does remain sequestered
with the Kahler potential ansatz (\ref{guess}), and
it is only the indirect effects of warping and a non-vanishing
$F_T$ which lead to scalar masses.
Finally, the masses are directly proportional to the radion
auxillary expectation value, $F_T$.


Since the auxillary component of the $T$ modulus, $F_T$, is
crucial in determining the scale for the resulting soft scalar
masses it is instructive to consider its magnitude including the
effects of stabilization and vanishing cosmological constant.
To see this consider the supergravity potential including
the effects of hidden sector supersymmetry breaking.
The derivative of the potential with
respect to $T$, assuming a canceled cosmological constant is
\ba
0= V^{\prime} &=& e^{K} \left( K^{T T^*}
F^* _T F^{\prime} _T +K^{T T^*} F_T K_{T T^*} W^*  + K^{T T^* \prime}
F^*
_T F_T
\right. \nonumber \\
& & \left.
+ K^{\Sigma \Sigma^* \prime} |F_{\Sigma}|^2 -3 W^* W^{\prime} \right)
\ea
where $\langle \Sigma \rangle \ll 1 $ is assumed, and
$^{\prime}= \partial / \partial T$,
and the superpotential is assumed to factorize as
$W=W(T)+W(\Sigma)$.
An extremum of the potential with
vanishing radion auxiliary component
would require $K^{\Sigma \Sigma^* \prime} |F_{\Sigma}|^2
=3 W^* W^{\prime}$. Combining this with the condition for
vanishing cosmological constant, $K^{\Sigma \Sigma^* } |F_{\Sigma}|^2
= 3 |W|^2$, implies $K^{\Sigma \Sigma^* \prime}/K^{\Sigma \Sigma^* }=
- K^{\prime}$ which is not satisfied for the no-scale potential
with the prefactor 3.
So one finds that $F_T=0$ is not an extremum of the potential
with hidden sector supersymmetry.
So assuming vanishing cosmological constant with $F_{\Sigma} \neq 0$,
then $F_T$ must be non-vanishing in the stable ground state.

The overall magnitude of
the radion auxiliary component
depends on whether the radion is stabilized in the
globally supersymmetric limit or only including supergravity effects.
First consider the case in which it is stabilized in the globally
supersymmetric limit
in the absence of hidden sector supersymmetry
breaking.
In this case $\langle W^{\prime} \rangle =0$ and $F_T=0$
in this limit.
So near the ground state minimum $\langle T \rangle = T_0$ the
superpotential can be expanded as $W={1 \over 2} M (T-T_0)^2
+ \cdots$. One can then verify that with
hidden sector supersymmetry breaking turned on
$W^{\prime}=0$ is no longer
a solution.
This is clear from the following equivalent expression
for $V^{\prime} $ obtained by substituting for $F_T^{\prime}$
\ba
0=V^{\prime} &= & e^{K} \left( K^{T T^*} F^* _T (W^{\prime \prime}
+ K_{TT} W + K_T W^{\prime})
\right. \nonumber \\
& &  \left.  + K^{T T^* } K_{T T^*} W^{*} F_T
+K^{T T^* \prime} |F_T|^2
+  K^{\Sigma \Sigma^* \prime }
|F_{\Sigma} | ^2 -3 W^* W^{\prime} \right)    ~.
\label{vprime}
\ea
Now suppose that $W^{\prime}=0$ which implies
$F_T = K^{\prime} W$. Then with $W \sim {\cal O}(m_{3/2})$,
all the remaining
terms in this equation are ${\cal O}(m^2_{3/2})$
except the one
involving
$W^{\prime \prime}$ which is ${\cal O}(m_{3/2})$.
For $M \gg m_{3/2}$ (which is the case if globally supersymmetric
dynamics dominates the stabilization of the radion) this term cannot
possibly cancel the
other terms. So $W^{\prime} \neq 0$ which
implies that $F_T \neq 0$ with the superpotential above.
In this limit with $M \gg m_{3/2}$
one then expects that an expression for the extremum
can be written as an expansion in terms of the
of the radion auxiliary component, derivatives of the superpotential,
and $\Delta \equiv
(T-T_0)/T_0$ in powers of $m_{3/2}$.
Having established that $F_T$ and $W^{\prime}$ are
both nonzero, one might guess that $F_T
\sim {\cal O}(m_{3/2})$.
An inspection of (\ref{vprime})
indicates that this is not possible for
essentially the same reasons that $W^{\prime}$
could not vanish. Namely, all the terms are
manifestly ${\cal O}(m^2_{3/2})$ except the
term $F^*_T W^{\prime \prime}$
which would be  ${\cal O}(m_{3/2})$.
In fact, solving (\ref{vprime}) for $\Delta$ implies
$T-T_0 =3 \langle W \rangle / 2 M \hbox{Re}(T)
+ \cdots$ which leads to
a vanishing radion auxiliary component at
lowest order. At next order one finds
$F_T \sim {\cal O}(m^2_{3/2}/M)$.
Thus radion stabilization in the global supersymmetric
limit leads to a nonzero but negligible radion
auxiliary component.
So in this case the soft scalar masses resulting from
warping would be insignificant.

Another possibility is that the radion is stabilized
only in the presence of supersymmetry breaking and only including supergravity
interactions.
Examples of this type have been studied \cite{lutysundrum,lutychacko}.
Inspecting (\ref{vprime})
one finds that a consistent solution
with the superpotential $\langle W \rangle$,
its derivatives, $\langle W^{\prime} \rangle$,
and the radion auxiliary component $F_T$, all
order ${\cal O}(m_{3/2})$ at the extremum is possible.
Of course,
the dependence of the radion auxiliary component
on the volume is model--dependent.
In the previous example with the radion stabilized
in the supersymmetric limit the obstruction to finding
such a solution appears to be that there
$W^{\prime \prime } \sim M$.
An example of a model in which the radion is stabilized
by the supersymmetry breaking and supergravity interactions
with $F_T \sim {\cal O}(m_{3/2})$
is provided by both bulk and
boundary gaugino condensation which in the presence of
hidden sector supersymmetry breaking stabilizes the radion
\cite{lutychacko}. In models of this type warping of
the bulk geometry generally induces tree--level visible
sector masses through the radion auxiliary component.

\subsection{Conformal Sequestering from Geometry}
\label{sec:conseq}

The question of what classes of models have vanishing or suppressed
tree-level soft masses even with hidden sector supersymmetry breaking
is an interseting one.
In these cases if the scalar masses do in fact vanish then
anomaly mediation can be important \cite{randallsundrum,lutyetal}.
As we have seen, vanishing tree-level masses do not generally
arise from BWSB, but can be achieved in specific models.
In particular, pure five--dimensional supergravity with
a flat interior separating end of the world branes
 gives vanishing tree-level masses
\cite{randallsundrum,lutysundrum},
at least by inheritance if this case can
be obtained by a string or M-theory orbifold as
argued in section \ref{sec:puresimple}.
Vanishing masses are not protected once the effects of
warping and radion stabilization are included, and tree-level
masses generally result, as discussed in the previous subsection.
However, the soft masses (\ref{masswarp}) for the pure five--dimesional
case do in fact vanish for arbitrary $F_T$ if the special
condition
\beq
(h_{\rm vis}^{\prime})^2 =
(1 + h_{\rm vis}) h_{\rm vis}^{\prime \prime}
\label{spcon}
\eq
is satisfied, where $h_{\rm vis}$ is the function parameterizing
the bulk warping in the Kahler potential ansatz (\ref{guess}).
This condition is satisfied not only for a flat internal metric, but also
for an internal metric which is a section of AdS space
with exponential warping
\beq
m_Q^2 =0 ~~~~~\hbox{for}~~~~~
1 + h_{\rm vis}(T + T^{\dagger}) = \left\{
\begin{array}{ccl}
1 & & \hbox{flat} \\
e^{-k(T+T^{\dagger})} & & \hbox{AdS}
\end{array}
\right. \eeq An exponential warping of this form occurs in the
supersymmetric version of the Randall-Sundrum model with two
branes \cite{randallsundrumi,lutysundrumads,baggerpok}.

The vanishing visible sector scalar masses which result with a
co-dimension one AdS bulk can probably be traced to the existence
in the five--dimensional bulk of Anti-de Sitter supersymmetry
which is related in four dimensions to superconformal symmetry
\cite{ls2001}. This relation can be made precise through the
AdS/CFT correspondence between a boundary four--dimensional
conformal field theory and a five--dimensional anti-de Sitter
space. This correspondence suggests that there may be a boundary
description entirely in terms of four--dimensional superconformal
field theory dynamics for the vanishing of scalar masses with a
five--dimensional pure AdS bulk \cite{ls2001}. Of course a precise
correspondence between a boundary field theory and large low
curvature bulk is only good at large $g^2N$. But it is still
interesting to explore the heuristic correspondence between the
geometric brane world example considered here and
four--dimensional field theories based solely on approximate
conformal symmetry.

Two classes of field theory models which are
approximately superconformal have been suggested
for suppressing visible sector
scalar masses \cite{nelsonstrassler,ls2001}.
Both rely on the observation that in global
supersymmetry the soft mass squared for the scalar component
of a chiral multiplet may be represented as the
highest component of the wave function factor treated
as a vector superfield.
$m^2 = Z|_{\theta^2 \bar{\theta}^2}$.
Now field theory dynamics which is approximately
conformal over some range of scales and couples
to the wave function factor results in an approximately
constant anomalous dimension $\gamma$ for the wave function
over this range, $Z=Z_0(\mu / \mu_0)^{ \gamma}$,
where $\mu$ is the renormalization group scale and $\mu_0$
refers to the scale below which the theory is approximately
conformal. As \cite{nelsonstrassler} demonstrates, if
all non-$U(1)_R$ symmetries are explicitly broken,
then the highest component, namely the soft mass,
likewise also has an approximately constant anomalous dimension
over this range.
This results in a potentially sizeable
suppression of the soft mass.

The first class of field theory models
proposed by Nelson and Strassler
utilize strong dynamics which is approximately conformal
over some range of scales and couples directly at tree-level
to some of the visible sector scalars \cite{nelsonstrassler}.
The second class of models proposed by Luty and Sundrum
assume a hidden sector which is strongly interacting
and conformal down to a low scale, and couples only to
the weakly coupled visible sector fields through non-renormalizable
operators represented by the wave function factor in the low energy
theory of the hidden sector \cite{ls2001}.

The AdS/CFT correspondence suggests a five--dimensional bulk
geometric interpretation of these models \cite{ls2001}. In the
AdS/CFT correspondence, five--dimensional anti-de Sitter space
with exponential warping corresponds to a conformal field theory
in the four--dimensional boundary theory. In the present case the
end of the world branes break the full anti-de Sitter symmetry
which in the boundary description correspond to UV and IR mass
scales of the theory. In this interpretation the end of the world
branes may be thought of as the UV and IR ends points of the
conformal renormalization group flow.
Interactions which couple the visible and hidden sectors in the
boundary field theory description, $(1/M^2)\int d^4
\theta~Q^{\dagger} Q \Sigma^{\dagger} \Sigma$ in global
supersymmetry, are suppressed by a large mass scale $M$ and so can
be thought of as generated by physics residing on the UV brane.
The mass scale $M$ then determines the position of the UV brane.

In the Nelson--Strassler models the strongly coupled approximately
conformal dynamics which couples directly to the standard model
corresponds in the bulk to the AdS warping.
In the Luty--Sundrum models it is the strongly coupled approximately
conformal hidden sector which corresponds in the bulk to the AdS warping.
In both cases the exponential warping in the AdS bulk
between the UV and IR branes corresponds
to the exponential suppression of the
soft mass operators $m^2 = Z|_{\theta^2 \bar{\theta}^2}$
in the boundary theory between the UV and IR limits
of the conformal renormalization group flow.
Since the boundary theories are only approximately conformal
over some range of scales the corresponding bulk geometry should only
be approximately AdS with some residual warping near the UV and
IR walls where conformal invariance is violated.
This deviation from AdS would presumably result
in only exponentially suppressed masses from
(\ref{masswarp}) in the bulk geometric
picture as required by the bulk/boundary correspondence.

In the Luty-Sundrum conformal suppression
mechanism
there is potentially a puzzle in
translating the
anomalous dimension effects of the hidden sector wave functions
in the boundary theory written
in globally supersymmetric language as a D-term quartic interaction between
the visible and hidden sectors, into
an operator written in
locally
supersymmetric language.
In particular, the boundary theory
suppression of D-term quartic interactions between
the visible and hidden sectors in global supersymmetry
applies in the locally supersymmtric
theory to the suppression of quartic
operators
in the IR in
which frame -- supergravity or Einstein frame?
Since the supergravity frame is manifestly supersymmetric
this seems the most natural choice.
And as discussed in section 2, vanishing of non-derivative
quartic couplings in this frame results in vanishing scalar masses.
A suppression in the IR of the couplings appearing in the
Kahler potential would with supersymmetry
breaking in the ${\em visible}$  sector lead
to
universal masses in the ${\em hidden}$ sector
of order the gravitino mass.
Since this would
not give a picture consistent
with the results of \cite{nelsonstrassler},
the suppression occurs in the supergravity frame.

In the case of a flat five--dimensional bulk discussed in
section \ref{sec:hwfivelimit}, vanishing tree-level scalar
masses are only obtained if the only vector boson in the
bulk is that of the minimal supergravity multiplet.
Additional bulk vector multiplets coupled to
brane matter
Cherns-Simons forms through a modified Bianchi identity
do give rise to tree-level masses from hidden sector supersymmetry
breaking.
Although an analogous calculation for the case of a five--dimensional
AdS with additional vector multipets coupled to brane matter
as above has not been completed, it should
be expected that also in
this case tree-level masses arise from integrating out the bulk vectors
(more on this below).

In the case of a flat interior
the existence of bulk gauge symmetries under which the five--dimensional
vector bosons transforms is closely related to the existence of
unbroken global symmetries which act on the brane matter and vice versa.
This is at least true in orbifold compactifications for which
bulk five--dimensional gauge symmetries arise microscopically
from higher dimensional diffeomorphisms are in one to one correspondence
with global symmetries on the D-brane which have the same origin.
And this correspondence is likely to hold also for more general
compactifications.
So in the flat interior case vanishing tree-level masses require
no gauge symmetries in the bulk (beyond that related through
the minimal supergravity multiplet) or equivalently that the
D-brane matter not possess any global symmetries.

Now in the AdS/CFT correspondence, on very general grounds
and by construction, global symmetries in the boundary theory
are realized as local gauge symmetries in the bulk.
In the boundary field theory description the
existence of an unbroken non-anomalous global symmetry has an
important effect on the mechanism of conformal suppression of
soft masses \cite{nelsonstrassler}.
The anomalous dimension for an exactly conserved global
current vanishes.
So for every Abelian global symmetry in the conformal
field theory
(which commutes with
supersymmetry -- see below) there
is one eigenvector (determined by the global charges) of the
(visible or hidden)
soft mass squared matrix which is not suppressed by
the conformal suppression mechanism.
In the AdS/CFT correspondence such global symmetries in the
conformal field theory are realized by local symmetries in the
bulk with concomitant vector bosons.

So in
the bulk AdS interpretation of
the field theory conformal suppression mechanism, vanishing
scalar masses require the absence of global symmetries acting
on the
conformal sector matter.
In the field theory interpretation, if the goal
is vanishing tree-level scalar masses even with hidden
sector supersymmetry breaing, the model building problem
is to find theories with the conformal suppression mechanism
but no global symmetries.
In the bulk language the problem is to find compactifications
which reduce to pure five--dimensional Anti-de-Sitter
supergravity with
end of the world branes and no bulk
vector multiplets.
Which is easier may depend on the model builder.

As a final comment about the bulk--boundary correspondence
between these pictures of conformal sequestering, note
that superconformal invariance of an $N=1$ four--dimensional boundary
theory requires the existence of an unbroken $U(1)_R$ symmetry.
This current is related by superconformal symmetry to the
dilation current, and since it does not commute with supersymmetry
it does not leave an eigenvector of the scalar mass squared
matrix unsuppressed.
In the bulk picture, anti-de Sitter supersymmetry requires
the existence of a single $U(1)$ gauge boson in the minimal
gravity multiplet, and couples on the boundary to the
$U(1)_R$ current.
And (at least in the flat interior case) does not give
non-derivative couplings between the branes which would lead to
tree-level soft masses.
So in both pictures the one Abelian symmetry which is required
by supersymmetry does not give
rise to soft masses, but additional ones would.

\section{Massive Moduli}
\label{sec:massmoduli}

The lowest order inherited Kahler potentials in all the brane world
scenarios presented here (with the exception of pure
five--dimensional supergravity) are not sequestered
and contain brane--brane non-derivative contact interactions
which give rise to unsuppressed tree-level visible sector
scalar masses from hidden sector supersymmetry breaking.
All these contain, in the absence of supersymmetry breaking,
four--dimensional massless moduli such as the $T_{i \bar j}$.
Although it seems a priori unlikely, one might wonder if stabilizing
these moduli in the four dimensional effective
theory could modify the form of the inherited Kahler potential
in such a way that tree-level masses are not induced and
that the sequestered intuition is recovered.
In this section we show that the opposite is the case.
Namely, without stabilizing the moduli the tree-level masses
vanish. This feature is not
restricted to the special sequestered
Kahler potential but applies the more general
class of no-scale Kahler potentials
and appears to be a consequence of the auxillary
equations of motion of the moduli. When the moduli
are stabilized however,
it is seen
that
the most naive expectation that with hidden sector
supersymmetry breaking the moduli can simply be replaced with
scalar expectation values is correct.

There are two possibilities for the scales associated with
stabilizing the moduli.
The first is that the moduli gain masses well below the compactification
scale.
The analysis of the previous sections within the low energy
four--dimensional theory
is then unaltered since the moduli are part of this theory,
and the Kahler potential and resulting patterns of
tree-level scalar masses are obtained assuming moduli
stabilization as detailed below.
Alternately one might hope that the moduli gain a mass much
larger than the compactification scale.
In this case, a different low energy theory might in principle
result.
This latter option seems unlikely, however, at least for moduli
which arise from dimensional reduction of supermultiplets
which are part of the higher dimensional theory and appear
in the effective theory above the compactification scale.
These moduli must be present in the bulk
space and are protected from gaining a mass parameterically
larger than the compactification scale by gauge or
supersymmetries
of the underlying theory.
Even for other moduli which do not arise directly from extended
supermultiplets, stabilization or projection
mechanisms generally give masses
at most of order the compactification scale.
At best it might be possible to consider compactifications
with more than one geometric compactification scale.
The minimal set of moduli in the low energy four--dimensional
theory are then just those that arise from extended supermultiplets
in the effective theory just above the lowest compactification scale.
Other moduli (or vector multiplets)
could gain masses at the higher compactification scales and
not appear in the five dimensional effective theory.
The most extreme possible example of this type is a compactification
of a fundamental theory of gravity
to pure five--dimensional supergravity at a mass scale well above
the compactification scale from five to four dimensions.
Since by assumption all moduli which might be present
in the five--dimensional theory are lifted by the higher mass
scale compactification, the bulk contains no additional
vector or hyper multiplets, and
the only modulus which survives
in the four--dimensional theory is the radion supermultiplet
which arises from fields in the five--dimensional supergravity
multiplet.




In order to investigate the influence of a moduli stabilizing
potential consider
for simplicity the
inherited Kahler potentials of the types discussed
above with only diagonal moduli.
These Kahler potentials may be written in the form
\beq
K= -\sum_{i=1}^n
p_i \ln \left( T_i+ T^{\dagger} _i
- \hbox{tr} {Q^{\dagger}_i Q _i}
-{\hbox{tr}\Sigma^{\dagger} _i \Sigma_i } \right),
\label{kahlerdef}
\eeq
where the sum of logarithms form
(\ref{kahlersumln}) is obtained for $n=3$ and $p_i=1$, while the no-scale
sequestered form (\ref{fivenoscale}) is obtained for $n=1$ and $p=3$.
Also assume that the moduli are stabilized by superpotential
interactions of the form
\beq
W= W(\Sigma) + W(T_i).
\label{supert}
\eeq
Together with the Kahler potential
this superpotential is assumed to
give a mass to both the real and imaginary components of $T_i$.
This may or may not require supersymmetry breaking.
In the Einstein frame
the scalar potential is given by
the standard formula
\beq
V = e^{K}(K^{i\bar{j}}F_i F^*_{\bar{j}} - 3 {\vert W \vert ^2})
= e^{K} \left(V_0(T_i,\Sigma_j,W)
+ \sum_i \lambda_i(T_j,W) Q^{\dagger} _i Q_i  + \cdots \right)
\label{softp}
\eeq
where in the second expression the dependence on the
visible scalars to second order in the fields has been
isolated, and
where the coefficients $\lambda_i$ are given below.
Up to a wavefunction renormalization the
coefficients of $Q^{\dagger}_i Q_i$ in the potential (\ref{softp})
are the scalar masses for those
fields.

There is an important subtlety here because the field
redefinition for $T_i$ (\ref{fieldredef}) also contains a dependence on
the scalars, and one might worry whether this also
contributes to their masses.
When the moduli are
massive this concern is spurious, and it
is straightforward
to see that this dependence does not
contribute to the scalar masses.
The worrisome field redefinition is
\beq
T_i= \langle T_i \rangle + \delta T_i
\eeq
where $\langle T_i \rangle$ is the expectation value of the
moduli at the minimum of the potential (if it exists), including
possible shifts due to supersymmetry breaking,
and
\beq
\delta T_i = g_i(\delta R_j) + i  a_i + {1 \over 2}
{Q^{\dagger}_i Q_i }
\label{dt}
\eeq
where $g_i$ is a polynomial function of the geometric moduli,
as for example in (\ref{fieldredef}) and (\ref{typeifd}).
Since moduli stabilization and a vanishing
cosmological
constant is assumed, the action for $T_i$ begins
at ${\cal O}((T_i -\langle T_i \rangle)^2)$. But
a contribution to
the scalar mass from the field redefinition
(\ref{dt})
could only come from
$(\partial V_0 / \partial T_j ) (T_j- \langle T_j \rangle)$ which vanishes.
Note that  it is the chiral
component $T_i$ which is stabilized,
instead of the geometric variable $R_i$. This is
the only assumption consistent with $N=1$ $d=4$ supersymmetry.
Thus to compute the soft mass with
a stabilizing potential for the moduli one may simply
freeze the moduli at their expectation values.
Using the usual supergravity
potential,
the Kahler potential in (\ref{kahlerdef}),
the superpotential in (\ref{supert}), and assuming
$\langle \Sigma_i \rangle$ is small compared with the
four--dimensional Planck scale,
then with stabilized moduli
the scalar masses are given by
\beq
m^2 _i = m^2 _{3/2} \left(1 - {2 \hbox{Re}T_i \over p^2 _i
} {\vert F_{\Sigma_i}
\vert ^2  \over \vert W \vert ^2  }
- {4 (\hbox{Re}T_i)^2 \over p^2_i} {\vert F_{T_i} \vert ^2
 \over \vert W \vert ^2 } \right) ~,
\label{smass2}
\eeq
where the gravitino mass in Einstein frame is
$m^2_{3/2} =  e^{K} \vert W \vert ^2 $, and where here
the unconventional notation
$F_I \equiv \partial_I  W + (\partial_I K ) W$ is employed
which differs from the standard notation by a factor of
$e^{K/2} K^{i \bar{j}}$.
Note that with vanishing auxillary component for the
moduli (\ref{smass2}) agrees with the formulae found in
Section 2.1 and Appendix B where for the Kahler potential
(\ref{kahlerdef}) the moduli
were simply replaced by their expectation
values.
But before concluding that the masses are non-vanishing when the
moduli are stabilized,
it is necessary to check that (\ref{smass2})
is not identically zero once the cosmological constant is canceled.
The condition for the cosmological constant to vanish is found to be
\beq
0 =
\sum_i \left( {2 \hbox{Re}T_i \over p_i} \vert F_{\Sigma_i}
\vert ^2 + {4 (\hbox{Re}T_i)^2 \over p_i}
{\vert F_{T_i} \vert ^2 }
\right)
-3 {\vert W \vert ^2  }
\label{cosmocan}
\eeq
and this is not in general equivalent to (\ref{smass2}).
However, there are some special situations in which
(\ref{cosmocan}) and (\ref{smass2}) equivalent, some of
which
have
been detailed in Section 2.1 and Appendix B.
By including an auxillary component for the diagonal
moduli the results presented here slightly extend the
discussion found in these other sections.
In the five--dimensional supergravity brane world model
there is only the single radion modulus, $T$, for which $p=3$.
In this case, the inherited no-scale Kahler potential without warping
gives vanishing visible sector scalar masses even for $F_T \neq 0$.
For three moduli $T_i$, with $i=1,2,3$, it is seen by inspection
that the scalar masses
continue to vanish if both the stabilization and
supersymmetry breaking preserve an $S_3$ symmetry.
For generic patterns of moduli values and
supersymmetry breaking on the distant brane,
however, the equation for the scalar
mass does not vanish when the cosmological constant vanishes.

Returning to the issue of moduli stabilization,
we note that the expression for the soft masses
above are
dramatically
modified if the moduli do not have any superpotential
interactions. In \cite{lutysundrum}
it was demonstrated that for the sequestered
Kahler potential the absence of superpotential
interactions for the radion implies that the
conformal compensator always vanishes.
It is not too difficult to extend their results to include
the more general no-scale Kahler potentials (\ref{kahlerdef}).
To see this, note that without stabilizing the moduli
there are
additional contributions to the soft masses from the term
$\partial V_0 / \partial T_j (T_j- \langle T_j \rangle)$.
Operationally it is then more convenient to
substitute the field redefinition (\ref{dt}) into the  scalar
potential. Now for instance, the inverse metric for the hidden sector
fields is independent of $Q^{\dagger} _i Q _i$. Visible scalar masses
only come from terms involving $F_{T_i}= -p_i W_0 /R_i$ where
$R_i= 2 \hbox{Re} \langle T_i \rangle$ with  $\langle T_i \rangle$
defined in
(\ref{dt}),
and terms involving $F_{Q_i} = p_i |W| Q^{*}/R_i$.
In this case with vanishing superpotential interactions for
the moduli and the no-scale Kahler potentials (\ref{kahlerdef}) the
scalar potential for the visible sector fields
to quadratic order is proportional to
\beq
 {1 \over p_i}
\left(\begin{array}{cc}
F_{T_i} & F_{Q_i} \nonumber \\
\end{array} \right)
 \left(
\begin{array}{cc}
R^2_i   + Q^{\dagger}_i Q_i R_i & Q^{\dagger}_i R_i  \nonumber \\
Q_i  R_i   & R_i  \nonumber \\
\end{array}
\right) \left( \begin{array}{c}
F_{T_i}^{\dagger} \nonumber \\
F^{\dagger} _{Q_i} \nonumber \\
\end{array} \right)
\eeq
where there is no sum on repeated indices.
Using the auxillary components given
above it is readily confirmed
that the soft scalar masses vanish. This demonstrates
that
without superpotential
interactions or additional Kahler corrections for the moduli,
general non-vanishing auxillary components
do not lead to
soft scalar masses.
But as argued in the
previous paragraph, these conclusions
do not apply when there are superpotential interactions
that stabilize the moduli.

Finally, one might wonder why (\ref{smass2}) does not
contain any contributions suppressed by
the radion masses. The moduli couple to both hidden and visible sectors,
and in integrating out these states one might expect
corrections to the Kahler potential
suppressed by these masses.
But the operators that are generated are not of
the form that would lead to tree-level masses after
supersymmetry breaking.
This was demonstrated
in a model with a single volume modulus \cite{lutysundrum}.
But here we see more generally that integrating out any number of
moduli in the low energy theory does not affect the form of the
tree-level visible sector masses.
Instead the operators which are generated
include a number of supersymmetric covariant derivatives
which, with hidden sector supersymmetry breaking,
do not contribute directly to scalar masses.
This form for the effective operators which are generated by
integrating out moduli
is readily confirmed by either a component or supergraph calculation.

\section{Gaugino Masses}
\label{sec:gauginomass}

Gaugino masses require the breaking of both
supersymmetry and $R$-symmetry.
Both these symmetries are broken by a gauge kinetic
function auxiliary expectation value
\beq
\int d^2 \theta ~ {\cal F} \left(S,T,{ \mu /  (\Phi M_{\rm reg}}) 
 \right) W^{\alpha} W_{\alpha},
\eq
where $\mu$ is the infrared renormalization scale.
The gauge kinetic function
${\cal F}(S,T,  \mu / (\Phi M_{\rm reg}))$ 
depends on the dilaton $S$, the various T
moduli, and at the loop-level
on the conformal compensator superfield,
$\Phi$, through the regulator mass scale $M_{\rm reg}$.
If hidden sector brane fields $\Sigma$
are charged under global or gauge symmetries
then the visible sector gauge kinetic function is at least bilinear
in these fields.
If the hidden sector scalar expectation values are small
compared with the fundamental Planck scale,
any hidden sector auxiliary expectation values give
rise directly to visible sector gaugino masses which
are suppressed by powers of hidden sector scalar expectation
values over the fundamental scale, and therefore unimportant.
In this case, only the auxiliary components for the
dilaton, $S$, moduli, $T$, or conformal compensator, $\Phi$,
can give important direct contributions to the visible sector
gaugino masses.


The ratio of gaugino to scalar masses depends crucially on
the origin of the auxiliary component within the gauge kinetic
function.
One-loop anomaly mediated contributions to the gaugino masses arise
from the conformal compensator auxiliary expectation
value, $F_{\Phi}$, which, as discussed in section \ref{section1},
can be induced indirectly by hidden sector supersymmetry breaking.
If supersymmetry breaking is isolated in matter fields on the
hidden sector brane, then as argued above, tree-level scalar
masses in general arise from interactions between
the branes, except in very special circumstances.
In the scenario with supersymmetry breaking isolated on the
hidden sector brane
the gaugino masses are then a loop factor
smaller than the scalar masses.
Although model dependent,
avoiding experimental bounds on gaugino masses of
${\cal O}(50-100~ \hbox{GeV})$ from
direct searches then generally requires
some tuning to obtain radiative electroweak symmetry breaking.
This occurs
because the scalar masses, including that of the scalar
Higgs, are somewhat larger than the electroweak scale as
implied by the large ratio between scalar and gaugino masses.
This is also the case for moduli dominated supersymmetry
breaking, $F_T \neq 0$, which gives rise to tree-level scalar
masses and one-loop gaugino masses through threshold effects.
Dilaton dominated supersymmetry breaking, $F_S \neq 0$, however,
gives rise to tree-level visible sector
masses for both gauginos and scalars.
{}From the criterion of avoiding tuning of electroweak symmetry
breaking by obtaining both scalar and gaugino masses at the
same order,
dilaton dominated supersymmetry breaking, or a combination
of dilaton, moduli, and hidden sector breaking, then seems most
natural.


\section{Gaugino Mediation}
\label{sec:gauginomed}

Gaugino mediation is a variant of BWSB
in which Standard Model gauge multiplets reside in
(at least a subspace of) the bulk, while
standard model quarks and leptons are confined to a visible sector
brane \cite{gauginoa,gauginob}.
Supersymmetry breaking takes place on a hidden sector brane
which is physically separated from the visible brane in a compact
manifold. 
Since the gauge field multiplets reside in the bulk of
the compact manifold and are in direct physical
contact with the hidden sector supersymmetry breaking brane,
the gauginos feel supersymmetry breaking directly
at tree level and obtain masses at the compactification
scale of order \cite{gauginoa}
\beq
m_{1/2}^2 \sim {F \over V_g} \sim
    { V^{1/2} \over V_g } m_{3/2}^2,
\eq
where $V_g$ is the volume of the subspace
within the compact manifold in which the
gauge field multiplets reside,
$F$ is the supersymmetry breaking auxiliary expectation
value on the hidden sector brane,
and from (\ref{gravitinomass}) the gravitino mass is
$m_{3/2} \sim F / V^{1/2}$.
Note that for $V_g \sim V$ the gaugino masses are parameterically
smaller than the gravitino mass.
If the Higgs multiplets also reside in the bulk and
are in contact with hidden sector brane,
Higgs sector mass squared parameters
receive direct tree level contributions at the compactification
scale of order
\beq
m_h^2 \sim {F^2 \over V_h} \sim {V \over V_h} m_{3/2}^2,
\eq
where $V_h$ is the volume of the subspace within the compact
manifold in which the Higgs multiplets reside.
The scalar squarks and sleptons confined on the
visible sector brane do not couple directly
to the hidden sector brane and therefore have been
argued to receive masses squared only
radiatively from bulk gauge multiplets at one loop \cite{gauginoa,gauginob}.
At the compactification scale these
contributions to scalar masses are
\beq
m^2 \sim {g_4^2 m_{1/2}^2 \over 16 \pi^2}
   \sim {V \over V_g^2} {g_4^2 m_{3/2}^2 \over 16 \pi^2 },
\label{gmedmass}
\eq
where $g_4$ are the four dimensional gauge couplings.
In order to obtain the observed values of the
four dimensional gauge couplings the compactification volume
should be not much larger than the fundamental scale.
For one extra dimension a compactification radius of order the
inverse GUT scale is generally assumed.

The gaugino mediated contributions (\ref{gmedmass}) to the
scalar squark and slepton radiative masses at the compactification
scale
are suppressed by a loop factor relative to the gaugino masses.
This leads to the expectation that,
from a low energy point of view,
gaugino mediation amounts effectively to
no-scale boundary conditions in which the squark and slepton masses
nearly vanish at the messenger scale while gaugino masses
and perhaps Higgs
sector parameters are non-vanishing \cite{gauginoa,gauginob}.
If this is the case, the dominant contribution to the squark
and slepton masses squared comes from renormalization group evolution
between the compactification
and electroweak scales.
This yields $m^2 \sim m_{1/2}^2$
at the electroweak scale.
Theses contributions do not violate flavor.

However, as discussed above,
with brane world supersymmetry breaking
scalar masses squared generically
arise at tree level from volume suppressed couplings
between the visible and hidden sector branes,
\beq
m^2 \sim {F^2 \over V} \sim m_{3/2}^2.
\label{treegaugino}
\eq
A specific realization of gaugino mediation has not
yet been presented in a string or M theory background.
String or M-theory models in which
the gauge groups arise from ADE singularities
on a sub-space with visible and hidden sector
matter fields localized at physically separated singularities
within this sub-space might give a realization of gaugino
mediation within a consistent theory which contains gravity.
The expectation
that tree-level scalar masses (\ref{treegaugino}) generically
arise should hold in such models.
These contributions are not suppressed compared
with the gravitino mass and parameterically dominate over the
gaugino mediated contributions (\ref{gmedmass}) by a loop
factor and at least one power of the volume.
In addition, as discussed above
these contributions are generally not universal.
So gaugino mediation through the bulk alone is not
sufficient to solve the supersymmetric sflavor problem.
Additional assumptions are required to solve the sflavor problem
such as flavor symmetries.
Dominance of the gaugino mediated contributions to scalar masses
might might arise if one had a
model such as
a 
flat five dimensional
realization of string or M-theory
with very small or vanishing tensions (so as to avoid warping
of the internal geometry). Dominance of gaugino
mediation would also occur in
models where the
volume of the subspace in which the gauge multiplets
propagate is much smaller than the volume of
the gravitating space.


\section{The Supersymmetric Flavor Problem}
\label{sec:flavor}

Virtual squark and slepton processes can in principle
generate low energy quark and lepton flavor violating processes
at levels much larger than allowed by current experimental bounds.
This seems to imply that sufficent structure exists in the squark
and slepton soft mass matrices to suppress these potentially
dangerous supersymmetric contributions to flavor violation.
Within any senario for transmitting supersymmetry breaking to the
squark and slepton fields it is then worth addressing what
assumptions, if any, are required to avoid excessive sflavor violation.

Within a given supersymmetry breaking mediation
scenario, if squarks and sleptons with the same gauge
quantum numbers acquire non-universal masses
at the messenger scale, then there is potentially
a sflavor problem.
This is because the squark and slepton
eigenvectors associated with the non-universal masses
define directions in flavor space.
If these are not aligned or proportional to the
eigenvectors defined by the quark and lepton masses
to a sufficient degree dangerous sflavor violation can arise.
Without additional assumptions about underlying flavor symmetries
this is of course not in general quaranteed.

In the string and M-theory BWSB models presented here,
such as the orbifold examples, fields with the same gauge
quantum numbers do generally acquire non-universal masses.
So without additional assumptions about flavor symmetries
in the underlying theory, BWSB scenarios alone do not appear to
be free of the supersymmetric flavor problem.
However, it is worth noting that, as discussed in section \ref{sec:puresimple},
if a direct compactification of string/M-theory
to pure five--dimensional supergravity
with end of the world branes exists, it would presumably
inherit at lowest order the no-scale form of the Kahler
potential for the untwisted states from the underlying $N=4$ Khaler potential,
and give vanishing tree-level masses.
This should probably be considered an interesting property of
a given model, rather than a general feature of BWSB.

In most of the models presented here
multiple copies of chiral matter arise as the
result of discrete or continuous internal geometric symmetries.
Copies of visible sector chiral matter with the same gauge quantum numbers
are by definition different flavor generations.
So in these models
flavor is closely related to internal geometric symmetries.
The appearance of soft masses which
do not commute with flavor can be traced to
the existence of fields in the bulk which in the underlying
theory necessarily also transform under these geometric symmetries.

>From the low energy effective field theory point of view it
might seem that including bulk fields which transform under flavor
is the only way in which to obtain non-universal masses.
However, as the models presented here also illustrate,
the soft masses depend not only on which
bulk fields are present, but also on the form of the
bulk--brane matter couplings
which are suppressed by the fundamental scale of theory.
>From an effective field theory point of view these couplings
are restricted only by four--dimensional $N=1$ supersymmetry
(and possibly anomaly cancellation).
So even with the minimal supergravity multiplet in the bulk,
an effective field theory analysis of BWSB requires additional
assumptions about flavor symmetries at the fundamental scale in order
to avoid dangerous sflavor violation in general.






\section{Conclusions}
\label{sec:conclusions}

Brane worlds provide an interesting scenario in which
to realize hidden sector supersymmetry breaking.
As discussed here, even though supersymmetry breaking
can be isolated on a hidden sector brane which is not in
direct physical contact with the visible sector brane,
unsuppressed tree-level squark and slepton masses are generally
obtained. This is illustrated in a number of models.
In those models with more than one generation the masses
are generically non-universal. In addition, these
models typically contain tachyons. These can be removed
by a projection, but this requires a correlation between
the projection and the pattern of hidden sector
supersymmetry breaking.

The origin of the non-derivative interactions which couple
the branes and gives rise to tree-level visible sector
masses is easy to understand
in the underlying theory as arising from exchange of bulk
supergravity fields between the branes.
It is important to note that for total space-time dimensions
greater than five, bulk fields in the minimal supergravity
multiplet are sufficient to give rise to non-derivative
unsuppressed brane--brane couplings even in flat backgrounds.
Higher dimensional supersymmetry guarantees the existence
of these bulk states which therefore can not simply be ignored by fiat.
In addition, corrections to the leading tree-level masses
determined by the form of the inherited Kahler potential are not
suppressed by additional powers of the compactification volume
as has been claimed and can be significant.
This is consistent with the expectation that the inherited
Kahler potential is not protected in any way in the low energy theory.

Since the tree-level soft masses are not generally degenerate,
additional assumptions about flavor symmetries and their breaking are
required to ensure proportionality or alignment
in order to avoid dangerous sflavor changing effects.
So contrary to previous expectations,
BWSB
alone does not give a solution to the supersymmetric
flavor problem.
In this sense the compactification model building and phenomenology of
BWSB is similar
to standard hidden sector supersymmetry breaking.
Since tree-level scalar masses are generic, brane world realizations
do not in general provide a robust rationale for anomaly
mediated supersymmetry breaking.
It may be possible, however, to construct specific models in which
tree-level masses vanish and anomaly mediation is important.
The most natural setting for this would seem to be
compactifications of a consistent microscopic theory
to pure five--dimensional supergravity with
end of the world branes, although none are known at present.
Another possibility is a Horava-Witten compactification of
M-theory on a Calabi-Yau fibration with $h^{1,1}=1$.
The lowest order inherited Kahler potential for these cases is
the sequestered no-scale form which gives vanishing tree-level
soft masses.
For more general compactifications in which the inherited
Kahler potential is the sum of logarithms form, unbroken
flavor symmetries in the hidden sector can also give vanishing
tree-level masses; but this of course requires additional
assumptions about flavor symmetries.
In either case, corrections to the inherited Kahler potential,
which generally give rise to tree-level soft masses,
would have to be tuned to be small by moving to some corner
of moduli space and/or ensuring that the brane tensions vanish,
presumably by some discrete symmetry.
On top of this, additional interactions have to be assumed
which lift the tachyonic right--handed slepton of anomaly
mediation (although these might be provided by
the corrections to the inherited Kahler potential -- however
in this case anomaly mediation does not give the dominant contribution
to scalar masses).
All of these model assumptions required to achieve
anomaly mediation could also likely be achieved in a
standard hidden sector supersymmetry scenario without
reference to a brane world picture.

Scenarios for transmitting supersymmetry breaking to
the visible sector seem most natural if the scalars and
gauginos receive masses which are of the same order
without any tuning of parameters.
In the brane world realization tree-level masses for
scalars are naturally obtained.
Tree-level masses for gauginos require an auxiliary component
for the dilaton.
So the most natural BWSB
scenario seems to be
either dilaton domination, or one in which the dilaton
acquires an auxiliary expectation value comparable
to those in the hidden sector.
Dilaton domination gives, at leading order, universal
tree-level contributions to visible
sector scalar masses.
However, even in this case which is flavor blind at lowest order,
as we have seen, flavor violating
corrections to the dilaton Kahler potential are not
necessarily small, in particular in the Horava-Witten
theory with standard gauge coupling unification.
So additional assumptions about flavor symmetries seem to
be required.
This also has implications for the standard dilaton
domination scenario in the language of weakly coupled string theory.
Phenomenological
analyses \cite{louisnir} have assumed that scalar degeneracy
and therefore alignment is not
likely to be better than ${\cal O} (\alpha_{\rm GUT})$.
This is just barely
large enough to explain the suppression of flavor violating
processes in the kaon system.
However, in the strongly coupled Horava-Witten limit
of heterotic string theory
the non-universal flavor violating couplings of the dilaton arise
classically at ${\cal O}(T/S)$.
On both phenomenological and theoretical
grounds, one expects that $T/S$ is not much smaller than unity.
So even though these corrections are non-perturbative from the
heterotic string point of view, they are not likely to be small
numerically.
And therefore again, additional assumptions about flavor symmetries
seem to be required.

In a top--down approach to models of nature, one could hope that
large classes of models or vacua within an underlying theory
might have generic features which can eventually be confronted
experimentally at low energies.
If so, then these classes, and perhaps ultimately the underlying theory,
can be considered to be predictive and testable.
Even though there are many problems that string/M-theory
can not address with our present level of understanding,
certain questions regarding the mechanism of supersymmetry
breaking and resulting patterns of squark and sleptons masses
can be addressed with present technology.
One might have hoped that BWSB vacua within string/M-theory
might have provided an example of a class of models with
universal generic features by predicting universal
flavor conserving squark and slepton masses as the result
of physically separating the hidden and visible sectors within
an internal manifold.
Unfortunately,
for a rather large class of models this does not appear to be the case.
The superpartner spectrum and magnitude of sflavor violation
appears to be very model dependent.
So just as for standard perturbative string theory with hidden
sector supersymmetry breaking, one is reduced to
presenting predictions (in principle) for
specific models rather than for broad classes.
The latter would clearly be preferable.
Perhaps future scenarios for realizing supersymmetry breaking
within string/M-theory will provide general patterns or features
which are predictive and testable.

\noindent

{\bf Acknowledgements:}

\noindent

We would like to thank Z. Chacko, C. Csaki, S. Kachru, and J. Polchinski
for discussions.
The work of A.A., M.D., and M.G. was
supported in part by a grant from the US
Department of Energy.
The work of S.T. was supported in part by the US National
Science Foundation under grants PHY98-70115 and PHY99-07949,
the Alfred P. Sloan Foundation, and Stanford University through
the Fredrick E. Terman Fellowship.


\section{Appendix A: M-theory Orbifold Compactifications and
their Inherited Kahler Potentials}
\label{app:orbifold}


On possibility for obtaining brane world models with $N=1$
supersymmetry from M-theory is to consider
Horava-Witten orbifold backgrounds $S^1 / Z_2 \times {\cal M}$
where ${\cal M}$ is a
six--dimensional $N=1$ preserving compact orbifold of $T^6$.
If the $S^1/Z_2$ interval length is large, $R_{11} \gsim \ell_{11}$,
a brane world background with end of the world branes results.
In the weakly coupled limit, $R_{11} \ll \ell_{11}$,
modular invariance of one-loop string amplitudes in the
resulting perturbative heterotic string theory
gives certain consistency conditions on ${\cal M}$ discussed
below, and also
in general requires
the existence of additional twisted states which reside
at fixed points of ${\cal M}$.
In the absence of an underlying theory of M-theory orbifolds,
we content ourselves with orbifold compactifications
of this type which are modular invariant in the weakly coupled
heterotic limit. We will assume that some of
the perturbative
heterotic models exist at strong coupling
with a geometric description.
Then for these models
the lowest order Kahler potential for the untwisted states
is inherited from the $N=4$ Kahler potential obtained from
compactification on $T^6$.
As discussed in section \ref{hw10d} a low energy supergravity analysis
implies that this lowest order inherited form
for the untwisted states
should receive small corrections in the brane world limit
as long as the compact volume is
large in eleven-dimensional Planck units.
Our discussion below applies to these classes of models,
again assuming they are consistent compactifications
of M-theory.

In an orbifold construction the states which survive in the low
energy theory are invariant under the orbifold action.
In general this action is non-trivial in both compact geometric
directions as well as in the gauge group of the underlying theory.
In addition, twisted states which reside at orbifold fixed
points also appear in the low energy theory.
For M-theory orbifold backgrounds $S^1/Z_2 \times {\cal M}$, a subset of
the $E_8 \times E_8^{\prime}$ M-theory
twisted gauge supermultiplets which reside on the end of the world
visible and hidden sector branes
survive in the low energy theory,
$Q_i \subset {\bf 248}_i \in E_8$ and
$\Sigma_i \subset {\bf 248}_i^{\prime} \in E_8^{\prime}$ respectively,
where $i=1,2,3$ labels the internal complex coordinates
of ${\cal M}$. From the weakly coupled heterotic string point of view these
fields are untwisted states, and will be referred to as such below.
The lowest order tree-level Kahler potential for these
states is inherited directly from the $N=4$ Kahler
potential (\ref{tdeteqn}) by simply removing non-invariant states.
There are in general additional twisted states in the low energy theory
which reside at fixed points of ${\cal M}$.
The dependence of the Kahler potential on these twisted states
is not restricted by extended symmetries since ${\cal M}$ preserves
only $N=1$ supersymmetry.
Here we focus only on the untwisted
$Q_i$ and $\Sigma_i$ visible and
hidden sector fields which do inherit a lowest order Kahler potential
from the ten--dimensional theory.


For simplicity we consider orbifold backgrounds ${\cal M}$ which are
symmetric Abelian orbifolds of $T^6$.
An Abelian orbifold group $\Gamma$ acts on
the three complex planes of $T^6$ by
\beq
z_i \rightarrow e^{2 \pi i r_i} z_i.
\eq
Modular invariance of one-loop string amplitudes
requires that \cite{polchinski}  
\beq
| \Gamma | \sum _i r_i =0 ~\hbox{mod} ~2
\label{modone}
\eq
where $| \Gamma |$ is the order of the orbifold group.
The existence of an unbroken $N=1$ supersymmetry requires that
\beq
 \sum _i r_i =0
\eq
for which the modular invariance condition (\ref{modone})
is then automatically satisfied.

To describe the action of $\Gamma$ in the gauge groups in the
present case, it is
convenient to consider the subgroup $SU(8) \subset SO(16) \subset E_8$
and likewise for $E_8^{\prime}$.
The actions on
$\lambda \in SU(8)$  and
$\lambda^{\prime} \in SU(8)^{\prime}$
are
\begin{eqnarray}
\lambda_a & \rightarrow e^{2 \pi i \beta_a} \lambda_a
  \nonumber  \\
\lambda_a^{\prime} & \rightarrow e^{2 \pi i \beta_a^{\prime} }
\lambda_a^{\prime}
  \label{orbgauge}
\end{eqnarray}
where $a=1,\dots,8$.
In the weak coupling limit, modular invariance requires that \cite{polchinski}
\beq
| \Gamma | \sum_a \beta_a = | \Gamma |
   \sum_a \beta^{\prime}_a = 0 ~\hbox{mod} ~2
\eq
and
\beq
| \Gamma | \left[  \sum_a
(\beta_a^2 + \beta_a^{\prime 2}) -
  \sum_i r_i^2 \right]  =  \left\{
\begin{array}{lll}
0 ~\hbox{mod} ~2 & &  | \Gamma |~ \hbox{even} \\
0 ~\hbox{mod} ~1 & &  | \Gamma |~ \hbox{odd} \\
\end{array}
\right.
\eq
The orbifold 
action (\ref{orbgauge})
breaks the gauge groups to a subgroup, but does not reduce
the rank for Abelian $\Gamma$.

One of the simplest orbifold constructions begins with
three $T^2$ tori, each of which
preserves a $Z_3$ symmetry $z_i \rightarrow e^{2 \pi i /3} z_i$.
A $Z_3$ orbifold twist consistent with this symmetry,
modular invariance, and preserving $N=1$ supersymmetry is
$r_i=(1,1,-2)/3$.
This orbifold leaves invariant an $S_3$ geometric symmetry
since the twist on each plane is actually identical in this
case, $z_i \rightarrow e^{2 \pi i /3} z_i$.
The so--called standard embedding involves also a gauge twist
by an element of $SU(3) \subset E_8$ given by
$\beta_a = r_i \delta_{ia}$, and
$\beta_a^{\prime} = 0$.
This leaves an unbroken subgroup $SU(3) \times E_6 \times E_8^{\prime}$.
The visible sector untwisted matter fields, $Q_i$,
which are invariant under the orbifold projection
transform as $( {\bf 3}, {\bf 27} )_i \in SU(3) \times E_6$
for $i=1,2,3$.
The untwisted sector then has 3 generations of
$( {\bf 3}, {\bf 27} ) \in SU(3) \times E_6$.
There are additional twisted states which cancel gauge anomalies.
For this orbifold there are no hidden sector matter fields
since all the $\Sigma_i$ are non-invariant.
So this particular orbifold
is not a useful model for hidden supersymmetry breaking.
However, because of the $S_3$ symmetry
it does provide a simple example in which
there are states in different
sectors $i \neq j$ with the same gauge quantum numbers.
So off--diagonal combinations of fields,
$Q_i Q_{\bar j}^{\dagger}$, are gauge invariant and
can appear in the Kahler potential.
The lowest order
tree-level Kahler potential inherited from (\ref{tdeteqn})
for the untwisted states of this orbifold is then \cite{polchinski}
\beq
K= -\ln \hbox{det}(T_{i \bar j} + T^{\dagger}_{i \bar j} -
\hbox{tr} Q_i Q_{\bar j}^{\dagger} )
-\ln(S + S^{\dagger} ),
\label{nohidden}
\eeq
where the trace is over the gauge quantum numbers.
Note that even though the orbifold only leaves invariant an
$S_3 \times U(1)_R \subset SU(4)$ $R$-symmetry, the inherited
Kahler potential possess an accidental $SU(3) \times U(1)_R$ global
symmetry since it depends only on two derivative terms
in the ten--dimensional theory.
Higher order corrections to the full Kahler potential
in the four--dimensional theory
would of course break the accidental
continuous flavor symmetry to the discrete $S_3 \subset SU(3)$.

In the perturbative string limit this
orbifold may be deformed to a smooth Calabi-Yau manifold by turning
on blow up modes which resolve the fixed points,
giving a Calabi-Yau with $h^{1,1}=36$ and $h^{2,1}=0$.
The blow up modes break the $SU(3)$.
There are in total 36 generations of ${\bf 27} \in E_6$,
of which 27 come from the twisted sector.
It would be useful to understand the M-theory geometric lift of this
compactification.

A variant of the above $Z_3$ orbifold
that does possess hidden sector matter
involves the same spacetime and visible sector gauge
twists, $r_i=(1,1,-2)/3$ and
$\beta_a  =r_i \delta_{ia}$, but with the hidden sector gauge
twist
$\beta^{\prime}_a = (2,2,2,0^5)/3$, where the exponent
indicates the multiplicity of the component.
This is modular invariant, preserves $N=1$ supersymmetry,
and has an $S_3$ symmetry.
This choice of gauge twists
is not unique, since the same spacetime twist
can be accompanied by other
modular invariant gauge twists in both the hidden
and visible sectors.
Orbifolds based upon this $Z_3$ spacetime twist
thus represent a class of models.
The form of the Kahler potential for the untwisted
states will be common to this class, and all
of these models will have in the untwisted
sector 3 (or 0) generations because of the $S^3$ symmetry.
For the choice of gauge twists above
the visible sector matter content
is identical to the previous example and in particular
the untwisted sector has 3 generations.
The hidden sector unbroken gauge group with this
gauge twist
is $E^{\prime} _6 \times SU(3) ^{\prime} \subset E_8^{\prime}$.
And in this case there is hidden sector matter
$({\bf 3},{\bf 27})_i \in
E^{\prime} _6 \times SU(3) ^{\prime} \subset E_8^{\prime}$ for $i=1,2,3$.
Because of the $S_3$ symmetry, off--diagonal combinations of
both visible and hidden sector fields, $Q_i Q_{\bar j}^{\dagger}$
and $\Sigma_i \Sigma_{\bar j}^{\dagger}$, are gauge invariant and
can appear in the Kahler potential.
The lowest order
tree-level inherited Kahler potential or supergravity $f$
function for the untwisted states
of this class of $Z_3$ orbifolds
is then just given by (\ref{tdeteqn})
or (\ref{nfourf}) respectively,
\beq
K = - \ln \det \left(T_{\bar{i} j }+ T^{\dagger}_{\bar{i} j}
- \hbox{tr} Q^{\dagger}_{\bar{i}} Q_j - \hbox{tr}
\Sigma^{\dagger}_{\bar{i}} \Sigma_j \right)
 -\ln(S + S^{\dagger} )
 \label{ztkahler}
\eeq
In this example each sector $i$ has only
one chiral matter field
appearing in the visible and hidden untwisted
sectors. For more generic choices
of the gauge twists, however,
there will appear in each sector $i$ states
with different gauge quantum numbers according to
which matter visible sector states
$Q_i \subset {\bf 248}_i \in E_8$ and hidden sector states
$\Sigma_i \subset {\bf 248}_i^{\prime} \in E_8^{\prime}$
survive the orbifold projection. In these more
generic examples the trace appearing inside the determinant
in the inherited Kahler potential includes a sum over these states.
But we remphasize that the form of the
Kahler potential and
the existence of
3 (or 0) generations is determined by the spacetime
twist, with the choice of gauge twist
determining the representation
content.

The Kahler potential (\ref{ztkahler}) is not of the sequestered form and
has non-derivative interactions between the visible and hidden sector
branes.
It gives rise to tree-level masses from hidden sector supersymmetry breaking
as discussed in Appendix B.
However, the lowest order
visible sector scalar masses arising from hidden sector
auxiliary expectation values with classes of orbifolds
such as the ones given above
which preserve an $S_3$ geometric symmetry inherit the
$\hbox{Tr}~m^2=0$ sum rule from the $N=4$ Kahler potential
discussed in Appendix B.
The associated visible sector tree-level tachyon(s) imply that
such $S_3$ symmetric orbifolds can not be phenomenologically
viable in the absence of large corrections which would
lift the tree-level tachyon(s).
Even though this particular class of orbifolds are not realistic
for the reasons given above, they do illustrate that the
sequestered intuition is not generally valid for brane world
models which preserve $N=1$ supersymmetry.

It is also possible to consider orbifolds that reduce the
form of the inherited
Kahler potential to a sum of logarithms rather than a
logarithm of a determinant.
This is illustrated in a class of
$Z_6$ orbifolds with the
space-time twist $r_i=(1,2,-3)/6$.
This is modular invariant, preserves $N=1$ supersymmetry,
but is not invariant under any discrete symmetry which
interchanges the three planes.
Because there is no geometric symmetry between the planes,
only the diagonal moduli, $T_i \equiv T_{i \bar i}$ survive
the orbifold projection.
For the gauge twists, consider the case
$\beta_a = (3^2,2,1^2,0^3)/6$ and $\beta^{\prime}_a = (3^2,2,1^4,0)/6$,
where again the exponents indicate the multiplicity of the component.
These twists satisfy the modular invariant conditions.
This choice of gauge twists
is not unique, since the same spacetime twist
can be accompanied by other
modular invariant gauge twists in both the hidden
and visible sectors.
Orbifolds based upon this $Z_6$ spacetime twist
thus represent a class of models.
With the gauge twists given above the unbroken gauge groups are
$SU(4) \times SU(2) \times SO(4) \times U(1)^2 \subset E_8$
and $SU(4)^{\prime} \times SO(4)^{\prime}
\times U(1)^{\prime 3} \subset E_8^{\prime}$.
With these twists there is matter in both the visible
and hidden sector since some of the $Q_i$ and $\Sigma_i$
are invariant.
However, because there is no discrete geometric symmetry,
off--diagonal combinations of fields $Q_i Q^{\dagger}_j$ and
$\Sigma_i \Sigma_i^{\dagger}$ are not guaranteed to be gauge
invariant since states in different $i \neq j$ sectors
generally have different gauge quantum numbers.
In fact no gauge invariant off--diagonal terms exist
if a given representation under the unbroken
subgroup
arises only once. This is always the case
if the representations in the low energy theory
are obtained by projecting an adjoint representation
of a larger group in the higher dimensional theory.
One may explicitly verify this, although it is important
to keep track of all the $U(1)$ charges to verify this.
In this model the untwisted sector has many different
gauge quantum numbers but only one generation of each.

The lowest order
tree-level Kahler potential for the untwisted states of the
class of $Z_6$ orbifolds given above
without gauge invariant off-diagonal
combinations of fields inherited from (\ref{tdeteqn}) is then
\beq
K = - \sum^3 _{i=1} \ln(T_i + T_i^{\dagger}
- \hbox{tr} Q_{i}^{\dagger} Q_{i} -
  \hbox{tr} \Sigma_{i}^{\dagger} \Sigma_{i})
 - \ln (S + S^{\dagger}),
\label{appkahlersimp}
\eeq
where the traces are over both gauge quantum numbers
and are different for each $i$.
This Kahler potential is invariant only under $U(1)_R$ since
the visible and hidden sector
representations and multiplicities are distinct for each $i$.
This Kahler potential
is not of the sequestered form.
The non-derivative couplings between the visible and
hidden sector branes implied by (\ref{appkahlersimp})
give rise to unsuppressed
tree-level scalar masses as discussed in the Appendix B.

The previous class of examples has many untwisted
states, but due to the spacetime twist
there is only one generation in the untwisted
sector for each gauge quantum number.
Next we
consider a class of orbifolds that is a
hybrid of the previous two examples such that
in the untwisted
sector there are two generations.
This is illustrated
in a $Z_6$ orbifold with spacetime twist
$r_i=(1,1,-2)/6$, which is
modular invariant and preserves $N=1$.
In addition we consider gauge twists in the hidden
and visible sectors chosen to be modular invariant and provide untwisted
hidden and visible sector chiral matter. Since
the spacetime twist preserves a $S_2$ permutation
symmetry, states in the untwisted sector
with $Q_{i}$ for $i=1,2$ are guaranteed
to have the same gauge quantum numbers.
This class of models then has two generations in the
untwisted sector.
Consequently,
off-diagonal contributions $Q^{\dagger} _i Q_{\bar{j}}$
for $i,j=1,2$ are guaranteed to be gauge invariant.
The lowest order tree-level
Kahler potential for the untwisted states
inherited from (\ref{tdeteqn}) is then
\beq
K = - \ln \det_{i=1,2} \left(T_{i \bar{j}}+
T^{\dagger}_{i \bar{j}}  -
\hbox{tr} Q^{\dagger} _i Q_{\bar{j}} - \hbox{tr} \Sigma^{\dagger} _i
\Sigma _{\bar{j}} \right)
- \ln \left(T_{3 \bar{3}} + T^{\dagger}_{3 \bar{3}} -
\hbox{tr} Q^{\dagger} _3 Q_{\bar{3}} -
   \hbox{tr} \Sigma^{\dagger}_3
 \Sigma_3
\right)  ~.
\eeq
This is also not of the sequestered form.
The soft mass spectrum for the two generations
with hidden sector supersymmetry
breaking is generically non-universal, as discussed in Appendix B.


\section{Appendix B: Soft
Scalar Masses from Inherited Kahler
Potentials}
\label{app:inherit}

The spectrum of visible sector soft scalar masses arising from
hidden sector supersymmetry breaking depends on the four--dimensional
Kahler potential.
For any theory geometrically
embedded in a higher dimensional supersymmetric theory,
the lowest order tree-level soft masses depend on the form
of the Kahler potential inherited from the underlying theory which
necessarily has extended supersymmetry.
In many cases the
lowest order inherited tree-level Kahler metric for the
matter fields is
either quaternionic if the underlying
microscopic theory has 8 supercharges or
flat if it has 16 supercharges.
At lowest order the inherited Kahler potential for
the untwisted states in an orbifolded theory is obtained
directly from that of the underlying theory with extended
supersymmetry by simply truncating to the light fields which
survive in the four--dimensional theory,
as discussed in Appendix A.

The tree-level spectra of visible sector soft masses arising from
the inherited Kahler potentials possess some
special features
as described below.
In general, unacceptable
visible sector tachyons arise with hidden sector supersymmetry
breaking, but are absent in certain classes of compactifications.
These classes however depend on the pattern of
hidden sector supersymmetry breaking, and cannot be chosen a priori.
In models with more than one generation in the
untwisted sector the soft masses are generically
non-universal.
For definiteness we focus on microscopic
theories which have the matter content of a weakly coupled
string theory in ten dimensions in a background
which preserves at lowest order 16 supercharges.
However, since the form of the inherited Kahler potential only
depends on the existence of extended supersymmetries of
the underlying background, the results are more generally applicable,
to, for example, the untwisted states
appearing in Horava-Witten orbifold backgrounds or D-brane
theories on an orbifold.

Consider first the case in which
all the visible and hidden sector
fields, geometric moduli, and dilaton survive in the four--dimensional
theory.
While this is not a realistic compactification since the
surviving states form $N=4$ multiplets, it is instructive in
illustrating the origin of relations among the visible sector
scalar spectrum in more realistic examples discussed below
with fewer surviving states.
The low-energy four--dimensional theory at the level of two-derivative
terms has an $SU(4)$ global symmetry
which is inherited from the $R$-symmetry of the microscopic theory.
The fermions in both the visible and hidden sector transform
as ${\bf 4} \in SU(4)$, and the scalars as ${\bf 6} \in SU(4)$.
In the $N=1$ low energy description only an
$SU(3) \times U(1)_R \subset SU(4)$ is manifest.
Under this subgroup the visible and hidden sector $N=1$
chiral multiplets transform as complex
${\bf 3}_+  \in SU(3) \times U(1)_R$,
the $T_{i \bar j}$ moduli chiral multiplets transform as complex
${\bf 8}_0 \oplus {\bf 1}_0 \in SU(3) \times U(1)_R$,
the $T_{ij}$ moduli chiral multiplets transform as
${\bf 6} \oplus \overline{\bf 3}$,
and the dilaton chiral multiplet
is invariant.
In the $N=1$ description the $SU(3)$ subgroup is a manifest
global flavor symmetry of the low energy theory at the two--derivative
level.
The four--dimensional inherited Kahler potential, in this case,
is fixed by the $N=4$ supersymmetry to be
\cite{kahlera,kahlerb,polchinski}
\beq
K= -\ln \hbox{det}(T_{i \bar j} + T^{\dagger}_{i \bar j} -
\hbox{tr} Q_i Q_{\bar j}^{\dagger} -
\hbox{tr} \Sigma_i \Sigma_{\bar j}^{\dagger}) -\ln(S^{\dagger} +
S),
\label{nfourkahler}
\eeq
where the traces are over the visible and hidden sector gauge
groups respectively,
$i$ and $\bar j$ are global $SU(3)$ flavor
fundamental and anti-fundamental indices respectively,
and the dependence on the $T_{ij}$ moduli is suppressed.
Note that
this is manifestly
invariant under the $SU(3) \times U(1)_R$ global flavor symmetry.

In order to exhibit a simple expression for the visible
sector scalar masses arising from the Kahler potential
(\ref{nfourkahler}) consider the case in which
only diagonal scalar moduli acquire an expectation value,
$ \langle T_{i \bar i} \rangle \neq 0$, while
the off--diagonal moduli vanish,
$\langle T_{i \bar j} \rangle  = \langle T_{i j} \rangle =0$.
If all the $T_{i \bar j}$ moduli are stabilized by
superpotential interactions and have vanishing auxiliary
components, $\langle F_{T_{i \bar j}} \rangle =0$, the visible sector scalar
masses arising from hidden sector auxiliary components,
$F_i \equiv \langle F_{\Sigma_i} \rangle $,
are easily computed from the
relevant terms in the $N=1$ supergravity potential
\beq
V \supset \hbox{exp}(K)
\left( F_i K^{i \bar j} F_{\bar j} - 3 |W|^2 \right),
\eq
where $K_{i \bar j} \equiv \partial_i \partial_{\bar j} K$,
$K^{i \bar j}$ is the inverse metric, and
$F_i \equiv \partial_i W + K_i W$.
The canonically normalized
visible sector scalar mass squared matrix
in the $3 \times 3$ $SU(3)$ flavor space,
in this case, is
\beq
m^2_{i \bar{j}} ={m ^2_{3/2} \over | W|^2} \left( | W|^2 \delta _{i \bar{j}}
- 2 \sqrt{\hbox{Re}T_{i} \hbox{Re}T_{j}}
F_{i} F^*_{\bar j} \right)
\label{appmassmatrix}
\eeq
with no sum on repeated indices, and
where $2 \hbox{Re} T_{i} \equiv
  \langle T_{i \bar i} + T_{i \bar i}^{\dagger} \rangle$,
and $F_i F_{\bar j}^* \equiv \hbox{tr} F_i F_{\bar j}^*$ for
each $i \bar j$.
In deriving this expression the cosmological constant
is assumed to vanish by the relation
\beq
\sum_i 2 \hbox{Re}T_{i} |F^2_{i} | -3 |W|^2 =0
\label{appcc}
\eeq
and the gravitino mass squared with this condition is
\beq
m^2_{3/2} \equiv \hbox{exp}(K) \vert W \vert ^2
\eq
The eigenvalues of (\ref{appmassmatrix})
are in general non-vanishing and non-degenerate.
The trace of the mass squared matrix (\ref{appmassmatrix})
vanishes,
\beq
\hbox{Tr} ~m^2 =0,
\eeq
as the result of the vanishing cosmological
constant condition (\ref{appcc})
and $SU(3)$ flavor invariance which implies that the
$Q_i$ multiplicities in the visible
sector are identical for each $i$, and likewise for
the $\Sigma_i$ in the hidden sector.
Note that the Tr is only over visible sector scalars.
This of course implies that there is at least one
visible sector tree-level tachyon for any non-vanishing
hidden sector auxiliary component.

The form of the visible sector scalar spectrum arising from
the Kahler potential (\ref{nfourkahler})
is illustrated in the simplest case
where only one of the hidden sector fields
has an auxiliary expectation value.
The mass squared eigenvalues for
$F_{1} \neq 0$, $F_{2} = F_{3} =0$,
but still allowing for arbitrary $T_{i \bar{i}}$,
are
\beq
m^2_{Q_i} = m^2_{3/2} (-2,1,1).
\label{eigensimp}
\eeq
This pattern of breaking preserves a $SU(2) \subset SU(3)$
flavor symmetry which enforces the two-fold degeneracy
of the eigenvalues.
Another simple case is the situation in which hidden
sector fields acquire a $SU(3)$ diagonal auxiliary component,
$F_{i} = F$, and that the moduli have an $SU(3)$ diagonal
value, $T_{i \bar i} = T$ and are stabilized with vanishing
auxillary components.
Then from (\ref{appmassmatrix}) and (\ref{appcc}) the visible sector mass
squared matrix, in this case, is
\beq
m^2_{i \bar{j}} =
m^2_{3/2} \left( \begin{array}{ccc}
$0$ & $-1$ & $-1$ \nonumber \\
$-1$ & $0$ & $-1$ \nonumber \\
$-1$ & $-1$ & $0$ \\
\end{array}
\right) ~.
\label{app2n4mass2}
\eeq
This $S_3 \subset SU(3)$ symmetric matrix has
eigenvectors which transform as ${\bf 2} \oplus {\bf 1} \in S_3$
with the same eigenvalues as (\ref{eigensimp}).
This may be understood by
noting that the $SU(3)$ flavor symmetry of
the Kahler potential (\ref{nfourkahler}) may be used
to rotate the three $F_i$ auxiliary components into one component,
$F_{1}$ say,
which is the case previously discussed just above.


Consider next the case in which the only states
which survive compactification in the low-energy four--dimensional theory
are the diagonal geometric moduli, $T_i \equiv T_{i \bar i}$,
the dilaton, $S$, and some of the visible and hidden sector fields,
$Q_i$ and $\Sigma_i$.
In addition, restrict attention to cases in which
the off--diagonal combinations of visible and hidden
sector fields, $Q_i Q_{\bar j}^{\dagger}$ and
$\Sigma_i \Sigma_{\bar j}^{\dagger}$, are not gauge invariant,
and therefore do not appear in the Kahler potential.
This situation arises in some of the orbifold compactifications
discussed in section \ref{sec:hwsimple} and Appendix A.
The lowest order tree-level
Kahler potential inherited from (\ref{nfourkahler}) for this
truncated set of states is
\beq
K = - \sum_{i=1}^3 \ln(T_{i} + T_{\bar i}^{\dagger}
- {\rm tr}Q_{i} Q_{\bar i}^{\dagger} -
  {\rm tr}\Sigma_{i} \Sigma_{\bar i}^{\dagger} )
   -\ln(S^{\dagger} + S)
\label{appdiagkahler}
\eeq
where again the traces are over visible and hidden sector gauge
groups respectively.
Note that for a generic compactification
the visible and hidden sector gauge groups, matter
representations, and multiplicities are not necessarily the same.
So the Kahler potential is then
only invariant under a $U(1)_R \subset SU(4)$.
Allowing for arbitrary hidden sector
auxiliary components $F_i \equiv F_{\Sigma_i}$,
the canonically normalized visible sector scalar masses squared
arising from (\ref{appdiagkahler}) with vanishing cosmological
constant and stabilized moduli with vanishing auxillary
components are
\beq
m^2 _i = {m^2_{3/2} \over | W|^2} ( |W|^2 - 2 \hbox{Re} T_i |F_i|^2 )
\label{appmassdiag}
\eeq
with no sum on the repeated index and
where 
$|F_i|^2 \equiv {\rm tr}F_i^*F_i$
is the sum of the hidden sector auxiliary components squared
for given $i$.
The scalar masses (\ref{appmassdiag}) do not vanish in general.
Note that all the visible sector fields $Q_i$ with the same $i$ are
degenerate with mass squared $m^2_i$,
but are not generically degenerate with the $Q_j$ fields
for $i \neq j$.
In fact, in this case,
there are just three possible eigenvalues, $m^2_{i=1,2,3}$.
The multiplicities of each eigenvalue, however,
will not be the same in general
since the multiplicities of the visible sector fields $Q_i$
are not necessarily the same for each $i$.
Assuming there is at least one matter field $Q_i$ for
each $i$ -- that is, there are no empty sectors --
a sum rule for these masses similar to that found in the
$N=4$ case discussed above in which all fields survive
is obtained.
In this case, the sum of the three eigenvalues (\ref{appmassdiag})
vanishes as the result of the vanishing cosmological
constant condition (\ref{appcc})
\beq
 m^2_1 + m^2_2 + m^2_3 =0 ~.
\label{appmasstrace}
\eeq
However, since the multiplicities for each $i$ are not necessarily
the same, $\hbox{Tr}~m^2 \neq 0$ in general.
The condition (\ref{appmasstrace}) implies that in
this case there are also generically tree-level tachyons in the visible
sector.
Tachyons are avoided in compactifications in which all visible
sector matter
fields $Q_i$ in the $i$-th sector(s) with negative mass squared
eigenvalue(s) are projected out of the low-energy theory.
Since tachyonic states occur in $i$-th sectors
where $\Sigma_i$ has an auxillary component
$| F_{\Sigma_i} |^2 > |W|^2 /(2 \hbox{Re}T_i)$,
avoiding tachyons requires a correlation between the orbifold projection
and the
direction of supersymmetry breaking. This could
be achieved in an orbifold compactification where
the projections in
the visible sector and hidden sector are
anti-correlated. That is, empty visible
sectors $i$ correspond to non-empty hidden sectors
and vice versa.

A special case of the condition (\ref{appmasstrace})
resulting from the Kahler potential (\ref{appdiagkahler})
is obtained for hidden
sector fields which acquire diagonal auxiliary components,
$F_{i} = F$, and scalar moduli which have an $S_3$ symmetric expectation
value, $T_{i} = T$.
In this case, the vanishing cosmological constant condition
(\ref{appcc}) reduces to
$2 \hbox{Re} T |F|^2  - | W |^2 =0$.
This implies that all the scalar masses squared
eigenvalues (\ref{appmassdiag}) vanish,
\beq
m_i^2 =0 ~.
\eeq
This differs from the spectrum (\ref{eigensimp}) obtained from
the Kahler potential (\ref{nfourkahler}) with the same diagonal moduli
and auxiliary components.
The difference traces to the fact that unlike (\ref{nfourkahler}),
the Kahler potential (\ref{appdiagkahler}) does
not preserve a $SU(3)$ flavor symmetry.
The hidden sector auxiliary expectation values can
therefore not be rotated into a single flavor.
For the case of diagonal hidden sector auxiliary components
the mass eigenvalues (\ref{appmassdiag}) are necessarily identical.
The condition (\ref{appmasstrace}) then in turn implies that the scalar masses
in fact vanish in this case.
The lowest order vanishing of the visible sector masses
may also be understood as a remnant of the underlying
$N=4$ Kahler potential (\ref{nfourkahler}) since the diagonal elements
of (\ref{app2n4mass2}) vanish, and by assumption the off-diagonal elements
have been projected out of the low-energy four--dimensional theory.

Consider next the case where the only states that survive
in the low enery theory are moduli $T_{i \bar{j}}$ for $i=1,2$,
$T_{3 \bar{3}}$, and visible and hidden sector chiral matter
$Q_{i}$, $\Sigma_{i}$. Further, suppose there is a
$S_2$ permutation symmetry
for $i=1,2$, so that the low-energy theory has two generations
of $Q_{i}$ and $\Sigma_{i}$ for $i=1,2$,
and states $Q_3$ and $\Sigma_3$
which have different quantum numbers from the first two generations.
The visible and hidden sector gauge groups do not have to be
identical in general.
A class of orbifold examples leading to this type of matter
content is presented in the Appendix A.
The lowest order tree-level Kahler potential
inherited from (\ref{nfourkahler}) by this set of states is
\beq
K = - \ln \det_{i=1,2} \left(T_{i \bar{j}}+
T^{\dagger}_{i \bar{j}}  -
\hbox{tr} Q^{\dagger} _i Q_{\bar{j}}
- \hbox{tr} \Sigma^{\dagger} _i
\Sigma _{\bar{j}} \right)
- \ln \left(T_{3 \bar{3}} + T^{\dagger}_{3 \bar{3}} -
\hbox{tr} Q^{\dagger} _3 Q_{\bar{3}} - \hbox{tr}
\Sigma^{\dagger} _3 \Sigma_3
\right)  ~.
\eeq
Assuming that only $T_{i \bar{i}}$ acquire expectation values
and that all the $T_{i \bar{j}}$ moduli
are stabilized with vanishing auxillary components,
then with
hidden sector supersymmetry breaking
the soft masses are
\beq
m^2_{i \bar{j}}=
m ^2_{3/2} \left( \delta _{i \bar{j}}
- 2 \sqrt{\hbox{Re}T_{i\bar{i}} \hbox{Re}T_{j\bar{j}}}
{F_{i} F^*_{\bar j} \over |W|^2} \right)
\eeq
for $i,j=1,2$, and
\beq
m^2_3 = m^2_{3/2} (1-2 \hbox{Re}(T_3)
{|F_{\Sigma_3}|^2 \over |W|^2 }) ~.
\eeq
The mass eigenvalues satisfy the sum rule ${\rm Tr}~m^2=0$,
and are given by
\beq
m^2= m^2_{3/2}(1,-2+x_3,1-x_3)
\eeq
where $x_3 \equiv 2 \hbox{Re}T_3 |F_{\Sigma_3}|^2/|W|^2$ is
in the range $0 \leq x_3 \leq 3$.
It is important
to note that the first two states have the same
gauge quantum numbers. Inspecting the mass eigenvalues
indicates that with
hidden sector supersymmetry breaking the
two generations are not degenerate unless hidden sector
supersymmetry breaking is isolated in the third hidden generation.
If the visible sector soft masses are non-vanishing one of the scalars in the
first two generations is tachyonic.

As a final example
consider the restrictive class of examples
in which visible and hidden sector gauge groups,
matter representations, and multiplicities are the same for
each $i$ sector.
The inherited Kahler potential is then a special case of the
sum of logarithms form (\ref{appdiagkahler}) given above.
Each logarithm term in the sum in the Kahler potential
is then identical.
The Kahler potential (\ref{appdiagkahler}) is then invariant
under a $S_3 \times U(1)_R \subset SU(4)$ global symmetry.
If the discrete $S_3$ symmetry is gauged, then in addition
to the dilaton, only a single overall volume modulus
$T = {1 \over 3} \sum_i T_i$, and a single set of visible and hidden sector matter
fields, $Q =  {1 \over 3} \sum_i Q_i$ and
$\Sigma =  {1 \over 3} \sum_i \Sigma_i$ survive in the
truncated theory.
The lowest order tree-level
Kahler potential for the untwisted
states inherited from (\ref{nfourkahler})
in this case is
\beq
K = - 3 \ln(T + T^{\dagger}
- {\rm tr}Q Q^{\dagger} -
  {\rm tr}\Sigma \Sigma^{\dagger} )
   -\ln(S^{\dagger} + S).
\label{appnoscale}
\eeq
This is the sequestered no-scale Kahler potential.
It follows from the special diagonal case
of the sum of logarithm form discussed above
that non-vanishing
hidden sector auxiliary components, $F_{\Sigma} \neq 0$,
give rise to vanishing tree-level visible sector scalar masses
\beq
m^2 =0.
\eeq
The lowest order vanishing of the visible sector soft masses,
in this case, is also a result of inheritance from the
$N=4$ Kahler potential (\ref{nfourkahler}) of the underlying
microscopic theory since only the $S_3$ symmetric combination
of the diagonal elements of the mass squared matrix
(\ref{app2n4mass2}) remain in the low-energy four--dimensional theory.

The lowest order tree-level
visible sector scalar spectra which arises from the inherited
Kahler potentials discussed above have a number of interesting
features.
For generic compactifications and hidden sector supersymmetry breaking,
visible sector tachyons in general arise at lowest order.
In the absence of large corrections which lift
these phenomenologically unacceptable tachyons,
specific compactifications may project out these dangerous visible sector
states.
For orbifold compactifications of string theories in ten dimensions
this generally requires completely projecting visible
sector matter fields out of one or two planes of the six--dimensional
internal manifold. In addition, the supersymmetry
breaking must occur in hidden sector fields with internal
components along the same planes in order to not give
tachyonic masses to the remaining visible sector states.
This provides an important previously overlooked
criterion for compactification model builders.
Another feature is that visible sector scalars are not
generically degenerate. This was illustrated in models
with 2 and 3 generations.
In some cases degeneracy does result from hidden sector auxiliary
expectation values which are invariant under an unbroken flavor symmetry.
But, in this case, it is the unbroken
flavor symmetry which of course enforces
degeneracy, rather than the specific form of the inherited
Kahler potential.
In the no-scale sequestered case, degenerate, and in fact vanishing,
lowest order scalar masses do result.
But, in this case, degeneracy follows simply from the fact that
by definition only
flavor singlet states survive in the low-energy theory.
Note also that the lowest order no-scale Kahler potential arises in only
a very restrictive and special class of backgrounds,
and is therefore not a generic feature of standard compactifications
or brane world scenarios.

The tree-level sum rules for scalar masses squared for the
inherited Kahler potentials can be traced to the fact that
the Kahler metric of the underlying parent theory with 16 supersymmetries
is necessarily flat.
In more general examples in which the relevant underlying microscopic
theory has only 8 supersymmetries the parent quaternionic Kahler metric
would imply less restrictive sum rules for the inherited Kahler potentials.

Finally, it is important to note that the inherited form of the
Kahler potentials are only valid at lowest order tree-level in the truncated
low-energy theory and are not protected in any way from corrections.
Properly integrating out heavy states in general modifies
the full Kahler potential of the low-energy theory.
This is true of both standard compactifications in which quantum
corrections can be important, as well as
brane world scenarios in which tree-level bulk interactions between branes
can be important.

\end{document}